\documentclass{aa}
\usepackage{textcomp}
\usepackage{graphicx}
\usepackage{amsmath}
\usepackage{amssymb}
\usepackage{bm}
\usepackage{upgreek}
\usepackage{IEEEtrantools}
\usepackage{multirow}
\usepackage[dvipsnames]{xcolor}
\usepackage[colorlinks=true,allcolors=Blue]{hyperref}
\usepackage{txfonts}
\usepackage{multirow}
\usepackage[normalem]{ulem} 
\usepackage{longtable}

\newcommand{\sect}[1]{\text{Sect.~\ref{#1}}}
\newcommand{\fig}[1]{\text{Fig.~\ref{#1}}}
\newcommand{\tab}[1]{\text{Table~\ref{#1}}}

\newcommand{\code}[1]{\texttt{#1}}
\newcommand{\multitd}{\code{Multi3D}}
\newcommand{\scate}{\code{SCATE}}

\newcommand{\balder}{\code{Balder}}
\newcommand{\blue}{\code{Blue}}
\newcommand{\mtd}{\textlangle3D\textrangle}
\newcommand{\marcs}{\code{MARCS}}

\newcommand{\cobold}{\code{CO$^{5}$BOLD}}
\newcommand{\stagger}{\code{Stagger}}
\newcommand{\convec}{\code{Convec}}
\newcommand{\atmo}{\code{ATMO}}
\newcommand{\bifrost}{\code{Bifrost}}

\newcommand{\teff}{T_{\mathrm{eff}}}

\newcommand{\lggf}{\log{gf}}

\newcommand{\lgeps}[1]{\log{\epsilon_{\mathrm{#1}}}}

\newcommand{\dex}{\mathrm{dex}}

\newcommand{\nm}{\mathrm{nm}}

\begin{document} 

\title{The chemical make-up of the Sun: A 2020 vision}
\author{
M.~Asplund\thanks{\email{martin.b.asplund@gmail.com}} 
\and
A.~M.~Amarsi\inst{1}
\and
N.~Grevesse\inst{2,3}
}
\institute{
Theoretical Astrophysics, Department of Physics and Astronomy, 
Uppsala University, Box 516, SE-751 20 Uppsala, Sweden 
\and
Centre Spatial de Li\`ege, Universit\'e de Li\'ege, avenue Pr\'e Aily, 4031 Angleur-Li\`ege, Belgium
\and
Space Sciences, Technologies and Astrophysics Research (STAR)
Institute, 
Universit\'e de Li\`ege, All\'ee du 6 ao\^ut, 17, B5C, 4000 Li\`ege, Belgium 
}

   \date{Received: January 29, 2021; accepted: May 3, 2021}
   
\abstract
   {The chemical composition of the Sun is a fundamental yardstick in astronomy, relative to which essentially all cosmic objects are referenced. As such, having accurate knowledge of the solar elemental abundances is crucial for an extremely broad range of topics. 
   }
   {We reassess the solar abundances of all 83 long-lived elements, using highly realistic solar modelling and state-of-the-art spectroscopic analysis techniques coupled with the best available atomic data and observations.  
   }
   {The basis for our solar spectroscopic analysis is a 3D radiative-hydrodynamical model of the solar surface convection and atmosphere, which reproduces the full arsenal of key observational diagnostics. New complete and comprehensive three-dimensional (3D) spectral line formation calculations taking into account of departures from local thermodynamic equilibrium (non-LTE) are presented for Na, Mg, K, Ca, and Fe using comprehensive model atoms with reliable radiative and collisional data. Our newly derived abundances for C, N, and O are based on a 3D non-LTE analysis of permitted and forbidden atomic lines as well as 3D LTE calculations for a total of 879 molecular transitions of CH, C$_2$, CO, NH, CN, and OH. Previous 3D-based calculations for another 50 elements are re-evaluated based on updated atomic data, a stringent selection of lines, improved consideration of blends, and new non-LTE calculations available in the literature. 
   For elements where spectroscopic determinations of the quiet Sun are not possible, the recommended solar abundances are revisited based on complementary methods, including helioseismology (He), solar wind data from the Genesis sample return mission (noble gases), sunspot observations (four elements), and measurements of the most primitive meteorites (15 elements).}
   {Our new improved analysis confirms the relatively low solar abundances of C, N, and O obtained in our previous 3D-based studies: $\lgeps{C}=8.46\pm0.04$,  $\lgeps{N}=7.83\pm0.07$, and  $\lgeps{O}=8.69\pm0.04$. Excellent agreement between all available atomic and molecular indicators is achieved for C and O, but for N the atomic lines imply a lower abundance than for the molecular transitions for unknown reasons. The revised solar abundances for the other elements also typically agree well with our previously recommended values, with only Li, F, Ne, Mg, Cl, Kr, Rb, Rh, Ba, W, Ir, and Pb differing by more than 0.05\,dex. The here-advocated present-day photospheric metal mass fraction is only slightly higher than our previous value, mainly due to the revised Ne abundance from Genesis solar wind measurements:  $X_{\rm surface}=0.7438\pm0.0054$, $Y_{\rm surface}=0.2423\pm 0.0054$, $Z_{\rm surface}=0.0139\pm 0.0006$, and $Z_{\rm surface}/X_{\rm surface}=0.0187\pm 0.0009$. Overall, the solar abundances agree well with those of CI chondritic meteorites, but we identify a correlation with condensation temperature such that moderately volatile elements are enhanced by $\approx 0.04$\,dex in the CI chondrites and refractory elements possibly depleted by $\approx 0.02$\,dex, conflicting with conventional wisdom of the past half-century. Instead, the solar chemical composition more closely resembles that of the fine-grained matrix of CM chondrites with the expected exception of the highly volatile elements.} 
   {Updated present-day solar photospheric and proto-solar abundances are presented for 83 elements, including for all long-lived isotopes. The so-called solar modelling problem -- a persistent discrepancy between helioseismology and solar interior models constructed with a low solar metallicity similar to that advocated here -- remains intact with our revised solar abundances, suggesting shortcomings with the computed opacities and/or treatment of mixing below the convection zone in existing standard solar models. The uncovered trend between the solar and CI chondritic abundances with condensation temperature is not yet understood but is likely imprinted by planet formation, especially since a similar trend of opposite sign is observed between the Sun and solar twins. 
}   

\keywords{Sun: abundances -- Sun: photosphere -- Sun: atmosphere -- line: formation --  Meteorites, meteors, meteoroids -- Sun: helioseismology}

\maketitle

\section{Introduction}
\label{s:introduction}

The pursuit of detailed knowledge of the elemental abundances in the Solar System has played a critical role in the development of modern astronomy. Since the chemical make-up of the planets is not expected to precisely reflect the composition of the nebula from which the Sun and planets formed, there are two methods to determine the proto-solar abundances, both of which were initiated a century ago: solar spectroscopy \citep{1925PhDT.........1P,1929ApJ....70...11R} and mass spectroscopy of primitive meteorites \citep{1922NW.....10..918G,10007418258}. The two techniques each have their advantages and are thus highly complementary. Indeed a common approach is to combine the meteoritic and photospheric measurements to estimate the Solar System's composition, starting from the early influential works by \citet{1948ZA.....24..306U} and \citet{1956RvMP...28...53S}, who provided a large amount of abundance data for the seminal papers on the origin of the elements by \citet{1957RvMP...29..547B} and \citet{evolution1957nuclear}. More recently, the widely used literature compilations by \citet{1989GeCoA..53..197A}, \citet{1993oee..conf...15G}, \citet{1998SSRv...85..161G}, \citet{2003ApJ...591.1220L,2019arXiv191200844L}, \citet{2009LanB...4B...44L}, and \citet{2014pacs.book...15P} adopted this approach with a preference for the meteoritic values for most elements on account of their typically higher precision\footnote{An indication of the incredible impact the solar elemental abundances have across astronomy and beyond is the fact that \citet{1989GeCoA..53..197A} is among the ten most cited astronomy articles of all time with more than 8000 citations to date according to ADS, or more than 12000 citations in Google Scholar.}. Due to severe depletion of the highly volatile elements -- H, He, C, N, O, Ne, Ar, Xe, and Kr -- meteorites cannot, however, be relied upon to measure the bulk Solar System composition, for which the Sun is indispensable. There is also no guarantee that the abundances of other elements in the most primitive meteorites, the CI chondrites, truly reflect those of the Sun; indeed, this long-held assumption is not correct, as we show here.

The photospheric abundances are certainly not without shortcomings themselves. While often referred to as observed elemental abundances, the chemical composition of the Sun cannot be directly measured from the observed solar spectrum. Inferring the chemical make-up of a star requires a realistic model for the stellar atmosphere and of the transfer of radiation coupled with all atomic processes that shape the predicted emergent radiation to be compared with the observed spectrum. By necessity, the derived solar (and stellar) abundances are model dependent and therefore are not more reliable than the models from which they are inferred. Traditionally, quantitative solar spectroscopy has been done by means of one-dimensional (1D) model atmospheres: either theoretical models in radiative and convective equilibrium \citep[e.g.][]{1993KurCD..13.....K,2008A&A...486..951G} or semi-empirical models in which the atmospheric temperature stratification is deduced from observations of the continuum centre-to-limb variation and of line depths, for example \citep[e.g.][]{1974SoPh...39...19H,2001ApJ...558..830A,2006ApJ...639..441F}. Neither type of 1D model attempts to describe the temporal variations or horizontal inhomogeneities in the atmosphere, and as such they include at best only a rudimentary description of convective energy transport \citep[e.g. mixing length theory,][]{1958ZA.....46..108B}.

A huge body of influential and careful work on determining the solar chemical composition has been carried out over several decades using such 1D solar atmosphere models \citep[e.g.][]{1960ApJS....5....1G,1968MNRAS.138..143L,1978MNRAS.182..249L,1973A&A....27...29G,1999A&A...347..348G,1976Sci...191.1223R,1991A&A...249..545H,1995A&A...296..217B}.
This has often been done in close collaboration with atomic physicists measuring or computing transition probabilities and other necessary input data, which is a highly successful synergy 
\citep[e.g.][]{1969A&A.....2..446G,1975SoPh...44..257A,1981ApJ...248..867B,1991A&A...249..539B,1992A&A...259..301H,2000A&AS..142..467B,2001ApJ...563.1075L,2013ApJS..205...11L,2003ApJ...584L.107J,2006A&A...456.1181L,2007A&A...472L..43B,2019ApJS..243...33D}.
These abundance analyses have mostly been within the framework of local thermodynamic equilibrium (LTE), thus assuming the Boltzmann, Saha, and Guldberg-Waage distributions for the atomic and molecular level populations and the Planck function for the radiation source function \citep[e.g.][]{2003rtsa.book.....R}. While LTE is for many spectral lines that form in the solar photosphere a reasonable approximation, its validity cannot be taken for granted. For some elements, departures from LTE for the line formation in 1D model atmospheres (here: 1D non-LTE) and their impact on the inferred solar abundances have been explored 
\citep[e.g.][]{1984A&A...130..319S,1991A&A...245L...9K,1994PASJ...46...53T,2003A&A...407..691K,2011A&A...528A..87M,2011A&A...528A.103L,2012MNRAS.427...27B,2014ApJ...782...80S}.
Unfortunately, such statistical equilibrium calculations have still not been performed for the majority of the elements, rendering their solar abundances prone to systematic errors 
\citep{2005ARA&A..43..481A,2012MNRAS.427...27B,2018A&ARv..26....6N}.

With increased computational power, it has become possible to perform 3D (magneto-)hydrodynamical simulations of the outer convective and atmospheric layers in the Sun and other stars 
\citep[e.g.][]{1990A&A...228..155N,1996A&A...313..497F,1999A&A...346..111L,1999A&A...346L..17A,2005A&A...429..335V,2007A&A...469..687C,2010NewA...15..460M,2013A&A...557A..26M,2013ApJ...769...18T,2013A&A...558A..48B}.
In such modelling, the equations for the conservation of mass, momentum, and energy are solved together with the radiative transfer equation in a 3D simulation box covering a small but representative volume of the star.  Besides shedding crucial light on, for example, the nature of convective energy transport 
\citep[e.g.][]{1998ApJ...499..914S},
the excitation and damping of solar and stellar oscillations
\citep[e.g.][]{2015A&A...583A.112S,2019ApJ...880...13Z},
and the dynamo-generation of stellar magnetic fields 
\citep[e.g.][]{2007A&A...465L..43V,2014ApJ...789..132R},
such 3D simulations are eminently suited as model atmospheres for the prediction of the emergent radiation spectrum and thus for solar abundance analyses 
\citep[e.g.][]{1995A&A...302..578K,2000A&A...359..743A,2005A&A...431..693A,2001ApJ...556L..63A,2001AIPC..598...23H,2008A&A...488.1031C,2011SoPh..268..255C}.
Because convection emerges naturally in a self-consistent manner, there is no need to invoke the mixing length theory. These 3D hydrodynamical models are parameter-free in the sense that there is no tuning of any parameters involved to improve agreement with observations, yet their granulation properties, spectral line shapes, continuum centre-to-limb variations, and spectral energy distributions agree extremely well with observations; the choice of, for example, numerical resolution and sub-grid treatment does not impact the predictions 
\citep[e.g.][]{2000A&A...359..669A,2009LRSP....6....2N,2013A&A...554A.118P}.
Importantly, this is achieved without invoking any micro- and macro-turbulence parameters since the convective motions and stellar oscillations introduce the necessary Doppler shifts  together with the (anti-)correlations in temperature and velocities in granules and intergranular lanes to explain the line strengths, widths, shifts and asymmetries \citep{2000A&A...359..729A}.
Clearly, the most recent 3D hydrodynamical solar models, such as the one employed herein, are highly realistic and far superior to any 1D model. 

Today it is also feasible to carry out 3D non-LTE line formation calculations with comprehensive model atoms involving the relevant radiative and collisional processes for solar/stellar abundance purposes \citep[e.g.][]{2003A&A...399L..31A,2004A&A...417..751A,2007A&A...473L..37C,2013A&A...554A..96L,2015A&A...583A..57S,2016MNRAS.455.3735A,2019A&A...624A.111A,2019A&A...631A..80B}.
Such statistical equilibrium computations, while computationally demanding\footnote{The 3D non-LTE calculations are dominated by the formal solution of the radiative transfer equation. While in 1D LTE the synthesis of a single spectral line requires radiative transfer solutions along a ray for $\sim 100$ frequencies and a handful of angles; in 3D non-LTE, this typically has to be done for $\sim 10^{4}$ frequencies involving all populated atomic levels and transitions, $\sim 25$ angles, $\sim 10^{4}$ atmospheric columns, and $20$--$50$ iterations to converge the rate equations for statistical equilibrium and for $\sim 5$ temporal snapshots, that is, $>10^9$ equivalent 1D LTE calculations.},
make optimal use of the highly realistic 3D hydrodynamical stellar surface convection models and self-consistently describe departures from the LTE within the trace-element assumption. The non-LTE effects can typically be expected to be exacerbated in 3D compared to 1D modelling
\citep{2005ARA&A..43..481A}.
The field has benefited tremendously from the increased availability of accurate atomic data for the vast number of radiative and collisional cross-sections required, including the previously highly uncertain inelastic collisions with electrons and with neutral hydrogen, which are of paramount importance in the atmospheres of late-type stars \citep[e.g.][]{2016A&ARv..24....9B}.
It is truly remarkable that \underline{all} free parameters -- mixing length parameters, microturbulence, macroturbulence, Uns\"old enhancement factor for pressure broadening, Drawin scaling factor for H collisions, etc -- which have hampered quantitative stellar spectroscopy for decades, are now finally obsolete. Together with vastly improved stellar modelling and transition probabilities, this has drastically improved the accuracy and precision of the inferred stellar chemical compositions. 

\citet{2009ARA&A..47..481A} presented the first fully 3D-based analysis of the solar photospheric elemental abundances with non-LTE calculations (usually based on 1D model atmospheres) for the most important elements; earlier 3D determinations for many elements were reported in for example 
\citet{2000A&A...359..743A,2004A&A...417..751A,2005ASPC..336...25A,2005A&A...431..693A}.
It was also the first time all spectroscopically available elements were analysed in a homogeneous manner and with the systematic uncertainties quantified in detail (see also 
\citealt{2015A&A...573A..26S,2015A&A...573A..25S} and
\citealt{2015A&A...573A..27G} for further details). 

Here, we discuss further improvements to our solar analysis, including new 3D non-LTE calculations for many more elements, better atomic data, and refined line selection, which have led to a revision of the abundances of numerous elements. We argue that our study represents a state-of-the-art determination of the solar chemical composition, yielding the most reliable results available today. Complemented with precise laboratory measurements of the compositions of the most pristine meteorites 
\citep[e.g.][]{2014pacs.book...15P},
the new solar abundances enable us to uncover subtle abundance differences with condensation temperatures of the elements between the Sun and CI chondrites, which reflect the complicated processes of star and planet formation. For convenience, we also provide estimates for the proto-solar (isotopic) abundances where the effects of atomic diffusion, nuclear burning, and radioactive decay have been taken into account, as well as the present-day and proto-solar mass fractions of H, He, and heavy elements.

\section{Present-day photospheric solar abundances}
\label{s:photosphere}

\begin{figure}
    \begin{center}
        \includegraphics[width=9.5cm]{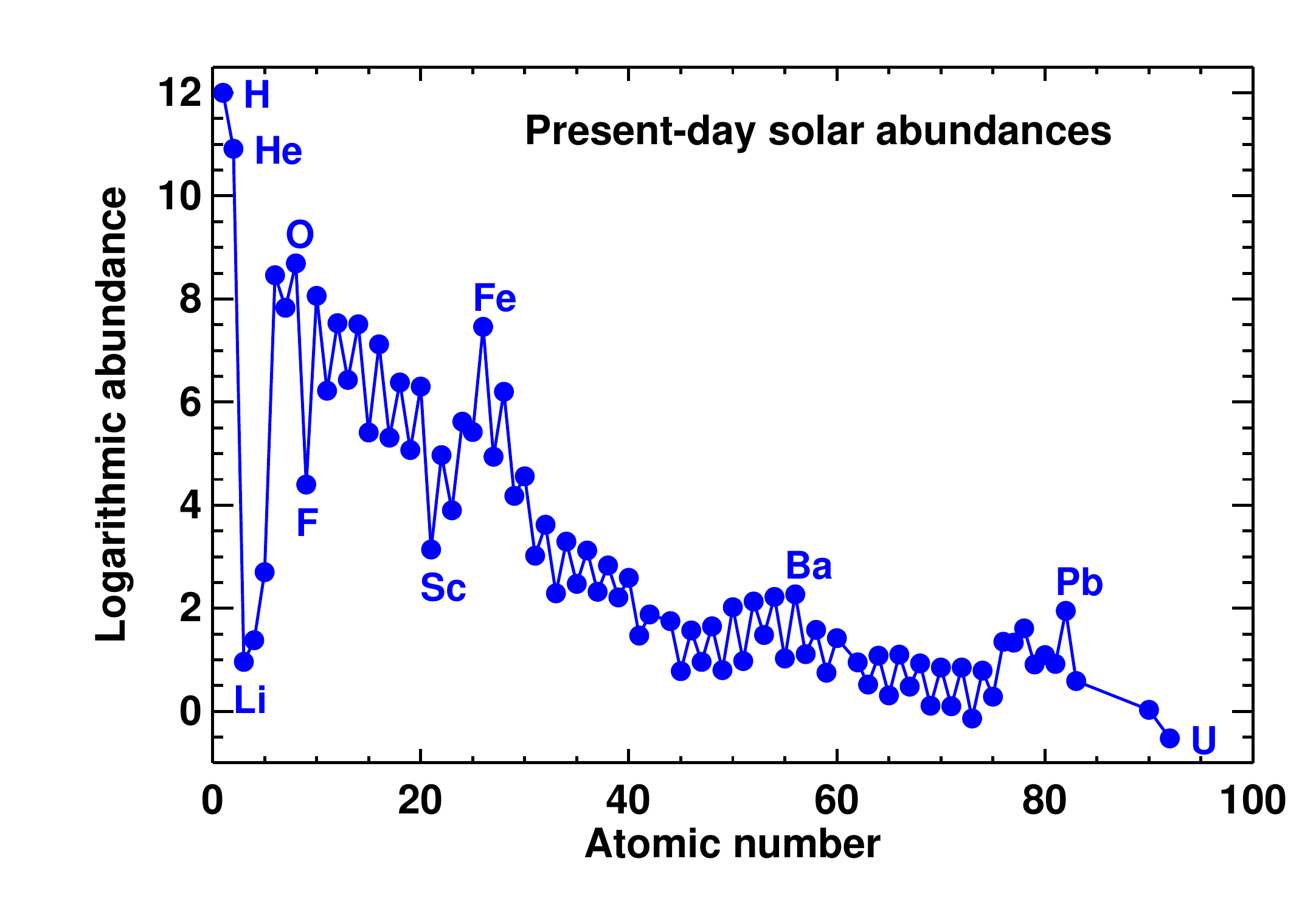}
        \caption{Present-day solar photospheric logarithmic abundances (by number) as a function of atomic number, with H defined as $\lgeps{H} \equiv 12.00$. For elements for which no photospheric determination has been possible the CI chondritic abundance is shown, corrected for the identified correlation with condensation temperature (Sect. \ref{s:Tcond}). H and He are the most abundant elements on account of being produced in the Big Bang with some contribution to He from stellar H-burning. Prominent peaks occur around O, Fe, Ba, and Pb, while elements with an even atomic number have higher abundances than neighbouring odd elements (so-called odd-even effect) as a result of stellar evolution and nuclear physics (minimum nuclear binding energy and nuclei magic numbers). Li, F, and Sc have particularly low abundances relative to nearby elements on account of them being odd elements, having relatively low binding energy, and not being part of the main nucleosynthesis production channels in stars. We note that the unstable elements Tc ($Z=43$) and Pm ($Z=61$) are not shown.}
        \label{f:sun_present}
    \end{center}
\end{figure}

The recommended solar photospheric elemental abundances presented in Table \ref{t:abund} and shown in Fig. \ref{f:sun_present} represent, in our opinion, the most reliable measurements available today, obtained with state-of-the-art analysis techniques and the best atomic and molecular data. All abundances have well-quantified and justified uncertainties and supersede those given in our previous summaries of the solar chemical composition 
\citep[e.g.][]{2005ASPC..336...25A,2009ARA&A..47..481A}. Table \ref{t:analysis_summary} intercompares the main ingredients of this study and our previous 3D-based solar abundance analyses in terms of 3D solar modelling and spectral line formation, showing that most recent progress has concerned non-LTE calculations, as described further below.

\begin{table*}[t!]
\caption{Summary of main ingredients of our 3D-based solar abundance analyses.
\label{t:analysis_summary}}
\centering
\smallskip
\begin{tabular}{l|c|c|c}
\hline
\noalign{\smallskip}
Ingredient & \citet{2005ASPC..336...25A} & {\citet{2009ARA&A..47..481A}} & This work \\
\hline
\noalign{\smallskip}
\noalign{\smallskip}
{\bf 3D model atmosphere: } & & & \\
Code & \convec{} & \stagger{} & \stagger{} \\
Code reference & \citet{1998ApJ...499..914S} & \citet{Nordlund:1995} & {\citet{2013A&A...557A..26M}} \\
Input abundances & \citet{1989GeCoA..53..197A}, & \citet{2005ASPC..336...25A} & {\citet{2009ARA&A..47..481A}}\\
                 & $\lgeps{He} =10.92$, $\lgeps{Fe} =7.50$ & & \\
Equation of state & \citet{1988ApJ...331..815M} & \citet{2013ApJ...769...18T} & \citet{2013ApJ...769...18T} \\
Continuous opacities & {\citet{1975A&A....42..407G}} & {\citet{2008A&A...486..951G}} & {\citet{2010A&A...517A..49H}} \\
Line opacities & {\citet{1993KurCD..13.....K}} & {\citet{2008A&A...486..951G}} & {\citet{2008A&A...486..951G}} \\
Numerical resolution & $200^2 {\rm x} 82$ & $240^3$ & $240^3$ \\
Geometrical dimension & $6{\rm x}6{\rm x}3.5$\,Mm & $6{\rm x}6{\rm x}3.5$\,Mm & $6{\rm x}6{\rm x}3.5$\,Mm \\
Opacity bins & 4 (opacity) & 12 (opacity \& wavelength) & 12 (opacity \& wavelength)\\
Effective temperature $\teff\pm$s.d. & $5767\pm21$\,K & $5778\pm5$\,K & $5773\pm16$\,K \\
Continuum CLV & too steep $T(\tau)$ & good agreement w/ obs. & good agreement w/ obs. \\
\noalign{\smallskip} 
{\bf Spectral line formation: } & & &  \\
Full 3D non-LTE & Li, O & Li, O & Li, C, N, O, Na, Mg, Al,  \\
                &       &       & Si, K, Ca, Fe, Ba \\
3D LTE + 1D/\mtd{} non-LTE & C, N, Na, Mg, Al, K &  C, N, Na, Mg, Al, Si, S, K, & S, Sc, Ti, Cr, Mn, Co, Cu,  \\
                    &                     &  Ca, Sc, Ti, Cr, Mn, Fe, Co, & Zn, Rb, Sr, Pr, Eu, Pb, Th \\
                    &                     &  Cu, Zn, Sr, Ba, Eu, Pb, Th  & \\
3D LTE & Be, B, Si, P, S, Ca, Fe & Be, B, P, V, Ni, $Z\ge 31$ & Be, B, P, V, Ni, $Z\ge 31$  \\
1D LTE (literature) & remaining elements & F, Cl, In, Th & F, Cl, In, Th \\
\noalign{\smallskip}
\hline 
\end{tabular}
\end{table*}

The abundances are expressed on the traditional astronomical logarithmic abundance scale. Hydrogen is the natural reference element for solar (and stellar) spectroscopy, both because it is the most abundant element and because it provides the continuous opacity in the optical and infrared through the negative hydrogen ion H$^-$. The normalisation of the elemental number density $N_{\rm X}$ for an element X is defined as $\lgeps{X} \equiv \log \left( N_{\rm X}/ N_{\rm H} \right) + 12.00$, for historical reasons, hence $\lgeps{H} \equiv 12.00.$\footnote{This choice was supposedly to avoid having negative elemental abundances for the Sun \citep{1951csa..book.....C,1960ApJS....5....1G}.
While this is also true for the here recommended solar photospheric abundances -- barely, with $\lgeps{Th} =0.03\pm0.10$ being the smallest value -- several naturally occurring elements indeed have $\lgeps{} < 0$ in CI chondritic meteorites (Sect. \ref{s:meteorites}): Ne, Ar, Kr, Xe, Ta, and U.}

After first summarising some central ingredients in our analysis, in the following sub-sections we present a detailed discussion of the elements for which solar photospheric abundances have been newly analysed, compared with our previous spectroscopic studies of elements heavier than Ne
\citep{2015A&A...573A..26S,2015A&A...573A..25S,2015A&A...573A..27G}. In some cases, we simply updated their results in light of new atomic data (in particular, transition probabilities) or improved non-LTE abundance corrections, or we made a different choice in terms of the line list or weighting of the individual lines. 
For Li, C, N, O, Na, Mg, Al, Si, K, Ca, Fe, and Ba, our results are based on new 3D non-LTE calculations. We also discuss 
the elements that are not accessible by spectroscopy of the quiet Sun for which other methods to infer the solar abundances are required.

\subsection{General considerations of the analysis}
\label{s:general}

{\bf Model atmospheres: } 
As a default solar model atmosphere, we employed the same 3D hydrodynamic simulation of the solar surface convection computed with the \stagger{} code
\citep{Nordlund:1995,1998ApJ...499..914S,2013A&A...557A..26M},
as was done in our recent studies of 3D non-LTE line formation of C, N, and O in the Sun \citep{2018A&A...616A..89A,2019A&A...624A.111A,2020A&A...636A.120A}. This is an updated version of the \stagger{} solar model used in \citet{2009ARA&A..47..481A}, 
\citet{2015A&A...573A..26S,2015A&A...573A..25S}, 
and \citet{2015A&A...573A..27G}, but the differences in photospheric structure are minor.   Detailed 3D radiative transfer with line-blanketing through the opacity binning technique 
\citep{1982A&A...107....1N,2018MNRAS.475.3369C}
is included under the assumption of LTE to compute the necessary radiative heating and cooling rates in the photosphere. The resulting temperature stratification is highly realistic, as is evident from comparisons with the observed continuum centre-to-limb variation
\citep{2018A&A...616A..89A}; this was also the case with the 3D model used in \citet{2009ARA&A..47..481A}. The predictions of our 3D solar models are in excellent agreement with all key observational diagnostics 
\citep[e.g.][]{2013A&A...554A.118P}. The resulting effective temperature of the 24\,hr simulation sequence is $5773\pm16$\,K (standard deviation), indistinguishable from the solar value of $5772$\,K \citep{2016AJ....152...41P}.  The 3D \stagger{} models employ an updated version of the so-called MHD (named after the authors) equation of state
\citep{1988ApJ...331..815M,2013ApJ...769...18T}, continuous and line opacities from the \marcs{} package 
\citep{2008A&A...486..951G,2010A&A...517A..49H},
and the solar chemical composition of \citet{2009ARA&A..47..481A}. No magnetic fields were included.   From the $240^3$ simulation grid covering 6\,Mm horizontally and 3\,Mm vertically, a resampled 3D grid covering the photosphere (1\,Mm vertically) was interpolated before the spectral line formation calculations for the abundance determinations to improve the numerical accuracy and to save computational time without sacrificing accuracy; the grid resolutions of both the original and interpolated solar models are sufficient to obtain the correct spectral line broadening from convective motions and oscillations \citep{2000A&A...359..669A,2000A&A...359..729A,2013A&A...554A.118P}. 

To quantify various systematic uncertainties, we also made use of several 1D model atmospheres: the temporally and spatially averaged 3D model (referred to here as \mtd{});
a 1D hydrostatic \atmo{} model with mixing length theory convection and identical equation of state and opacities to the 3D model 
\citep[][Appendix A]{2013A&A...557A..26M};
a 1D hydrostatic model computed with the widely used \marcs{} code 
\citep{2008A&A...486..951G};
and the 1D semi-empirical 
\citet{1974SoPh...39...19H}
model, which has traditionally been the model of choice for solar abundance analyses for the past half-century.  

{\bf Spectral line formation: } 
For key atoms and ions, 
the results are based on 3D non-LTE calculations, using the MPI-parallelised \balder{} code 
\citep{2018A&A...615A.139A}, our extensively modified version of \multitd{} \citep{2009ASPC..415...87L_short}.
For \ion{Li}{I}, \ion{C}{I}, 
\ion{N}{I}, \ion{Al}{I,} and \ion{Si}{i}/\ion{Si}{ii,} these results were adopted from recently published studies
\citep{2021MNRAS.500.2159W,2019A&A...624A.111A,2020A&A...636A.120A,2017A&A...607A..75N,2017MNRAS.464..264A}, while for 
\ion{O}{I}, \ion{Na}{I}, \ion{Mg}{I}/\ion{Mg}{II},
\ion{K}{I}, \ion{Ca}{I}/\ion{Ca}{II}, and \ion{Fe}{I}/\ion{Fe}{II}, they are based on new calculations.
We also adopted 3D non-LTE literature results for \ion{Ba}{ii}, which are based on \multitd{} calculations
\citep{2020A&A...634A..55G}.

\balder{} iteratively solves the 3D (or 1D) radiative transfer equation coupled with the equations for statistical equilibrium  (i.e. the atomic level populations fulfil $\partial n_{\rm i}/\partial t =0$)  under the assumption that the departures from LTE for an element do not feed back onto the atmospheric structure, nor do they impact the line formation of other elements. 
\balder{} employs a comprehensive equation of state, background continuum, and line opacities
calculated with the \blue{} code
as described in \citet{2016MNRAS.463.1518A},
with the present work adopting the occupation probability formalism 
to better model line opacities near the 
series limits
\citep{1994A&A...282..151H}.
The 3D non-LTE radiative transfer is solved iteratively for typically 5-10 snapshots of the 3D solar model in $26$ directions
using short characteristics \citep{2013A&A...549A.126I}.
The number of frequencies for which the radiative transfer equation needs to be computed to adequately resolve the bound-free and bound-bound transitions varies among elements, but for our most complex model atoms (Ca and Fe) it amounts to \textbf{$20000$} and \textbf{$45000$}, respectively. 
After convergence, a final formal solution is carried out for the emergent vertical intensity and for the spectral lines of interest, using a long characteristics solver and higher frequency resolution for accuracy.

Other atomic and ionic lines, as well as all molecular lines, are based on 3D LTE spectrum synthesis using the \scate{} code 
\citep{2010A&A...517A..49H},
for the most part as presented in our earlier papers
\citep{2015A&A...573A..26S,2015A&A...573A..25S,2015A&A...573A..27G}. New 3D LTE calculations were adopted for molecular species of C, N, and O. When no 3D non-LTE calculations had been performed, we relied on our own or published 1D non-LTE abundance corrections, if possible;
this procedure is not fully consistent but is expected to be more realistic than the LTE results.  Many elements, in particular those beyond the Fe peak, unfortunately still lack non-LTE studies, requiring us to adopt the 3D LTE results. It is emphasised that no micro- and macro-turbulence enter the 3D calculations 
\citep{2000A&A...359..729A};
for the comparison 1D spectrum syntheses, we assumed a microturbulence of 1\,km\,s$^{-1}$ \citep[e.g.][]{1978A&A....70..537H,2013MSAIS..24...37S}.

{\bf Atomic data: } 
The computation of the theoretical solar spectrum hinges critically on the availability of accurate atomic data. We have scoured and carefully assessed the literature and databases for the most up-to-date and most reliable input data required for our solar modelling: transition probabilities, hyperfine splitting, spectral line broadening, partition functions, excitation, ionisation, and dissociation energies, etc \citep[e.g.][]{1995A&AS..112..525P,2016A&A...588A..96B,2021A&A...645A.106H}. 
In most cases, experimental data have been selected over theoretical data, with the NIST Atomic Spectra Database\footnote{\url{https://physics.nist.gov/PhysRefData/ASD/lines_form.html}} \citep{2019APS..DMPN09004K}
providing much of the quantitative assessment of the uncertainties in $gf$-values. For the solar photosphere, the continuous opacities are well determined, except possibly in the UV, and hence they are not a major concern. For the non-LTE calculations, we also made extensive use of the bound-bound and bound-free radiative data from the Opacity Project, the Iron Project, and other dedicated atomic calculations 
\citep[e.g.][]{1987JPhB...20.6363S,1993A&A...279..298H,2005MNRAS.360..458B},
as well as theoretical cross-sections for collisions with electrons and with neutral hydrogen 
\citep[e.g.][]{1992cccc.rept.....B,1995CoPhC..85....1B,2006CoPhC.174..273Z,2013JPhB...46k2001Z,2016PhRvA..93d2705B,2017A&A...606A..11B,2017A&A...606A.147B}. Fortunately the situation today in terms of the necessary atomic and molecular data for solar/stellar spectroscopy is significantly improved compared to just a decade ago thanks to the tireless contributions from a relatively small group of atomic physicists to whom we are greatly indebted. 

{\bf Observations: } 
To derive the solar elemental abundances, the predicted solar spectrum was compared to the observed high-resolution disc-centre intensity spectrum of the quiet Sun with a preference for the Li\`ege (named after its location of production, not observations which were done at Jungfraujoch) visual solar atlas 
\citep{1973apds.book.....D}
due to its low degree of contamination of telluric absorption lines. We also made use of other solar atlases for the disc-centre intensity 
\citep{1999SoPh..184..421N,2005MSAIS...8..189K}, the disc-integrated flux \citep{1984sfat.book.....K,2006astro.ph..5029K,2016A&A...587A..65R}, as well as observations in the infrared 
\citep{delbouille1981photometric,1989hra1.book.....F,1996ApOpt..35.2747A,2010JQSRT.111..521H}
as needed and for evaluation purposes; disc-centre intensity spectra are in general preferable to flux to disentangle the contribution of blends. In some cases, we also employ the high-resolution optical centre-to-limb observations of \citet{2009A&A...508.1403P} and 
\citet{2015A&A...573A..74S} to verify the correctness of our line formation calculations. All of these observed solar spectra have extremely high resolving power (typically equivalent of $R \equiv \lambda /\Delta \lambda > 200,000$) and signal-to-noise (mostly $S/N > 1000$). We either perform a line profile fitting using standard $\chi^2$-techniques or match the observed and computed equivalent widths, depending on the line and possible presence of  blends. We allowed our own local continuum placement around the spectral lines in question, as appropriate. \citet{2016A&A...590A.118D} and \citet{2016A&A...587A..65R} examined the differences between some of these atlases and overall find excellent agreement; our results are not significantly affected by the particular choice of observations except for a few individual lines. 

\begin{table*}[t!]
\caption{Present-day solar photospheric abundances together with those in CI chondrites \citep{2014pacs.book...15P}, renormalised to the solar Si abundance (Sect. \ref{s:meteorites}). Elements that are commented on are those for which the adopted proto-solar abundances (Sect. \ref{s:protosolar}) are not based on spectroscopy of the solar photosphere; we caution that the differences between solar and CI abundances show a trend with condensation temperature (Sect. \ref{s:Tcond}), which should be taken into account when estimating the Solar System composition.
\label{t:abund}}
\begin{tabular}{rlrrl|rlrrl}
\hline
\noalign{\smallskip}
Z      &  Element & Photosphere & CI chondrites & Comments & Z      &  Element & Photosphere & CI chondrites & Comments \\
 \noalign{\smallskip}
 \hline
 \noalign{\smallskip}
1       &        H      &       $       12.00   \pm             0.00    $         &       $       8.22    \pm             0.04    $       &       definition      &         44      &       Ru      &       $       1.75    \pm     0.08    $         &       $       1.77    \pm             0.02    $       &               \\
2       &       He      &       $       10.914  \pm             0.013   $         &       $       1.29    \pm             0.18    $       &       helioseismology &         45      &       Rh      &       $       0.78    \pm     0.11    $         &       $       1.04    \pm             0.02    $       &               \\
3       &       Li      &       $       0.96    \pm             0.06    $         &       $       3.25    \pm             0.04    $       &       meteorites         &       46      &       Pd      &       $       1.57    \pm     0.10    $         &       $       1.65    \pm             0.02    $       &               \\
4       &       Be      &       $       1.38    \pm             0.09    $         &       $       1.32    \pm             0.03    $       &               &         47      &       Ag      &       $       0.96    \pm     0.10    $         &       $       1.20    \pm             0.04    $       &               \\
5       &        B      &       $       2.70    \pm             0.20    $         &       $       2.79    \pm             0.04    $       &               &         48      &       Cd      &               &       $       1.71    \pm         0.03    $       &       meteorites      \\
6       &        C      &       $       8.46    \pm             0.04    $         &       $       7.39    \pm             0.04    $       &               &         49      &       In      &       $       0.80    \pm     0.20    $         &       $       0.76    \pm             0.02    $       &               \\
7       &        N      &       $       7.83    \pm             0.07    $         &       $       6.26    \pm             0.06    $       &               &         50      &       Sn      &       $       2.02    \pm     0.10    $         &       $       2.07    \pm             0.06    $       &               \\
8       &        O      &       $       8.69    \pm             0.04    $         &       $       8.39    \pm             0.04    $       &               &         51      &       Sb      &               &       $       1.01    \pm         0.06    $       &       meteorites      \\
9       &        F      &       $       4.40    \pm             0.25    $         &       $       4.42    \pm             0.06    $       &               &         52      &       Te      &               &       $       2.18    \pm         0.03    $       &       meteorites      \\
10      &       Ne      &       $       8.06    \pm             0.05    $         &       $       -1.12   \pm             0.18    $       &       solar wind    &       53      &        I      &               &       $       1.55    \pm         0.08    $       &       meteorites      \\
11      &       Na      &       $       6.22    \pm             0.03    $         &       $       6.27    \pm             0.04    $       &               &         54      &       Xe      &       $       2.22    \pm     0.05    $         &       $       -1.95   \pm             0.18    $       &       nuclear physics \\
12      &       Mg      &       $       7.55    \pm             0.03    $         &       $       7.53    \pm             0.02    $       &               &         55      &       Cs      &               &       $       1.08    \pm         0.03    $       &       meteorites      \\
13      &       Al      &       $       6.43    \pm             0.03    $         &       $       6.43    \pm             0.03    $       &               &         56      &       Ba      &       $       2.27    \pm     0.05    $         &       $       2.18    \pm             0.02    $       &               \\
14      &       Si      &       $       7.51    \pm             0.03    $         &       $       7.51    \pm             0.01    $       &               &         57      &       La      &       $       1.11    \pm     0.04    $         &       $       1.17    \pm             0.01    $       &               \\
15      &        P      &       $       5.41    \pm             0.03    $         &       $       5.43    \pm             0.03    $       &               &         58      &       Ce      &       $       1.58    \pm     0.04    $         &       $       1.58    \pm             0.01    $       &               \\
16      &        S      &       $       7.12    \pm             0.03    $         &       $       7.15    \pm             0.02    $       &               &         59      &       Pr      &       $       0.75    \pm     0.05    $         &       $       0.76    \pm             0.01    $       &               \\
17      &       Cl      &       $       5.31    \pm             0.20    $         &       $       5.23    \pm             0.06    $       &               &         60      &       Nd      &       $       1.42    \pm     0.04    $         &       $       1.45    \pm             0.01    $       &               \\
18      &       Ar      &       $       6.38    \pm             0.10    $         &       $       -0.50   \pm             0.18    $       &       solar wind    &       62      &       Sm      &       $       0.95    \pm     0.04    $         &       $       0.94    \pm             0.01    $       &               \\
19      &        K      &       $       5.07    \pm             0.03    $         &       $       5.08    \pm             0.04    $       &               &         63      &       Eu      &       $       0.52    \pm     0.04    $         &       $       0.52    \pm             0.01    $       &               \\
20      &       Ca      &       $       6.30    \pm             0.03    $         &       $       6.29    \pm             0.03    $       &               &         64      &       Gd      &       $       1.08    \pm     0.04    $         &       $       1.05    \pm             0.01    $       &               \\
21      &       Sc      &       $       3.14    \pm             0.04    $         &       $       3.04    \pm             0.03    $       &               &         65      &       Tb      &       $       0.31    \pm     0.10    $         &       $       0.31    \pm             0.01    $       &               \\
22      &       Ti      &       $       4.97    \pm             0.05    $         &       $       4.90    \pm             0.03    $       &               &         66      &       Dy      &       $       1.10    \pm     0.04    $         &       $       1.13    \pm             0.01    $       &               \\
23      &        V      &       $       3.90    \pm             0.08    $         &       $       3.96    \pm             0.03    $       &               &         67      &       Ho      &       $       0.48    \pm     0.11    $         &       $       0.47    \pm             0.01    $       &               \\
24      &       Cr      &       $       5.62    \pm             0.04    $         &       $       5.63    \pm             0.02    $       &               &         68      &       Er      &       $       0.93    \pm     0.05    $         &       $       0.93    \pm             0.01    $       &               \\
25      &       Mn      &       $       5.42    \pm             0.06    $         &       $       5.47    \pm             0.03    $       &               &         69      &       Tm      &       $       0.11    \pm     0.04    $         &       $       0.12    \pm             0.01    $       &               \\
26      &       Fe      &       $       7.46    \pm             0.04    $         &       $       7.46    \pm             0.02    $       &               &         70      &       Yb      &       $       0.85    \pm     0.11    $         &       $       0.92    \pm             0.01    $       &               \\
27      &       Co      &       $       4.94    \pm             0.05    $         &       $       4.87    \pm             0.02    $       &               &         71      &       Lu      &       $       0.10    \pm     0.09    $         &       $       0.09    \pm             0.01    $       &               \\
28      &       Ni      &       $       6.20    \pm             0.04    $         &       $       6.20    \pm             0.03    $       &               &         72      &       Hf      &       $       0.85    \pm     0.05    $         &       $       0.71    \pm             0.01    $       &               \\
29      &       Cu      &       $       4.18    \pm             0.05    $         &       $       4.25    \pm             0.06    $       &               &         73      &       Ta      &               &       $       -0.15   \pm         0.04    $       &       meteorites      \\
30      &       Zn      &       $       4.56    \pm             0.05    $         &       $       4.61    \pm             0.02    $       &               &         74      &        W      &       $       0.79    \pm     0.11    $         &       $       0.65    \pm             0.04    $       &               \\
31      &       Ga      &       $       3.02    \pm             0.05    $         &       $       3.07    \pm             0.03    $       &               &         75      &       Re      &               &       $       0.26    \pm         0.02    $       &       meteorites      \\
32      &       Ge      &       $       3.62    \pm             0.10    $         &       $       3.58    \pm             0.04    $       &               &         76      &       Os      &       $       1.35    \pm     0.12    $         &       $       1.35    \pm             0.02    $       &               \\
33      &       As      &               &       $       2.30    \pm             0.04    $         &       meteorites      &       77      &       Ir      &                &       $       1.32    \pm             0.02    $       &       meteorites         \\
34      &       Se      &               &       $       3.34    \pm             0.03    $         &       meteorites      &       78      &       Pt      &                &       $       1.61    \pm             0.02    $       &       meteorites         \\
35      &       Br      &               &       $       2.54    \pm             0.06    $         &       meteorites      &       79      &       Au      &       $         0.91    \pm     0.12    $       &       $       0.81    \pm             0.05    $         &               \\
36      &       Kr      &       $       3.12    \pm             0.10    $         &       $       -2.27   \pm             0.18    $       &       solar wind    &       80      &       Hg      &               &       $       1.17    \pm         0.18    $       &       meteorites      \\
37      &       Rb      &       $       2.32    \pm             0.08    $         &       $       2.37    \pm             0.03    $       &               &         81      &       Tl      &       $       0.92    \pm     0.17    $         &       $       0.77    \pm             0.05    $       &               \\
38      &       Sr      &       $       2.83    \pm             0.06    $         &       $       2.88    \pm             0.03    $       &               &         82      &       Pb      &       $       1.95    \pm     0.08    $         &       $       2.03    \pm             0.03    $       &               \\
39      &        Y      &       $       2.21    \pm             0.05    $         &       $       2.15    \pm             0.02    $       &               &         83      &       Bi      &               &       $       0.65    \pm         0.04    $       &       meteorites      \\
40      &       Zr      &       $       2.59    \pm             0.04    $         &       $       2.53    \pm             0.02    $       &               &         90      &       Th      &       $       0.03    \pm     0.10    $         &       $       0.04    \pm             0.03    $       &               \\
41      &       Nb      &       $       1.47    \pm             0.06    $         &       $       1.42    \pm             0.04    $       &               &         92      &        U      &               &       $       -0.54   \pm         0.03    $       &       meteorites      \\
42      &       Mo      &       $       1.88    \pm             0.09    $         &       $       1.93    \pm             0.04    $       &               &                 &               &                                               &                                                 &               \\
\noalign{\smallskip}
\hline
 \end{tabular}
\end{table*}

{\bf Uncertainties: } 
Homogenously quantifying the uncertainties of elemental abundances, including systematic errors is almost as important as having accurate determinations of them. Concerning elements for which multiple spectral lines are available for the abundance analysis, we used the weighted standard error of the mean as the statistical uncertainty, the weights being our assessment of the quality of the lines. In cases when only one or two lines can be used, we estimated the statistical error from the goodness of the profile fitting or equivalent width measurement uncertainty. 

Following \citet{2009ARA&A..47..481A}, we devoted much effort to carefully investigating and evaluating possible systematic errors in the analysis, in particular related to the 3D model atmosphere and 3D non-LTE line formation.  To quantify the errors from shortcomings in the photosphere temperature structure we take half the differences between the derived abundances from the \mtd{} model and the \citet{1974SoPh...39...19H} model; as noted above, we know from the centre-to-limb variation that the \mtd{} model is a much closer match to the real Sun \citep{2009ARA&A..47..481A,2013A&A...554A.118P} with newer semi-empirical solar model atmospheres \citep{1998SSRv...85..161G,2001ApJ...558..830A} having $50-100$\,K lower temperatures than the \citet{1974SoPh...39...19H} model in the line-forming region, which is very similar to our 3D model. The impact of atmospheric inhomogeneities is evaluated from taking half the difference between the 3D and \mtd\ results, which is justified by the close resemblance between the predicted and observed solar granulation patterns. In general, the systematic uncertainty due to non-LTE line formation is estimated as half of the predicted non-LTE abundance correction (3D non-LTE $-$ 3D LTE, or 1D non-LTE $-$ 1D LTE) with a minimum error of 0.03\,dex when no non-LTE calculations are available; an exception is made for O for which the non-LTE effects can be accurately quantified using, for example, the centre-to-limb variation of the \ion{O}{i} 777\,nm triplet and therefore the uncertainties would otherwise be significantly overestimated. These three systematic errors are combined in quadrature with the statistical error to compute the total abundance error. 

For some elements, we tried to factor in uncertainties in the input atomic data but in general this has not been attempted due to those often being poorly known 
and the huge amount of atomic data required for non-LTE modelling. While impossible to know without even more sophisticated solar modelling, we expect our estimated uncertainties to be realistic and possibly conservative overall, although we caution that in a few cases, especially in the absence of any non-LTE calculations, the errors are likely underestimated; in cases where they were particularly striking, we artificially inflated the computed errors somewhat, as noted.

\subsection{Light elements: Lithium to beryllium}
\label{s:light}

Li, Be, and B were not analysed in our previous series of papers devoted to the solar chemical composition \citep{2015A&A...573A..26S,2015A&A...573A..25S,2015A&A...573A..27G} and are therefore discussed explicitly here.  
 
{\bf Lithium (Z=3):}
In practise, the solar Li abundance can only be determined from the very weak \ion{Li}{i} 670.8\,nm resonance line, an undertaking that is further complicated by the presence of several blending lines of CN, Fe, and other species. 
We adopted the recently revised solar Li abundance from \citet{2021MNRAS.500.2159W}, who presented new 3D non-LTE spectral line formation calculations for a wide range of stellar parameters, including for the Sun. Importantly, their more complete treatment of background line opacities in the UV compared with previous studies
\citep[e.g.][]{1994A&A...288..860C,2003A&A...399L..31A,2009A&A...503..541L,2013A&A...554A..96L, 2018A&A...618A..16H,2020A&A...638A..58M}
significantly changes the statistical equilibrium balance of \ion{Li}{i}, leading in general to a strengthening of the 670.8\,nm line. Using their new 3D non-LTE calculations and allowing for blends, \citet{2021MNRAS.500.2159W} derived a solar Li abundance of $\lgeps{Li} =0.96\pm0.03$ (statistical) $\pm 0.05$ (systematic) and a 0.09\,dex (23\%) reduction compared to the advocated value in \citet{2009ARA&A..47..481A}.

{\bf Beryllium (Z=4):}
Following \citet{1975A&A....42...37C}, it was widely believed that Be was depleted by a factor of two in the solar photosphere compared with CI chondrites. This conclusion was first challenged by \citet{1998Natur.392..791B} who demonstrated that missing UV opacity
\citep[e.g.][]{2001ApJ...546L..65B,2009ApJ...691.1634S}
in the stellar models leads to an underestimated solar Be abundance when relying on the \ion{Be}{ii} 313\,nm resonance lines. By calibrating the missing opacity in this wavelength region by enforcing the same  O abundance to be estimated from nearby UV OH lines with OH lines in the IR, they found a solar photospheric Be abundance consistent with the meteoritic value. Their conclusion was corroborated in a 3D LTE analysis by 
\citet{2004A&A...417..769A}, 
which we adopted ($\lgeps{Be} =1.38\pm0.09;$ statistical + systematic), suggesting no solar Be depletion. We caution, however, that unaccounted-for blends in this crowded UV region may have biased the inferred Be abundance upwards. Revisiting the topic with 3D non-LTE calculations and a careful assessment of blending lines and continuum placement together with a detailed evaluation of any missing UV opacity would be worthwhile to finally settle the issue, especially given the importance of accurately constraining the amount of mixing below the solar convection zone (Sect. \ref{s:sunvsci}).

{\bf Boron (Z=5):}
The solar B abundance can in practice only be inferred from the \ion{B}{i} 249.7\,nm resonance line, which makes the result quite uncertain due to the heavy crowding and possible missing opacity in this wavelength region. The most recent spectroscopic analysis is that of \citet{1999ApJ...512.1006C}, who inferred $\lgeps{B} =2.70\pm0.20$ from a 1D LTE spectrum synthesis, which we have consequently adopted here as the present-day photospheric abundance. 
While the minority ionisation stage and low excitation potential may suggest that the 249.7\,nm line should be susceptible to severe departures from LTE, this does not seem to be  the case for the Sun due to a somewhat fortuitous cancellation of non-LTE effects in 1D
\citep{1994A&A...286..169K,2010ApJ...713..458T}. No 3D non-LTE study for B exists to our knowledge.

\subsection{Carbon, nitrogen, and oxygen}
\label{s:cno}

\begin{table*}
\begin{center}
\caption{C, N, and O abundances in the solar photosphere as inferred from the available indicators and from different model atmospheres, with the combined statistical and systematic uncertainties in the final column.}
\label{t:cnoabund}
\begin{tabular}{l|cccc|cccc|c}
\hline
\noalign{\smallskip}
\multirow{2}{*}{Indicator} & \multicolumn{4}{c|}{Non-LTE} & \multicolumn{4}{c|}{LTE} & \multirow{2}{*}{Error} \\
 & 3D & \mtd{} & \marcs{} & HM & 3D & \mtd{} & \marcs{} & HM &  \\
\noalign{\smallskip}
\hline
\noalign{\smallskip}
[\ion{C}{I}] & $8.453$ & $8.424$ & $8.421$ & $8.432$ & $8.453$ & $8.424$ & $8.421$ & $8.432$ & $0.034$ \\ 
\ion{C}{I} & $8.470$ & $8.459$ & $8.456$ & $8.488$ & $8.480$ & $8.467$ & $8.462$ & $8.497$ & $0.038$ \\ 
C$_2$ Swan &  $-$ &  $-$ &  $-$ &  $-$ & $8.455$ & $8.473$ & $8.433$ & $8.520$ & $0.029$ \\ 
CH (A-X) &  $-$ &  $-$ &  $-$ &  $-$ & $8.459$ & $8.471$ & $8.393$ & $8.555$ & $0.048$ \\ 
CH (d$\upnu=1$) &  $-$ &  $-$ &  $-$ &  $-$ & $8.470$ & $8.503$ & $8.438$ & $8.573$ & $0.043$ \\ 
CO (d$\upnu=1$) &  $-$ &  $-$ &  $-$ &  $-$ & $8.487$ & $8.653$ & $8.606$ & $8.683$ & $0.101$ \\ 
CO (d$\upnu=2$) &  $-$ &  $-$ &  $-$ &  $-$ & $8.467$ & $8.681$ & $8.609$ & $8.743$ & $0.122$ \\ 
\noalign{\smallskip}
\hline
\noalign{\smallskip}
\ion{N}{I} & $7.767$ & $7.816$ & $7.804$ & $7.864$ & $7.780$ & $7.828$ & $7.813$ & $7.879$ & $0.035$ \\ 
NH (d$\upnu=0$) &  $-$ &  $-$ &  $-$ &  $-$ & $7.881$ & $8.032$ & $8.021$ & $8.079$ & $0.087$ \\ 
NH (d$\upnu=1$) &  $-$ &  $-$ &  $-$ &  $-$ & $7.921$ & $7.988$ & $7.914$ & $8.056$ & $0.050$ \\ 
CN (0-0) &  $-$ &  $-$ &  $-$ &  $-$ & $7.867$ & $7.964$ & $7.927$ & $8.016$ & $0.072$ \\ 
CN (d$\upnu\geq1$) &  $-$ &  $-$ &  $-$ &  $-$ & $7.899$ & $7.994$ & $7.959$ & $8.046$ & $0.078$ \\ 
\noalign{\smallskip}
\hline
\noalign{\smallskip}
[\ion{O}{I}] & $8.703$ & $8.688$ & $8.664$ & $8.705$ & $8.703$ & $8.688$ & $8.664$ & $8.705$ & $0.051$ \\ 
\ion{O}{I} & $8.686$ & $8.682$ & $8.657$ & $8.678$ & $8.805$ & $8.779$ & $8.732$ & $8.789$ & $0.030$ \\ 
OH (d$\upnu=0$) &  $-$ &  $-$ &  $-$ &  $-$ & $8.690$ & $8.792$ & $8.782$ & $8.837$ & $0.070$ \\ 
OH (d$\upnu=1$) &  $-$ &  $-$ &  $-$ &  $-$ & $8.707$ & $8.800$ & $8.728$ & $8.884$ & $0.063$ \\ 
OH (d$\upnu=2$) &  $-$ &  $-$ &  $-$ &  $-$ & $8.690$ & $8.756$ & $8.673$ & $8.848$ & $0.087$ \\ 
\noalign{\smallskip}
\hline
\noalign{\smallskip}
\end{tabular}
\end{center}
\end{table*}

The fourth, sixth, and third most abundant elements in the Sun by number,
C, N, and O are incredibly important to modern astronomy.
They also play a central role in the outstanding
solar modelling problem
\citep[e.g.][]{2008PhR...457..217B,2019ApJ...881..103Z}, wherein
a highly significant discrepancy exists pertaining to the
interior structure of the Sun as inferred precisely via
helioseismic inversions and standard solar models
(see \sect{s:helioseismology}).
O is the dominant source of opacity
at the most problematic region, namely the 
base of the convection zone, with C and N also providing
a non-negligible contribution there.
It has been noted
that higher C, N, and (especially) O 
abundances by $\approx 0.15\,\dex$ 
would alleviate the discrepancy 
with the sound speed \citep[e.g.][]{2009ApJ...704.1174P},
but this would not address all observational constraints \citep{2019A&A...621A..33B}. 
Adding to the controversy, 
all three elements form volatile compounds and are thus depleted in meteorites;
their abundances must be inferred through spectroscopy
of the solar photosphere.

Parallel to our work, the \cobold{} group have led a commendable
effort to measure the solar C, N, and O abundances from atomic lines,
similarly employing 3D hydrodynamic model atmospheres
\citep{2008A&A...488.1031C,2009A&A...498..877C,2010A&A...514A..92C,2013A&A...554A.126C,2015A&A...579A..88C,2015A&A...583A..57S}; they have not yet explored the molecular lines.
These studies tend to find slightly higher C, N, and O abundances 
by $0.03$ to $0.07\,\dex{}$ that
usually agree within the uncertainties \citep{2011SoPh..268..255C}.
These differences most likely do not originate in the
3D models or radiative transfer codes
themselves, but rather can be traced to our 
more conservative choice of unblended lines, the 
adopted oscillator strengths, and the measured equivalent widths, as 
has been demonstrated for C
\citep{2019A&A...624A.111A} and N
\citep{2020A&A...636A.120A}; moreover, unpublished
calculations 
have similarly indicated good agreement for the 
\ion{O}{I} $777\,\nm$ triplet when adopting exactly 
the same input data.

After submission of our article, we became aware of new 3D non-LTE line formation calculations of the \ion{O}{i} 777\,nm triplet and [\ion{O}{i}] 630\,nm forbidden line by \citet{Bergemann_2021_O} using 3D solar models computed with the \stagger{} and \bifrost{} codes, the latter with and without a chromosphere. Their results agree perfectly with those presented here and in \citet{2009ARA&A..47..481A}: $\lgeps{OI\,777} = 8.70 \pm 0.05$ and $\lgeps{[OI]\,630} = 8.71 \pm 0.03$, further supporting a low solar O abundance. Reassuringly, they find only a very minor effect ($\approx 0.01$\,dex) of a chromospheric temperature rise on these lines. 

An extension to using a 1D semi-empirical model atmosphere has recently been presented by 
\citet{2017A&A...600A..45C,2020A&A...643A.142C}, who derived a 3D solar model from the spectral inversion of three spatially resolved \ion{Fe}{I} lines. They applied it to the [\ion{O}{I}] $630.0\,\nm$ line in the Sun to infer $\lgeps{O}=8.80\pm0.03$.  While this is an interesting approach well worth pursuing further, 
we consider it premature to adopt their derived abundance until this empirical 3D solar model has been exposed to a variety of observational tests, which our theoretical 3D hydrodynamical models have successfully passed \citep{2009LRSP....6....2N,2013A&A...554A.118P}. In particular, it is not clear whether their model has the appropriate $T_{\rm eff}$ since it is neither an imposed constraint of the spectral inversion of the normalised line profiles, nor is it tested for afterwards. It remains to be seen whether consistent O abundances would be obtained from the [\ion{O}{I}], \ion{O}{I}, and OH lines.

\tab{t:cnoabund} shows the results of our reanalysis of 
all the available indicators of C, N, and O using our most recent 3D model atmosphere and the most up-to-date atomic and molecular data.
The atomic results are based on recent 3D non-LTE studies 
for C \citep{2019A&A...624A.111A}, 
N \citep{2020A&A...636A.120A},
and O \citep{2018A&A...616A..89A}; in the latter case, the calculations were extended to include the other available 
atomic indicators \citep[see also][]{amarsi2016three}.
The molecular results come from a new
homogeneous analysis of $879$ different lines,
with the background C, N, and O abundances
iterated to consistency \citep{amarsi_2021_cno}. 
Compared to \citet{2009ARA&A..47..481A}, the new recommended abundances are identical for N and O and only 0.03\,dex higher for C.

{\bf Carbon (Z=6):}  Of the three elements considered in this section, C has the largest number of available indicators
(\tab{t:cnoabund}) and is
arguably the best constrained.
The [\ion{C}{I}] result is drawn from the 3D non-LTE analysis of
the $872.7\,\nm$ line from \citet{2019A&A...624A.111A}, although
we note that this low-excitation line is well-reproduced in 3D LTE.
The \ion{C}{I} result is based on the $14$ lines also 
analysed in this 3D non-LTE study; here, the result has increased by $0.03\,\dex{}$ in light of new $gf$-values from large-scale atomic structure calculations by \citet{2021MNRAS.502.3780L}.
The molecular results are based on new 3D LTE calculations of
$39$ lines in the C$_{2}$ Swan system,
and CH lines which were divided into 
$51$ fundamental rovibrational (d$\upnu=1$) lines 
and seven electronic lines in the CH A-X system.
CO lines were also considered by
adopting the O abundance inferred from OH lines
for a particular model atmosphere
and including the error in this O abundance in the total error budget.
The CO lines were divided into two groups of rovibrational transitions,
including $28$ belonging to fundamental (d$\upnu=1$) bands
and $52$ to first overtone (d$\upnu=2$) bands.

The advocated solar C abundance is $8.46\pm0.04$.
This was evaluated as 
the weighted mean and error of the [\ion{C}{I}], 
\ion{C}{I}, C$_2$, CH, and CO results given in \tab{t:cnoabund};
to avoid giving these species too much weight, 
the rovibrational and electronic results for CH were first combined into a weighted mean ($8.46$),
as were the fundamental and first overtone results for CO ($8.48$).
The final result is a $0.03\,\dex{}$ increase over those stated in 
\citet{2009ARA&A..47..481A}. Nevertheless, the 
agreement between the different indicators found here is 
excellent, and noticeably improved over that earlier study. 
The reason for the new, slightly higher value can
primarily be attributed
to improvements to the atomic and molecular input data
([\ion{C}{I}] \citealt{2006JPhB...39.2159F};
\ion{C}{I} \citealt{2021MNRAS.502.3780L};
C$_2$ \citealt{2013JQSRT.124...11B};
CH \citealt{2014A&A...571A..47M};
CO \citealt{2015ApJS..216...15L}),
as well as to the employed molecular dissociation constants
\citep{2016A&A...588A..96B}.

{\bf Nitrogen (Z=7):}  Compared to C, there
are relatively few good indicators of the solar N abundance.
There are no [\ion{N}{I}] lines visible in the solar spectrum, 
and only five \ion{N}{I} lines were identified as being sufficiently
strong and unblended to be measured with confidence
\citep{2020A&A...636A.120A}.
The 3D non-LTE atomic result is combined 
with new 3D LTE results for molecular lines (\tab{t:cnoabund}).
This includes NH rovibrational lines that were divided into two
groups consisting of
$13$ pure rotational (d$\upnu=0$)
and $15$ fundamental (d$\upnu=1$) lines.
CN lines were also considered by
adopting weighted mean carbon abundances as
inferred from C$_2$, CH, and CO and 
for the different model atmospheres,
and including the error in this C
abundance in the total error budget.
The CN lines were separated into two groups consisting
of $59$ electronic lines in the 
$0-0$ band, which typically has the best data, and $463$ more lines with d$\upnu\geq1$ 
in various bands.

\tab{t:cnoabund} shows a discrepancy
of $0.13\,\dex$ between the atomic \ion{N}{I} result ($7.77$), and
the weighted mean ($7.90$) of the mean NH ($7.91$) and CN ($7.88$) results.
We can only speculate on where the inconsistency arises.
The \ion{N}{I} oscillator strengths were taken
from \citet{2002A&A...385..716T} and are expected to be
reliable at the $0.03\,\dex$ level based on rankings from NIST
(and verified with separate calculations as discussed in
Sect.~2.1 of \citealt{2020A&A...636A.120A}).
The oscillator strengths used here for NH and CN were taken from 
\citet{2014JChPh.141e4310B,2014ApJS..210...23B,2015JChPh.143b6101B},
however it is difficult to imagine both sets of
calculations having systematic offsets of $0.1\,\dex$.
Moreover, the dissociation energies of CN and NH are now well established \citep{2016A&A...588A..96B} and thus unlikely to be the reason behind the high molecular-based N abundance. 
We note that this discrepancy was already present
in \citet{2009ARA&A..47..481A},
albeit to a lesser degree, with the calculations
there based on an unpublished set of older molecular data
(A.J.~Sauval, private communication).

The advocated solar N abundance is $7.83\pm0.07$.
Owing to the atomic versus molecular discrepancy, this
was found by taking 
the unweighted mean of the atomic result together with the weighted mean of the molecular results.
The uncertainty is given by the standard error in the mean
(or equivalently, half the range between the two values).
The value is unchanged from that of
\citet{2009ARA&A..47..481A}; however, the stipulated uncertainty
is $0.02\,\dex{}$ larger.

{\bf Oxygen (Z=8):} The solar O abundance can be inferred
from a number of different atomic and molecular indicators
(\tab{t:cnoabund}).  For the atomic indicators, we carried out a
spectral line fitting analysis of the 
[\ion{O}{I}] $557.7\,\nm$, $630.0\,\nm$, and $636.4\,\nm$ lines,
and of the 
\ion{O}{I} $615.8\,\nm$, $777\,\nm$,
$844.7\,\nm$, and $926.9\,\nm$ lines, using the 3D non-LTE model
of \citet{2018A&A...616A..89A}.
The fitting method follows that presented
in \citet[][Chapter 4]{amarsi2016three}. Here,
the strengths of the CN and C$_2$ blending lines were fixed 
by adopting the line lists and abundances of
the earlier C and N analyses and the strength of the blending \ion{Ni}{I} $630\,\nm$ line calculated using the $gf$-value from 
\citet{2003ApJ...584L.107J} and the Ni abundance inferred
here ($\lgeps{Ni}=6.20$; \sect{s:fepeak}).
Ultimately, for [\ion{O}{I}] nearly all weight was given to the 
$630.0\,\nm$ line
after considering the relative difficulties in modelling 
the respective continua and blending contributions.
Similarly for \ion{O}{I}, most weight was given to 
the $777\,\nm$ triplet, because 
available centre-to-limb observations \citep{2009A&A...508.1403P}
allow us to correct for and constrain
errors in the cross-sections for 
inelastic collisions with neutral hydrogen (following \citealt{2018A&A...616A..89A}).
The molecular results are based on new 3D LTE calculations of
OH rovibrational lines, which were divided into three groups
consisting of $84$ pure rotational (d$\upnu=0$),
$50$ fundamental (d$\upnu=1$),
and $15$ first overtone (d$\upnu=2$) lines,
all giving highly consistent results.

The advocated solar O abundance is $8.69\pm0.04$.
Our new value was determined from 
the weighted mean and error of the [\ion{O}{I}], 
\ion{O}{I}, and OH results given in \tab{t:cnoabund},
albeit first combining the three rovibrational OH results
into a single weighted mean ($8.70$). 
An alternative value is obtained by taking the unweighted mean,
either of these three species, or of all five indicators in 
\tab{t:cnoabund}. This would put less weight on the 
strong \ion{O}{I} $777\,\nm$ triplet, but with both approaches
the value is increased by less than 
$0.005\,\dex$, since there is
excellent agreement between the different indicators.
Our recommended solar O abundance is unchanged from
\citet{2009ARA&A..47..481A}; however, the stipulated uncertainty
is $0.01\,\dex{}$ smaller.

Combined with our revised C abundance, the solar C/O ratio is $0.59\pm0.08$, a value which has steadily increased over the last few decades (e.g. $0.43$:  \citealt{1989GeCoA..53..197A}; $0.48$; \citealt{1991A&A...242..488G}; $0.49$: \citealt{1998SSRv...85..161G}; $0.50$: \citealt{2001ApJ...556L..63A,2002ApJ...573L.137A}; $0.54$:
\citealt{2004A&A...417..751A,2005A&A...431..693A}; $0.55$: \citealt{2009ARA&A..47..481A,2011SoPh..268..255C}; $0.56$: \citealt{2018A&A...616A..89A, 2019A&A...624A.111A}).  The Sun has a higher C/O ratio than most solar twins regardless of age \citep{2018ApJ...865...68B}, which could be a result of the Sun having migrated radially outwards by $2-3$\,kpc in the Milky Way disc since birth \citep{2012A&A...539A.143N}.

\subsection{Intermediate-mass elements: Sodium to calcium}
\label{s:intermediate}

{\bf Sodium (Z=11):}
The solar Na abundances presented in \citet{2009ARA&A..47..481A} and \citet{2015A&A...573A..25S} were based on a 3D LTE analysis 
corrected for non-LTE effects using 1D model atmospheres \citep{2011A&A...528A.103L}. Here, we present the first full 3D non-LTE calculations for Na in the Sun, employing the same 23-level Na model atom as in \citet{2011A&A...528A.103L}, which importantly contains a full quantum mechanical description for
inelastic \ion{Na}{i}+\ion{H}{i} collisions \citep{2010PhRvA..81c2706B,2010A&A...519A..20B}.
Experiments suggest these are sufficiently realistic for
non-LTE modelling \citep{2021ApJ...908..245B}.

A summary of the inferred abundances for the different lines using a variety of model atmospheres is provided in Table \ref{t:linelist_NaCa}. 
We employed the same five weak \ion{Na}{i} lines as well as transition data as in \citet{2015A&A...573A..25S}; the equivalent widths were corrected for minor contributions from blending lines.
The 3D non-LTE effects are minor ($\approx -0.03$\,dex) and very similar to those previously determined in 1D. 
Our recommended 3D non-LTE Na abundance, based on the unweighted mean of the five \ion{Na}{I} lines, becomes $\lgeps{Na} =6.22\pm0.01$ (statistical) $\pm 0.03$ (systematic), which is $0.01\,\dex{}$ larger than the value provided in \citet{2015A&A...573A..25S}.

{\bf Magnesium (Z=12):}
\citet{2015A&A...573A..25S} determined the solar photospheric Mg abundance based on 3D LTE line formation calculations coupled with non-LTE abundance corrections computed with different 1D model atmospheres. Their value,  $\lgeps{Mg} =7.59\pm0.04$, was only 0.01\,dex lower than the recommended abundance by \citet{2009ARA&A..47..481A}. 
The \ion{Mg}{i} and \ion{Mg}{ii} lines yielded fully consistent abundances in non-LTE, but they were significantly less so in LTE, highlighting the need for statistical equilibrium calculations for \ion{Mg}{ii} (non-LTE corrections are vanishingly small for \ion{Mg}{i} for the Sun). \citet{2015A&A...579A..53O} and \citet{2017ApJ...847...15B} carried out new 1D non-LTE calculations based on improved atomic data, but since their analyses were more focussed on other stars they have only very few lines in common with \citet{2015A&A...573A..25S}, most of their lines being much stronger in the Sun and less suitable for abundance purposes. 

For the first time, we carried out full 3D non-LTE calculations of Mg in the solar photosphere.
Our new atomic model consists of $48$ \ion{Mg}{i} levels,  $27$ \ion{Mg}{ii} levels, and the \ion{Mg}{iii} ground state; and $482$ bound-bound and $71$ bound-free radiative transitions,
drawing on new experimental and 
theoretical transition probabilities for \ion{Mg}{i} \citep{2017A&A...598A.102P}. We 
made use of realistic cross-sections for inelastic collisions with electrons from 
\citet{2017A&A...606A..11B} for \ion{Mg}{I}
and from \citet{2009JPhB...42v5002L} for \ion{Mg}{II},
and we used rate coefficients from \citet{2012A&A...541A..80B} and \citet{2016PhRvA..93d2705B} to describe inelastic collisions between \ion{Mg}{I} and neutral hydrogen.

A summary of the inferred abundances for the different lines using a variety of model atmospheres is provided in Table \ref{t:linelist_NaCa}.
The unweighted mean of eight \ion{Mg}{I} lines
gives $\lgeps{Mg}=7.56\pm0.01$ (statistical) in 3D non-LTE.
The systematic uncertainty as estimated from the 
3D LTE, \mtd{}, and HM models is less than $0.01\,\dex$ as these
all give highly consistent results for our \ion{Mg}{I} lines.
In 3D LTE, our \ion{Mg}{I} result is $0.04\,\dex$ smaller than 
that given in \citet{2015A&A...573A..25S} with
the new $gf$-values only having a minor effect ($-0.01$\,dex).
Instead, the difference stems
mainly from how pressure broadening was treated;
here, we used ABO theory 
\citep{1995MNRAS.276..859A,1997MNRAS.290..102B,1998MNRAS.296.1057B},
adopting new and extended tables (P.~S.~Barklem, private communication\footnote{\url{https://github.com/barklem/abo-cross}}).
Accordingly, we increased the systematic uncertainty in
the \ion{Mg}{I} result to $0.02\,\dex$.

The 3D non-LTE result from the unweighted mean of six lines of 
\ion{Mg}{II} is significantly smaller: 
$\lgeps{Mg}=7.52\pm0.03$
(statistical).  The \ion{Mg}{II} results have
a noticeably larger line-to-line scatter, beyond the uncertainties in the adopted $gf$-values, which are all classified A  or A$+$ ($\delta \log gf \le \pm 0.01$\,dex) by NIST \citep{2019APS..DMPN09004K}.
The systematic uncertainty as estimated from the 
3D LTE, \mtd{}, and HM models is $0.03\,\dex$ and is primarily
driven by the large departures from LTE, in particular for three of the lines.
Notably, we found that these lines are highly sensitive 
to the adopted recipe for inelastic collisions
between \ion{Mg}{II} and neutral hydrogen. Our results here are
based on the \citet{1968ZPhy..211..404D} recipe as formulated by
\citet{1993PhST...47..186L}, without any ad-hoc 
scale factors applied, for lack of a better approach.
Since the 3D LTE result is $\lgeps{Mg}=7.58$, 
more efficient collisions would help bring the 
\ion{Mg}{II} result into better agreement with \ion{Mg}{I}.
Another uncertainty concerns the choice of Mg abundance
for calculating the equation of state and hence electron pressures.
Here, this abundance was varied consistently with the line 
formation calculations.  
An alternative approach would fix the Mg abundance to the value
used to construct the model atmosphere ($\lgeps{Mg}=7.60$)
and vary the $gf$-values instead.   
Test calculations on the \mtd{} model suggest that 
this alternative approach could increase the overall \ion{Mg}{II} result by almost $0.02\,\dex{}$; the electron pressure does not affect \ion{Mg}{i} to first order since it is the minority ionisation stage in the solar photosphere.

Our recommended 3D non-LTE Mg abundance is the average of \ion{Mg}{i} and \ion{Mg}{ii}, weighted by the total statistical and systematic errors: $\lgeps{Mg}=7.55\pm0.03$.
A case could also be made for only adopting the \ion{Mg}{i} results given the additional complications with \ion{Mg}{ii,} but the final abundance would only be 0.01\,dex higher.

{\bf Aluminium (Z=13):}
\citet{2017A&A...607A..75N} recently performed full 3D non-LTE spectral line formation calculations for \ion{Al}{i} lines in the Sun,
based on their new model atom that includes 
realistic cross-sections for inelastic collisions with neutral hydrogen \citep{2013PhRvA..88e2704B}.
They employed the same 3D hydrodynamical \stagger{} model
solar atmosphere and 3D non-LTE radiative transfer code
\balder{} as we employed in this study.
Impressively, their solar modelling also agrees well with the observed IR emission lines of  \ion{Al}{i}. They adopted the same lines and line data as in \citet{2015A&A...573A..25S}, except that they excluded the 1089.1\,nm line due to telluric contamination, finding a solar photospheric Al abundance of $\lgeps{Al} =6.43\pm0.03$ (statistical + systematic). Here, we adopted the \citet{2017A&A...607A..75N} solar abundance, which is identical to the 3D LTE results of \citet{2015A&A...573A..25S}; for the Sun, departures from LTE are in fact largely unimportant for these five \ion{Al}{i} lines.

{\bf Silicon (Z=14):}
Here, we adopted the solar Si abundance from our recent analysis \citep{2017MNRAS.464..264A}, the first full 3D non-LTE line formation calculations performed for this element. The same atomic data and line weighting as in \citet{2015A&A...573A..25S} were used for the nine \ion{Si}{i} and one \ion{Si}{ii} lines. The inferred photospheric Si abundance ($\lgeps{Si} =7.51\pm0.03$) is identical to the values found by \citet{2009ARA&A..47..481A} and \citet{2015A&A...573A..25S}; this value is used to normalise the meteoritic abundance scale to the photospheric one, see Sect. \ref{s:ci}.

It should be noted that \citet{pehlivan_si} recently computed new $gf$-values for a large number of  \ion{Si}{i} and \ion{Si}{ii} lines, which are partly validated against experimental data. Had we adopted those for the seven Si lines in common with our study, the photospheric abundance would have increased to $\lgeps{Si} =7.53\pm0.03$ with a slightly reduced statistical scatter and a corresponding adjustment of the meteoritic abundances. We did, however, opt to retain the older experimental transition probabilities \citep{1973A&A....26..471G}, renormalised to the lifetimes of \citet{1980PhLA...76..125B} and \citet{1991PhLA..152..407O}), mainly since the new theoretical values imply a significant ionisation imbalance of 0.1\,dex between  \ion{Si}{i} and \ion{Si}{ii,} even in 3D non-LTE. The overall agreement with the meteoritic abundances further supports the \citet{1973A&A....26..471G} data (see Sect. \ref{s:sunvsci}).  Further theoretical and experimental work on Si would be very welcome.

{\bf Phosphorus (Z=15):}
We adopted the results from the 3D LTE analysis of \citet{2015A&A...573A..25S}: $\lgeps{P} =5.41\pm0.03$. 
To our knowledge, no non-LTE investigation of P exists in the literature, neither in 1D nor 3D. Due to the similarity between the weak \ion{P}{i} and \ion{S}{i} lines, the latter for
which non-LTE investigations do exist (see below), comparably small non-LTE effects could be expected for the two elements. 
The value is unchanged from \citet{2009ARA&A..47..481A}.

{\bf Sulphur (Z=16):}
The solar S abundance $\lgeps{S} =7.12\pm0.03$ is taken from \citet{2015A&A...573A..25S}
who included 1D non-LTE corrections ($-0.01$\,dex for optical lines but $-0.05..-0.10$ for the IR lines) from 
\citet{2005PASJ...57..751T} to their 3D LTE results. 
The value is unchanged from \citet{2009ARA&A..47..481A}.

{\bf Potassium (Z=19):}
The recommended solar K abundance in \citet{2009ARA&A..47..481A} was $\lgeps{K} =5.03\pm0.09$, which \citet{2015A&A...573A..25S} updated to $\lgeps{K} =5.04\pm0.05$. The latter result was based on a 3D LTE analysis of five weak lines (excluding the strong 766.5 and 769.9\,nm \ion{K}{i} resonance lines) together with 1D non-LTE abundance corrections taken from  \citet{1996PASJ...48..511T}
in the absence of consistent non-LTE calculations with a 3D solar model. 
\citet{2019A&A...627A.177R} recently presented a new 1D non-LTE analysis with a comprehensive model atom consisting of 
133 atomic levels of \ion{K}{i} plus the ground state of \ion{K}{ii} and 250 radiative bound-bound transitions. Particular attention was placed on calculating or adopting reliable cross-sections
for inelastic collisions, both with electrons,
and with neutral hydrogen. 
Their 1D non-LTE calculations with a theoretical hydrostatic \marcs{} model atmosphere using the equivalent widths of three lines (including \ion{K}{i} 769.9nm) yielded a solar abundance of $\lgeps{K} =5.11\pm0.01$ (standard deviation only).

Here, we present a full 3D non-LTE calculations for \ion{K}{I} using the same model atom as in \citet{2019A&A...627A.177R}.
A summary of the inferred abundances for the different lines using a variety of model atmospheres is provided in Table \ref{t:linelist_NaCa}.
Compared to \citet{2015A&A...573A..25S}, the $gf$-values have been updated to those recommended by NIST \citep{2019APS..DMPN09004K},
which in all cases are estimated to have an accuracy better than $1\%$. Furthermore, the spectral line profiles were re-examined and the equivalent widths remeasured, leading us ultimately to exclude the 404.4 and 580.2\,nm lines due to their substantial and uncertain blending.  Our recommended solar K abundance is thus based on the measured equivalent widths of three \ion{K}{i} lines: 693.9, 1177.0, and 1252.2\,nm. 
Although we did not consider the \ion{K}{i} 769.9\,nm resonance line for the final result due to its large line strength and very pronounced non-LTE effects ($\approx +0.2$\,dex), we note that its 3D non-LTE-based solar K abundance is fully consistent with those of the weak \ion{K}{i} lines.
Nevertheless, the departures from LTE are noticeable for the three retained \ion{K}{I} lines (3D non-LTE - 3D LTE $\approx -0.05$\,dex), which reinforces the importance of a consistent 3D non-LTE analysis as employed here.
Our recommended solar K abundance, based on the unweighted mean of the three \ion{K}{I} lines, is $\lgeps{K} =5.07\pm0.01$ (statistical) $\pm 0.03$ (systematic).

{\bf Calcium (Z=20):}
\citet{2015A&A...573A..25S} slightly revised the present-day photospheric Ca abundance of \citet{2009ARA&A..47..481A} to $\lgeps{Ca} =6.32\pm0.03$ with excellent agreement between the eleven \ion{Ca}{i} and five \ion{Ca}{ii} lines employed. These results were based on 3D LTE line formation calculations with our own 1D non-LTE abundance corrections calculated on the \mtd{} model atmosphere,
which range from $+0.02$ to $-0.03\,\dex$ for \ion{Ca}{i}, and
$0.00$ to $-0.07\,\dex$ for \ion{Ca}{ii}.
Recently, \citet{2017A&A...605A..53M} and 
\citet{2019A&A...623A.103O} presented new 1D non-LTE investigations that predict more severe, negative, non-LTE corrections for both species but more so for \ion{Ca}{II}, which would worsen the ionisation balance obtained in  \citet{2015A&A...573A..25S}.

In light of these issues, we carried out a detailed reanalysis of Ca, based on a consistent 3D non-LTE line formation study for the Sun.
Our new Ca model atom consists of  141 \ion{Ca}{i} levels,  46 \ion{Ca}{ii} levels, and the \ion{Ca}{iii} ground state, which are coupled radiatively by 2271 bound-bound and 135 bound-free transitions. We employed data for inelastic collisions with
electrons from \citet{2019PhRvA..99a2706Z} for \ion{Ca}{I}
and from \citet{2007A&A...469.1203M} for \ion{Ca}{II};
while for inelastic collisions with
neutral hydrogen we draw on data from \citet{2016PhRvA..93d2705B} 
and \citet{2018ApJ...867...87B} for \ion{Ca}{I} and \ion{Ca}{II,} respectively. 
\citet{2020A&A...637A..80O} reported a non-negligible influence of 
non-LTE effects on \ion{Mg}{I} influencing the statistical equilibrium
of Ca in the Sun. We therefore ran test calculations with the background Mg opacity treated in non-LTE; however, these were found to have a negligible impact on our results, at least for the Ca lines
adopted in our abundance analysis.

A summary of the inferred abundances for the different lines using a variety of model atmospheres is provided in Table \ref{t:linelist_NaCa}.
Compared to \citet{2015A&A...573A..25S}, we adopted the same line data and mostly the same lines, dropping only the $616.13\,\nm$ that is affected by the broad damping wings of the neighbouring \ion{Ca}{I} $616.22\,\nm$ line.
The unweighted mean results from the ten \ion{Ca}{I} lines
and the five \ion{Ca}{II} lines agree to better than $0.005\,\dex{}$, with a very small line-to-line scatter for each species. Our recommended solar Ca abundance is $\lgeps{Ca} =6.30\pm0.03$ (total).

\subsection{Fe-peak elements: Scandium to nickel}
\label{s:fepeak}

{\bf Scandium (Z=21):}
\citet{2015A&A...573A..26S} relied on a 3D LTE analysis of five \ion{Sc}{i} and nine \ion{Sc}{ii} lines coupled with non-LTE corrections computed with 1D model atmospheres to determine a solar Sc abundance of $\lgeps{Sc} =3.16\pm0.04$. With the new, highly accurate experimental transition probabilities of \citet{2019ApJS..241...21L}, the weighted average becomes $\lgeps{Sc} =3.14\pm0.01$ (statistical) $\pm 0.04$ (systematic)\footnote{\citet{2017MNRAS.472.3337P} also measured new lifetimes, which coupled with theoretical branching factors yield new rescaled semi-empirical \ion{Sc}{ii} oscillator strengths. Adopting those instead and excluding the highly discrepant 442.07\,nm line would yield a Sc abundance of  $\lgeps{Sc} =3.10\pm0.05$. Here, we preferred the arguably less uncertain experimental values of \citet{2019ApJS..241...21L}, which are available for all of our lines.}. It is noted that the departures from LTE are substantial for the \ion{Sc}{i} lines, but the weighted mean Sc abundance is largely driven by \ion{Sc}{ii}.  There is good agreement between the two ionisation stages, both of which give noticeably higher abundances than the CI meteoritic value (Sect. \ref{s:ci}). \citet{2019ApJS..241...21L} considered roughly twice as many Sc lines in their solar analysis, many of which are clean and have excellent atomic data; we postpone for later a study of these lines in conjunction with 3D non-LTE calculations and realistic collisional cross-sections. 

{\bf Titanium (Z=22):}
In their analysis of Fe-peak elements, \citet{2015A&A...573A..26S} already benefited from the highly accurate experimental transition probabilities for Ti from the Wisconsin atomic physics group (\ion{Ti}{i}: \citealt{2013ApJS..205...11L}; \ion{Ti}{ii}: \citealt{2013ApJS..208...27W}\footnote{\citet{2020A&A...643A.156L} presented a new set of theoretical \ion{Ti}{ii} $gf$-values, which are in good overall agreement with the \citet{2013ApJS..208...27W} experimental values, although they lead to significantly larger abundance scatter for our lines; we assess the latter source to be more reliable for our purposes.}). They also applied non-LTE abundance corrections calculated with 1D model atmospheres, which are significant for  \ion{Ti}{i} ($\approx +0.06$\,dex) but unimportant for \ion{Ti}{ii}. \citet{2015A&A...573A..26S} found a large and unexplained difference in Ti abundance from the two ionisation stages: $4.88\pm0.05$ versus $4.97\pm0.04$ from \ion{Ti}{I} and \ion{Ti}{II,} respectively; their recommended solar Ti abundance was a weighted average of the two, resulting in $\lgeps{Ti} =4.93\pm0.04$. It is noteworthy that all of the considered theoretical and semi-empirical 1D model atmospheres implied significantly larger \ion{Ti}{i} abundances than with the 3D \stagger\ model and more in line with the  \ion{Ti}{iI} results. This strong model atmosphere sensitivity suggests that the adopted 1D non-LTE corrections are not appropriate in 3D and that a consistent 3D non-LTE analysis would yield higher \ion{Ti}{i} abundances. In contrast to \citet{2015A&A...573A..26S}, we based our recommended solar Ti abundance solely on \ion{Ti}{ii} but increased the final error somewhat to reflect the discrepancy between the ionisation stages: $\lgeps{Ti} =4.97\pm0.05$ (total). A full 3D non-LTE study with realistic collisional cross-sections would be highly desirable; \citet{2013ApJS..205...11L} and \citet{2013ApJS..208...27W} list a large number of high-quality Ti lines with now excellent atomic data available.

{\bf Vanadium (Z=23):}
The recommended solar photospheric V abundance of \citet{2015A&A...573A..26S} was based on a 3D LTE analysis of \ion{V}{i} lines with an ad hoc non-LTE correction of 0.10\,dex across the board in the absence of actual statistical equilibrium calculations based on the behaviour of Sc and Ti in 1D model atmospheres. As a minority species, \ion{V}{i} is very sensitive to the atmospheric structure: the four considered theoretical and semi-empirical 1D model atmospheres imply a $0.07-0.18$\,dex higher abundance. Unfortunately, little guidance is available from the few available  \ion{V}{ii} lines due to their heavy blending, although nominally they suggest a $\approx 0.1$\,dex higher V abundance, both in 3D and 1D. Updating the \citet{2015A&A...573A..26S} results with the \citet{2014ApJS..215...20L} and \citet{2014ApJS..214...18W} experimental $gf$-values, the mean  \ion{V}{i}  abundance is only $+0.01$\,dex higher, while the \ion{V}{ii} result changes by $-0.02$\,dex \footnote{\citet{2016ApJS..224...35H} measured new experimental transitional probabilities but only for a subset of our solar diagnostics lines. On average, the difference with the \citet{2014ApJS..215...20L} values is small, $\approx +0.02$\,dex, and the line-to-line abundance scatter would increase somewhat. We argue that the \citet{2014ApJS..215...20L} data are preferable for our purposes (see also \citet{2017ApJS..231...18S}).}. We caution that our ad hoc \ion{V}{i} non-LTE correction may well be underestimated, and we strongly encourage the first non-LTE investigation of this element to be performed. In the meantime we recommend the value found by \citet{2015A&A...573A..26S} updated with the \citet{2014ApJS..215...20L} transition probabilities: $\lgeps{V} =3.90\pm0.08$ (total).

{\bf Chromium (Z=24):}
The solar photospheric Cr abundance of  \citet{2015A&A...573A..26S} was based on the weighted average of a large number of \ion{Cr}{i} and \ion{Cr}{ii} lines with 1D non-LTE abundance corrections. Updating this result with the new experimental \ion{Cr}{ii} transition probabilities of \citet{2017ApJS..228...10L} leads to the same mean abundance: $\lgeps{Cr} =5.62\pm0.04$.

{\bf Manganese (Z=25):}
A new solar Mn abundance analysis was presented by \citet{2019A&A...631A..80B} based on full 3D non-LTE calculations. Using the same 3D hydrodynamical solar model \citep{2013A&A...557A..26M} as for other elements and equipped with new ab initio inelastic \ion{Mn}{i}+\ion{H}{i} collisional cross-sections from \citet{2017A&A...606A.106B}
(see also \citealt{2020A&A...637A..28G}), they obtained $\lgeps{Mn} =5.52\pm0.04$. This is 0.10\,dex larger than the result of 
\citet{2015A&A...573A..26S}$(5.42\pm0.04$) based on 3D LTE model spectra combined with \mtd{} non-LTE abundance corrections; the two results do not overlap within the error bars.
The two studies have very few  \ion{Mn}{i} lines in common, making a detailed comparison difficult. The abundance difference cannot be traced to the adopted transition probabilities since the same sources are employed (mostly \citealt{2011ApJS..194...35D} and \citealt{2007A&A...472L..43B}). 
Departures from LTE are obviously very prominent, with the 1D non-LTE $-$ 1D LTE difference being $+0.07$\,dex.
according to \citet{2019A&A...631A..80B}. 
One may suspect that the 3D non-LTE effects and Mn abundances in \citet{2019A&A...631A..80B} have been somewhat overestimated due to incomplete UV line blanketing over-estimating the photo-ionisation rates and thus weakening the predicted \ion{Mn}{i} lines; a detailed comparison is further complicated by \citet{2019A&A...631A..80B} employing different non-LTE radiative transfer codes for their 1D and 3D analyses.

Until the situation has been clarified, we advocate the recommended photospheric Mn abundance of \citet{2015A&A...573A..26S} but with increased errors to reflect the remaining uncertainties associated with the non-LTE modelling: $\lgeps{Mn} =5.42\pm0.06$.
Some confidence in this result can be had by noting that \citet{2015A&A...573A..26S} obtained a value of $5.43$ from their consistent \mtd{} non-LTE analysis, only $0.01\,\dex{}$ larger. Our expectation is that, at least for atomic and ionic species in the disc-centre intensity spectrum of the Sun, the full 3D non-LTE result for Mn should be similar to the \mtd{} non-LTE abundance, as we have found in our investigations of Li, C, N, O, Na, Mg, Al, Si, K, Ca, and Fe. Nevertheless, we encourage the issue to be revisited with new 3D non-LTE calculations with realistic inclusion of blending lines both for the statistical equilibrium and for the \ion{Mn}{i} abundance diagnostic lines.

{\bf Iron (Z=26):}
In astronomy, Fe is often used as a proxy for the overall metallicity of a star on account of the element's relatively high abundance and the wealth of readily measurable lines. Iron is also an important opacity contributor in stellar interiors. In the Sun, many lines of both \ion{Fe}{i} and \ion{Fe}{ii} with accurate experimental atomic data are available for a spectroscopic analysis \citep[e.g.][]{1999A&A...347..348G,2000A&A...359..743A,2011SoPh..268..255C,2011A&A...528A..87M}. 
From a 3D analysis, \citet{2015A&A...573A..26S} obtained $7.45\pm0.04$ (including 1D non-LTE corrections of $\approx +0.01$\,dex) and $7.51\pm0.04$, from \ion{Fe}{i} and  \ion{Fe}{ii} lines, respectively,
leading to a weighted average of $\lgeps{Fe} =7.47\pm0.04$ (statistical + systematic). \citet{2017MNRAS.468.4311L} updated this to $\lgeps{Fe} =7.47\pm0.04$ based on fully consistent 3D non-LTE calculations for Fe that for the first time were based on a large model atom that included ab initio collisional cross-sections for  \ion{Fe}{i}+\ion{H}{i} \citep{2018A&A...612A..90B}.

We undertook a new analysis of \ion{Fe}{i} and \ion{Fe}{ii} using full 3D non-LTE calculations and improved atomic data.
Our new Fe model atom consists of $100$ \ion{Fe}{i} levels, $76$ \ion{Fe}{ii} levels, and the \ion{Fe}{iii} ground state, which are radiatively coupled through $4313$ bound-bound and $100$ bound-free transitions, amounting to $46000$ wavelength points.
We employed data for inelastic collisions with
electrons from \citet{2018ApJ...867...63W} for \ion{Fe}{I}
and from \citet{2015ApJ...808..174B} and
\citet{1995A&A...293..953Z} for \ion{Fe}{II};
while for inelastic collisions with
neutral hydrogen, we drew on data from \citet{2018A&A...612A..90B}
and \citet{2019MNRAS.483.5105Y} for \ion{Fe}{I} and \ion{Fe}{II,} respectively. Photoionisation cross-sections for \ion{Fe}{I} were sourced from \citet{2019PhRvA..99b3430Z}.
Recently, new accurate experimental transition probabilities for additional excellent solar \ion{Fe}{i} lines have been published \citep{2014ApJS..215...23D,2017ApJ...848..125B}, which we employed together with previous lines used by \citet{2015A&A...573A..26S} with the transition probabilities recommended by the Oxford \citep[e.g.][]{1995A&A...296..217B}, Hannover \citep[e.g.][]{1991A&A...249..545H}, and Wisconsin \citep[e.g.][]{1991JOSAB...8.1185O} groups\footnote{We did not utilise any of the IR \ion{Fe}{i} lines for which new experimental data are now available \citep{2013ApJ...779...17R} since these particular lines are typically too strong and/or blended in the Sun for a high-precision analysis.}.

\begin{figure}[t!]
    \begin{center}
        \includegraphics[width=9.5cm]{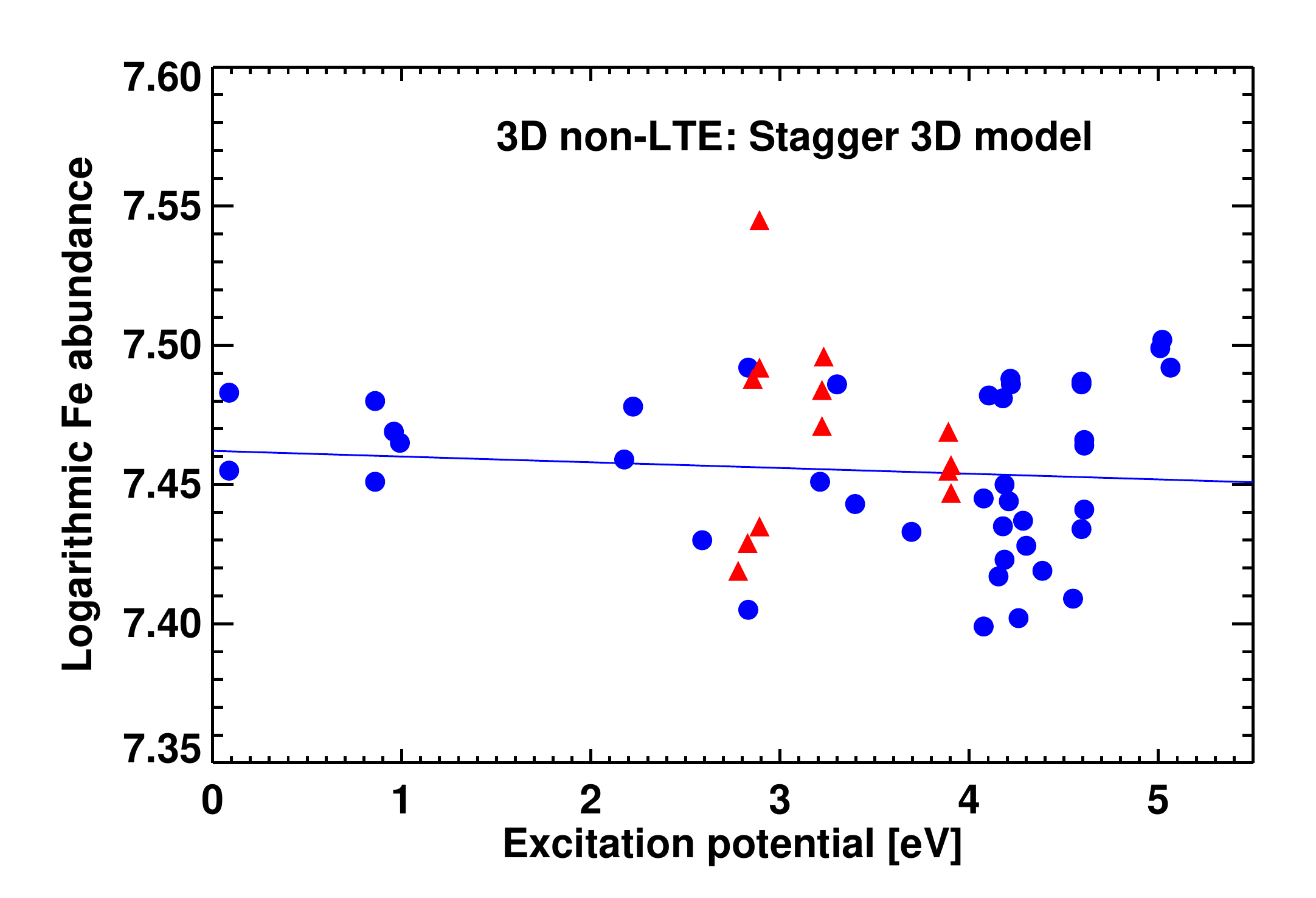}
        \includegraphics[width=9.5cm]{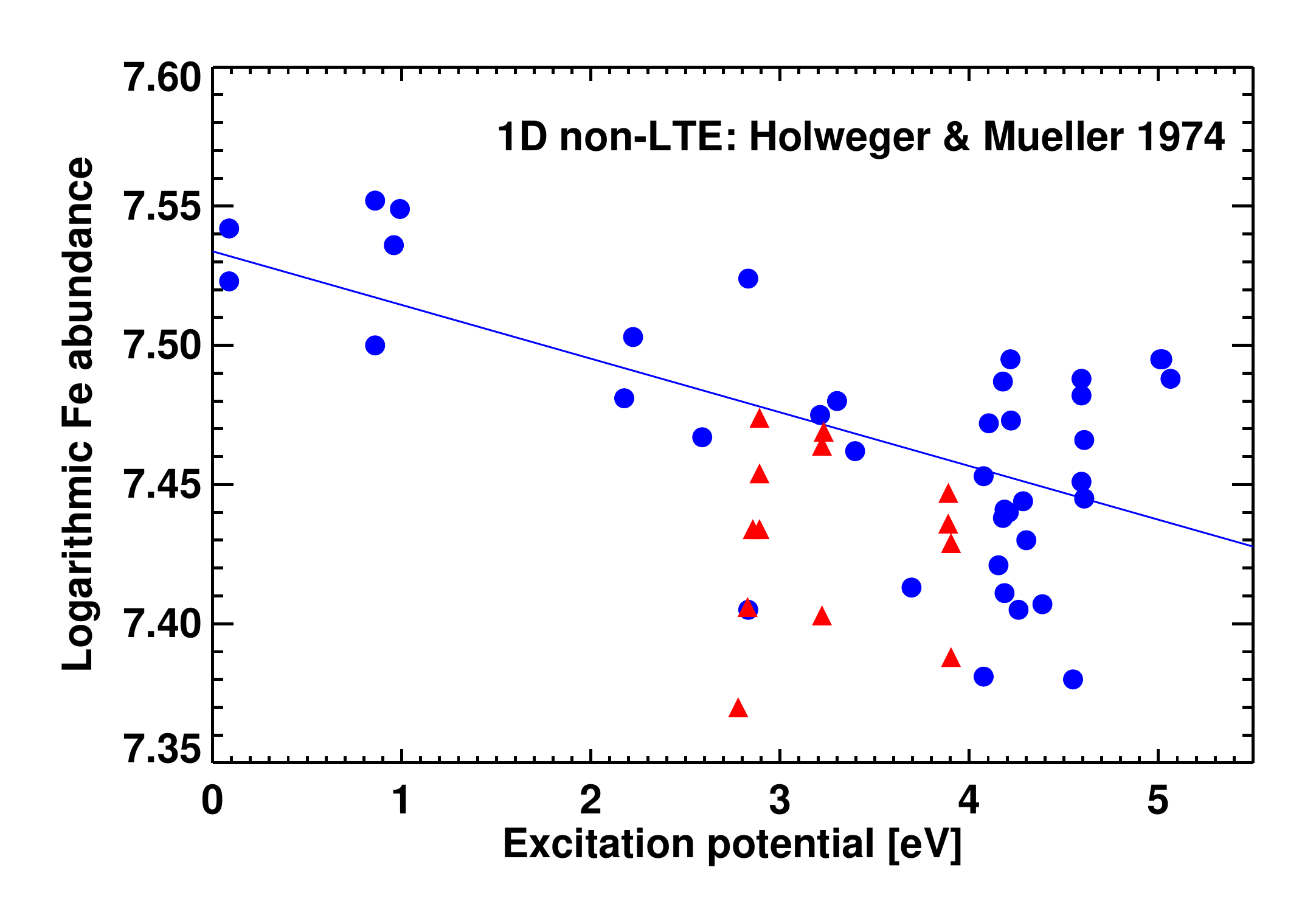}
        \caption{Inferred photospheric Fe abundances from detailed non-LTE calculations using the 3D hydrodynamical Stagger model ({\em Upper panel}) and the 1D \citet{1974SoPh...39...19H} model ({\em Lower panel}). The results for \ion{Fe}{i} and \ion{Fe}{ii} lines are shown as blue circles and red triangles, respectively. The solid lines denote the least-square fits to the \ion{Fe}{i} results. In 3D non-LTE, there is excellent excitation and ionisation balance for Fe, contrary to the 1D case; it is noted that the trend with excitation potential with the \citet{1974SoPh...39...19H} model is even steeper in LTE (see Table \ref{t:linelist_Fe}).}
        \label{f:fe}
    \end{center}
\end{figure}

A summary of the inferred abundances for the different lines using a variety of model atmospheres is provided in Table \ref{t:linelist_Fe}. We now obtain excellent agreement between the two ionisation stages: $7.46\pm0.04$ and $7.47\pm0.04$ from \ion{Fe}{I} and \ion{Fe}{II} respectively, based on unweighted means and
including statistical and systematic uncertainties. Departures from LTE are small for both species in the Sun. There are no significant dependencies with wavelength, excitation potential, line strength or provenance of the $gf$-values in 3D, in contrast to the results using the 1D semi-empirical \citet{1974SoPh...39...19H}  model atmosphere (Fig. \ref{f:fe}). With 1D theoretical models, the mean Fe abundance becomes unrealistically low ($\lgeps{Fe}<7.40$), while with the \mtd{} model, ionisation balance is not fulfilled, thus attesting to the need for full 3D non-LTE calculations. Our recommended photospheric Fe abundance is the average of the \ion{Fe}{I} and \ion{Fe}{II} results: $\lgeps{Fe} =7.46\pm0.04$ (total), a value which has hardly moved over the past two decades in spite of major improvements in 3D and non-LTE analysis techniques and better atomic data \citep{2000A&A...359..743A,2005ASPC..336...25A,2009ARA&A..47..481A,2012MNRAS.427...27B,2015A&A...573A..26S,2016MNRAS.463.1518A,2017MNRAS.468.4311L}.

{\bf Cobalt (Z=27):}
We update the \citet{2015A&A...573A..26S} results with the  $gf$-values from \citet{2015ApJS..220...13L} for the \ion{Co}{i} lines in common: $\lgeps{Co} =4.94\pm0.05$ (total). While the new Co abundance is only $0.01$\,dex higher, the reduced line-to-line scatter is a testament to the high accuracy of the new experimental data. As for most Fe-peak elements, one concern relates to the adopted non-LTE corrections, which may be prone to error as they stem from our 1D calculations, and mainly draw on classical and semi-empirical recipes for inelastic collisions with electrons and with neutral hydrogen.

{\bf Nickel (Z=28):}
In the absence of any non-LTE calculations for \ion{Ni}{i}, we adopt the 3D LTE based Ni abundance of \citet{2015A&A...573A..26S}, 
who used the accurate experimental transition probabilities from \citet{2014ApJS..211...20W}: $\lgeps{Ni} =6.20\pm0.04$.
Based on our detailed 3D non-LTE calculations for \ion{Fe}{i}, we expect the non-LTE effects for \ion{Ni}{i} to be similarly minor, but this requires confirmation.

\subsection{The heavy elements: Copper to uranium}
\label{s:heavy}

Relatively little work has been done recently to update the solar photospheric analysis by \citet{2015A&A...573A..27G} of the neutron capture elements. With the exception of the elements listed below, we adopt their recommended solar abundances. We note that many of these elements have few available spectral lines located in the crowded UV region, causing major problems in fitting the lines. Taking an even more conservative approach, we deemed some lines used by \citet{2015A&A...573A..27G} as too uncertain for a meaningful abundance determination even if very few, if any, alternative lines of the element are available. In some cases, we increased the error bars to attest to the challenges in working with heavily blended lines; often the blending lines are not even identified. The reader is furthermore cautioned that departures from LTE are largely unexplored territory for the heavy elements in the Sun. This deficiency can be expected to introduce significant systematic errors, quite possible at the level of $\pm 0.1$\,dex, in particular for many of the minority species.

{\bf Copper (Z=29):}
We retain the solar Cu abundance $\lgeps{Cu} =4.18\pm0.05$ of \citet{2015A&A...573A..27G}  despite it relying on old and relatively uncertain experimental oscillator strengths 
\citep[e.g.][]{1968ZA.....69..180K}
since the newer theoretical $gf$-values of 
\citet{2014ApJS..211...30L} imply an unacceptably large line-to-line abundance scatter. New laboratory data for this astrophysically important element would be welcome.

{\bf Germanium (Z=32):}
We reinspected the spectral line profile of the quite strong but blended \ion{Ge}{i} 326.95\,nm line and revised the equivalent width downwards relative to \citet{2015A&A...573A..27G} to $4.20\,\mathrm{pm}$.  Our 3D LTE result 
consequently shifts downwards to $\lgeps{Ge} =3.62\pm0.10$.

{\bf Rubidium (Z=37):}
The only two available \ion{Rb}{i} lines in the solar spectrum, the resonance lines at 780.03 and 794.76\,nm, are unfortunately both perturbed by other species. We revised the measured line strengths of \citet{2015A&A...573A..27G} (to 0.53 and 0.29\,pm, respectively) to better account for these blends and adjusted the remaining uncertainties to reflect the difficulties in fitting the weak \ion{Rb}{i} lines. The resulting 3D LTE abundance is slightly lower than before: $\lgeps{Rb} =2.44\pm0.08$ (total). Very recently, \citet{2020AstL...46..541K} performed 1D non-LTE calculations for the \ion{Rb}{i} resonance lines, employing ab initio data for inelastic collisions with neutral hydrogen
\citep{2018MNRAS.473.3810Y}. They found substantial departures from LTE ($\Delta \lgeps{Rb}  = -0.12$\,dex) and a solar Rb abundance consistent with the meteoritic value for the first time\footnote{Even more recently, \citet{Takeda_2021_Rb} performed a similar 1D non-LTE study, including for the Sun, where they found a less noticeable non-LTE correction: $-0.05$\,dex. Because they only considered the classical \citet{1968ZPhy..211..404D} formula with a scaling factor for the inelastic H collisions and no specification is given as to which recipe for the electron collisions is employed, we here give a preference to the results of \citet{2020AstL...46..541K}. Revisiting the solar Rb abundance issue with 3D non-LTE calculations would be desirable.}. Their study employed blending lines in the statistical equilibrium computations as well as Rb spectrum synthesis. 
Applying the new non-LTE corrections of \citet{2020AstL...46..541K} to our 3D LTE value above implies a solar Rb abundance of $\lgeps{Rb} =2.32\pm0.08$.

{\bf Rhodium (Z=45):}
We updated the \citet{2015A&A...573A..27G}  Rh abundance with the new experimental transition probabilities for \ion{Rh}{i} lines from \citet{2015MNRAS.450..223M}, while restricting the analysis to the 369.24\,nm line due to the heavy blending of the 343.49\,nm line. When also considering the uncertainties in the line strength and $gf$-values,  we obtain $\lgeps{Rh} =0.78\pm0.11$ (statistical + systematic). The derived photospheric Rh abundance remains substantially below the value measured in CI chondrites, which may indicate pronounced departures from LTE for this minority species in the 3D model that are not yet possible to account for. It is noted that 1D semi-empirical solar atmosphere models yield abundances in better agreement with the meteoritic evidence; the shallower temperature gradient of the \citet{1974SoPh...39...19H} model tends to result in smaller departures from LTE \citep{2005ARA&A..43..481A}, a phenomenon termed 'non-LTE masking' by \citet{1982A&A...115..104R}.

{\bf Palladium (Z=46):}
We reassessed the \ion{Pd}{i} lines at 324.27 and 340.46\,nm
and revised their equivalent widths upwards relative to 
\citet{2015A&A...573A..27G}, to $2.50\,\mathrm{pm}$ and
$3.58\,\mathrm{pm,}$ respectively. Our 3D LTE result 
consequently shifts upwards to $\lgeps{Pd} =1.57\pm0.10$, 
which is incidentally identical to the value recommended by \citet{2009ARA&A..47..481A}.

{\bf Barium (Z=56):}
Unfortunately, only a few spectral lines of Ba are available as abundance diagnostics, which furthermore are strong and saturated. The solar Ba abundances presented in \citet{2009ARA&A..47..481A} and \citet{2015A&A...573A..27G} were based on 3D LTE calculations to which 1D non-LTE abundance corrections were applied; naturally this approach is not fully self-consistent, but it was the only one available at the time for this element. Very recently, \citet{2020A&A...634A..55G} revisited the issue using full 3D non-LTE computations with the same \stagger{} 3D hydrodynamical solar model as employed elsewhere in our study. Their inferred solar Ba abundance of  $\lgeps{Ba} =2.27\pm0.01$ (statistical) $\pm 0.02$ (systematic) was obtained from a 1D LTE spectrum synthesis with a \marcs{} model atmosphere \citep{2008A&A...486..951G} to which abundance corrections from blending lines and 3D (i.e. the differences in abundances in 3D non-LTE to obtain the 1D LTE equivalent width) were applied separately; this is not fully consistent either, but it is a more realistic approach than that in \citet{2015A&A...573A..27G}. In addition to the three  \ion{Ba}{ii} lines employed by \citet{2015A&A...573A..27G}, \citet{2020A&A...634A..55G} also considered the heavily blended 614.17\,nm line and obtained consistent results.
It is noted that the 3D LTE abundances in the two studies differ substantially for unknown reasons.

Here, we adopt the \citet{2020A&A...634A..55G} 3D non-LTE based result but inflate their uncertainties slightly to account for possible systematic errors from the model atmosphere, non-LTE calculations, and their assumed microturbulence in 1D: $\lgeps{Ba} =2.27\pm0.05$ (total). A new study directly comparing the 3D non-LTE line profiles (including blends) with observations would be welcome for this important element, especially given the difference with the CI chondritic abundance ($\lgeps{Ba} =2.18\pm0.02$).

{\bf Praseodymium (Z=59):}
We apply the 1D non-LTE abundance correction ($+0.03$\,dex) for \ion{Pr}{i} from \citet{2009A&A...495..297M} to the 3D LTE results of \citet{2015A&A...573A..27G}, 
which in turn was based on 3D abundance corrections applied to the 1D spectrum synthesis of \citet{2009ApJS..182...80S}. The revised Pr abundance becomes $\lgeps{Pr} =0.75\pm0.05$.

{\bf Tungsten (Z=74):}
Compared to \citet{2015A&A...573A..27G}, we remove the weak $484.38\,\nm$ line owing to uncertainties in measuring its equivalent width.  Therefore, restricting the 3D LTE analysis of \citet{2015A&A...573A..27G} to the \ion{W}{i} 400.875\,nm line implies $\lgeps{W} =0.79\pm0.11$.

{\bf Osmium (Z=76):}
After inspecting the spectral line profile of the \ion{Os}{i} 330.16\,nm transition, we revise its equivalent width downwards relative to 
\citet{2015A&A...573A..27G}, to 
$0.86\,\mathrm{pm}$.  The resulting 3D LTE result is $0.05\,\dex{}$ lower: $\lgeps{Os} =1.35\pm0.12$.

{\bf Iridium (Z=77):}
We judge that the only detectable \ion{Ir}{i} line, at 322.08\,nm, is too badly blended and has too uncertain a continuum placement to allow us to make a meaningful solar Ir abundance measurement.

{\bf Gold (Z=79):}
Accounting for the uncertain continuum placement around the weak and badly blended \ion{Au}{i} 312.28\,nm line leads to a more conservative error estimate than in \citet{2015A&A...573A..27G}: $\lgeps{Au} =0.91\pm0.12$ (total).

{\bf Lead (Z=82):}
We update the solar Pb abundance based on our new measurement of the equivalent width (0.91\,pm) of the \ion{Pb}{i} 368.35\,nm, which results in $\lgeps{Pb} =1.95\pm0.08$, $0.03\,\dex{}$ larger than
in \citet{2015A&A...573A..27G}.

\subsection{Elements observed in sunspots: Fluorine, chlorine, indium, and thallium}
\label{s:halogen}

{\bf Fluorine (Z=9):}
While no spectral lines of F are detectable in the photospheric spectrum of the quiet Sun, there are features of the HF molecule in sunspot spectra that enable a determination of the solar F abundance. \citet{2014ApJ...788..149M} analysed the umbral atlas of \citet{2001sus..book.....W} with a 1D theoretical model atmosphere corresponding to an effective temperature of $\teff = 4250$\,K to estimate $\lgeps{F} =4.40\pm0.25$; most of the uncertainty stems from the choice of model atmosphere since the employed radiative-equilibrium model  without magnetic fields may not be appropriate for this particular sunspot umbra. Here, we recommend the value of \citet{2014ApJ...788..149M}, which represents a 0.16\,dex reduction compared to the older \citet{1969ApL.....4..143H} result adopted by \citet{2009ARA&A..47..481A}. 

{\bf Chlorine (Z=17):}
Similar to F, the photospheric Cl abundance can only be estimated from sunspot spectra since the HCl molecule requires cool temperatures; no atomic features of Cl are present in the solar spectrum. Recently, \citet{2016AJ....152..196M} revisited the previous HCl sunspot study of \citet{1972ApJ...175L..95H} with improved molecular data and observations to obtain a 0.19\,dex lower abundance: $\lgeps{Cl} =5.31\pm0.12$. 
Given the remaining model challenges, especially in finding a suitable model atmosphere for a sunspot of unknown $\teff$, we argue that the quoted uncertainty is likely overly optimistic, and we instead adopt $\pm 0.20$\,dex. A more accurate determination of the solar abundances of the halogen elements would be possible through the abundance ratio of another species with similar temperature dependence and spectral features with well-determined $gf$-values, such as FeH and MgH.

{\bf Indium (Z=49):}
Analyses of the \ion{In}{i} 451.13\,nm spectral line in the quiet Sun have traditionally yielded an In abundance much higher than the CI chondritic value (e.g. \citealt{2002SoPh..211....3B}: $\lgeps{In} =1.56\pm0.20$). \citet{2008MNRAS.384..370V} showed that this is likely due to an unknown blend and instead confirmed that the solar In abundance is close to the meteoritic value based on a sunspot spectrum: $\lgeps{In} =0.80\pm0.20$.
We adopt this value here, as we did in \citet{2009ARA&A..47..481A}.

{\bf Thallium (Z=81):}
The most recent determination of the solar Tl abundance dates back almost half a century. Updating with the current NIST-recommended transition probability for the only viable but heavily blended \ion{Tl}{i} 535.04\,nm line, the sunspot analysis of 
\citet{1972SoPh...26..250L} suggests $\lgeps{Tl} =0.92\pm0.17$, quite likely with an underestimated uncertainty. A more accurate determination would be possible from the abundance ratio with another similar species such as \ion{Zr}{i}.

\subsection{Indirect abundance determinations: Noble elements}
\label{s:noble}

The photospheric abundance of the noble gases cannot be inferred directly from spectroscopy due to the extremely high excitation potentials and thus minuscule level populations involved. Other indirect methods are therefore required.

{\bf Helium (Z=2):}
The solar He abundance can be derived from the corona, prominences, solar wind, solar-energetic particles, and the gas giant planets. However, the He/H ratio varies greatly depending on environment, from 0.01 to 0.09 
\citep[e.g.][]{2007A&A...471..315B,2014SoPh..289..977R}.
The Genesis sample return mission of the solar wind implies a ratio of about 0.05 \citep{2020M&PS...55..326H}, 
which is significantly lower than expected in the Sun as a result of Coulomb-drag fractionation.

Here, we instead rely on helioseismology, which enables an accurate determination of the He abundance in the solar convection zone and thus photosphere, which cannot be probed spectroscopically. 
The change in the adiabatic index $\Gamma_1= \left( \partial \ln P / \partial \ln \rho \right)_{\rm S}$ in the \ion{He}{ii} ionisation zone occurring at temperatures $T\approx 10^5$\,K (corresponding to a radius of $r \approx 0.98 R_\sun$) leaves an observable signature in the $p$-mode oscillation frequencies that depends on the He content (e.g. \citealt{2016LRSP...13....2B} and references therein). In \citet{2009ARA&A..47..481A}, we adopted the helioseismic He mass fraction value of $Y_{\rm surface}= 0.2485\pm0.0034$ from \citet{2004ApJ...606L..85B}; most of the quoted uncertainty stems from the choice of equation of state (OPAL: \citealt{2002ApJ...576.1064R} vs. MHD: \citealt{1988ApJ...331..815M}). Due to the overlapping higher ionisation zones of other abundant elements, especially O, there is a degeneracy with the adopted metal mass fraction $Z_{\rm surface}$ in the reference solar model for the helioseismic inversion: $\delta Y_{\rm surface}/\delta Z_{\rm surface} = -1.06$ and $-0.56$ 
with the OPAL \citep{2004ApJ...606L..85B} and MHD \citep{2006ApJ...646..560T} equation of states, respectively. 
Using a similar technique but with the improved SAHA-S3 equation of state, \citet{2013MNRAS.430.1636V} obtained maximum-likelihood estimates of $Y_{\rm surface}=0.240-0.255$ and $Z_{\rm surface}=0.008-0.013$. 

Taking the mean of the \citet{2004ApJ...606L..85B} and \citet{2013MNRAS.430.1636V} He abundances corrected to our recommended present-day photospheric composition ($Z_{\rm surface} = 0.0139$, \sect{s:protosolar}) implies $Y_{\rm surface}= 0.2423\pm0.0054$, which we adopt here. In terms of  the standard logarithmic scale, this corresponds to $\lgeps{He} =10.914\pm0.013$,  or He/H=0.082 by number density. 

{\bf Neon (Z=10):}
The absolute Ne abundance can be determined from the radiation originating in the high-temperature regions of the Sun (e.g. transition region, corona, flares) or in situ measurements of the solar wind and solar energetic particles. Both types of analyses are, however, complicated by the existence of the still poorly understood first ionisation potential (FIP) effect \citep[e.g.][]{2015LRSP...12....2L}; the abundances in the upper solar atmosphere and solar wind are modified relative to the photospheric values by a degree inversely dependent on the elemental ionisation potential. 
Arguably a more reliable approach, that we employ here, is to combine the abundance ratio of Ne ($\chi_{\rm ion}=21.56$\,eV) with an element with a similarly high ionisation potential and its photospheric abundance, under the assumption that the two elements experience the same, or at least a very similar, FIP effect. 

A common comparison element  is O ($\chi_{\rm ion}=13.62$\,eV) since it can be measured in conjunction with Ne in the upper atmosphere and solar wind, and its photospheric abundance can be determined spectroscopically (albeit with challenges, Sect. \ref{s:cno}); it is certainly conceivable that the difference in ionisation potential could still introduce elemental variations. \citet{2018ApJ...855...15Y} recently revisited the measured abundance ratios in the transition region of  the quiet Sun, finding a significantly higher Ne/O ratio than previously estimated: Ne/O=$0.24\pm0.05$ compared with $0.175\pm 0.031$ by \citet{2005A&A...439..361Y}. Most of this adjustment comes from improvements in the atomic data, resulting in changes to the ionisation and recombination rates. Together with our photospheric O abundance, this would imply a photospheric Ne abundance of $\lgeps{Ne} =8.08\pm0.09$, a 40\% increase compared with the value recommended in \citet{2009ARA&A..47..481A} using the \citet{2005A&A...439..361Y} Ne/O ratio.

\citet{2007ApJ...659..743L} inferred the absolute coronal Ne abundance from solar flare spectra:  $\lgeps{Ne} =8.11\pm0.12$. There are, however, variations between flares both in the absolute Ne abundances (e.g. $\lgeps{Ne} =7.92\pm0.03$ \citet{2015ApJ...800..110L} and in the Ne/O ratio during the solar cycle as a result of the FIP effect. It is expected that the results are the least susceptible to the FIP effect during solar minima, for which \citet{2015ApJ...800..110L} obtained Ne/O$=0.25\pm0.05$, in perfect agreement with the quiet Sun transition region determination by \citet{2018ApJ...855...15Y}. 
The Ne/O ratio is typically lower in solar energetic particles  
(e.g. $0.140-0.157\pm0.010$, \citealt{2020SSRv..216...20R}) and solar wind (e.g. $0.11-0.14$, \citealt{2013SSRv..175..125R}), which suggests the existence of additional elemental fractionation. 
Thus, inferences of the absolute solar composition directly from solar wind measurements \citep[e.g.][]{2016ApJ...816...13V} are likely prone to systematics
and are probably unreliable \citep[for further discussion see][]{2016MNRAS.463....2S}.

An independent probe of the Ne abundance is offered by the measurements from the Genesis sample return mission 
\citep{2019M&PS...54.1092B}.
Genesis offers a comparison with He, which has a more similar ionisation potential ($\chi_{\rm ion}=24.59$\,eV) to Ne than O has. The quoted $^4$He/$^{20}$Ne and $^{20}$Ne/$^{22}$Ne ratios from the Genesis study of \citet{2009GeCoA..73.7414H} together with the helioseismic He/H ratio in the convection zone lead to a solar Ne abundance of $\lgeps{Ne} =8.14\pm0.01$. When accounting for  correlations in the measured Ne/He and H/He ratios in different solar wind regimes, the abundance becomes $\lgeps{Ne} =8.06\pm0.03$ 
\citep{2019M&PS...54.1092B,2020M&PS...55..326H}, which is in excellent agreement with the above estimate using the \citet{2018ApJ...855...15Y} Ne/O ratio. In the absence of a proper understanding of the physical processes responsible for the FIP effect, it is  difficult to estimate the full uncertainties attached to this Genesis abundance. We tentatively assign half the difference between \citet{2009GeCoA..73.7414H} and \citet{2019M&PS...54.1092B} as a systematic error, which leads to our final recommended solar photospheric Ne abundance of $\lgeps{Ne} =8.06\pm0.03$ (statistical) $\pm 0.04$ (systematic). The excellent agreement with the coronal value by \citet{2018ApJ...855...15Y} lends strong support to this 0.13\,dex higher abundance than in \citet{2009ARA&A..47..481A} and consequently also the here advocated photospheric O abundance.

{\bf Argon (Z=18):}
\citet{2009ARA&A..47..481A} used the straight average from a variety of methods to arrive at a solar Ar abundance of $\lgeps{Ar} =6.40\pm0.13$ following the procedure outlined by \citet{2008ApJ...674..607L}: solar wind, solar flares and energetic particles, nuclear statistical equilibrium, Jupiter, and solar neighbourhood. Updating this exercise with new data and down-weighting the planetary and stellar data would lead to $\lgeps{Ar} =6.36\pm0.11$. 

Here, we instead opt for a single, high-precision measurement: the Genesis sample return of the solar wind 
\citep{2019M&PS...54.1092B}, in the same manner as our recommended Ne abundance was determined. Allowing for correlations between the measured H/He and $^{36}$Ar/$^4$He ratios, \citet{2019M&PS...54.1092B} and \citet{2020M&PS...55..326H} obtained $\lgeps{Ar} =6.38\pm0.08$ (statistical). We estimate that the remaining systematic errors are significant given the still poorly understood FIP effect  \citep{2015LRSP...12....2L} and the differences in ionisation potential between Ar ($\chi_{\rm ion}=15.76$\,eV) and the reference element He ($\chi_{\rm ion}=24.59$\,eV); it is noted that \citet{2020GeCoA.276..289M} have argued that the FIP fractionation still varies with ionisation potential even for $\chi_{\rm ion}>13$\,eV. We adopt $\lgeps{Ar} =6.38\pm0.10$ for the Sun.

{\bf Krypton (Z=36):}
Here, we base the recommended solar Kr abundance on the Genesis measurements of the solar wind by \citet{2020GeCoA.276..289M}.  With the quoted Kr and H fluences, one obtains  $\lgeps{Kr} =3.12\pm0.03$ (statistical). This value is preconditioned on the assumption that the FIP fractionation corrections are identical for Kr ($\chi_{\rm ion}=14.00$\,eV)  and H ($\chi_{\rm ion}=13.60$\,eV) even when individually the corrections are about a factor of three. Alternatively, using the Kr and Ar ($\chi_{\rm ion}=15.76$\,eV) fluences together with the above estimated solar Ar abundance leads to a slightly higher value with a greater statistical uncertainty: $\lgeps{Kr} =3.16\pm0.08$. \citet{2019M&PS...54.1092B} found a Kr abundance 0.2\,dex lower with the same Genesis data when using O ($\chi_{\rm ion}=13.62$\,eV) as the reference element, which may imply element-specific corrections for high-FIP elements beyond theoretical predictions \citep{2020GeCoA.276..289M}, or that the true photospheric O abundance is correspondingly much higher than advocated here
\citep{2017ApJ...851L..12L}.  In all likelihood, the uncertainty is dominated by systematic errors from the FIP effect, whichever method is chosen. We adopt $\lgeps{Kr} =3.12\pm0.10$ including estimated systematic errors.  

Traditionally, the solar Kr abundance has been estimated from the smooth variations in the abundances of nearby odd-mass number isotopes in CI chondrites (Sect. \ref{s:meteorites}) as predicted by the measured cross-sections for the slow neutron capture process 
\citep[e.g.][]{1998PhRvC..57..391W},
a method originally proposed by 
\citet{1947ZNatA...2..311S,1947ZNatA...2..604S}. With updated isotopic abundances, this leads to $\lgeps{Kr} =3.15\pm0.10$ (statistical). The true uncertainty is likely larger given that the odd-nuclei abundance smoothness has been questioned and that a more realistic distribution of neutron exposures than in this classical model leads to significantly different predictions for Kr 
\citep{1999ApJ...525..886A}. Nevertheless, it is reassuring that this completely independent method agrees well with the estimate based on the solar-wind.

{\bf Xenon (Z=54):}
The Xe abundance can be inferred via the photospheric Sm measurement combined with experimental neutron capture cross-sections under the assumption that the reaction flow has reached equilibrium. 
Using their measured cross-sections for the pure s-process isotopes, $^{130}$Xe and $^{150}$Sm \citet{2002PhRvC..66f4603R} obtained $\lgeps{Xe} =2.24\pm0.01$ (statistical). With our updated photospheric Sm abundance, we revise this estimate slightly to $\lgeps{Xe} =2.22\pm0.05$, including estimated systematic errors (adopting the CI chondritic Sm abundance instead to normalise the $s$-process production would imply a 0.01\,dex lower Xe abundance). The solar wind abundance inferred from Genesis data  \citep{2020GeCoA.276..289M} is significantly higher, $\lgeps{Xe} =2.42\pm0.05$ (statistical), suggesting a different FIP fractionation between Xe ($\chi_{\rm ion}=12.13$\,eV) and H, as also expected on theoretical grounds \citep{2017ApJ...851L..12L}.

\section{Comparison with meteoritic abundances}
\label{s:meteorites}

\subsection{Chemical composition of CI chondrites}
\label{s:ci}

Independent and complementary data on the overall chemical composition of the Solar System are offered by meteorites. The vast majority of the more than 50,000 meteorites found on Earth have undergone fractionation and alteration to varying degrees, leaving just a handful of the most primitive and undifferentiated carbonaceous chondrites, the so-called CI chondrites: the Alais, Ivuna, Orgueil, Revelstoke, and Tonk meteorites, named after the locations of the falls\footnote{Orgueil is  the largest CI chondrite weighing in at 14\,kg, followed by Alais (6\,kg) and Ivuna (0.7\,kg). In comparison, the Tonk (8\,g) and Revelstoke (1\,g) meteorites are tiny, and as a result very few mass spectroscopic analyses have been undertaken on them.}. On account of the overall similarities with the solar photospheric abundances, it has long been argued that the CI chondrites contain pristine samples of the original proto-solar nebula 
(e.g. \citealt{1952PhRv...88..248U}, \citealt{1968ode..conf..125C}; 
see \citealt{2019arXiv191200844L} for a historical overview). 
The aqueous alteration and thermal metamorphism of the textural and mineralogical structure all CI chondrites have experienced are not believed to have significantly changed its overall chemical composition, with the exception of the highly volatile elements 
that are strongly depleted \citep{2014pacs.book...15P}\footnote{Large CI chondrites such as Orgueil do display some heterogeneity due to parent body redistribution of elements, which contributes significantly to the statistical uncertainties of the meteoritic abundances \citep{2014pacs.book...15P}.}. 
Since these depleted elements are also the most cosmically abundant, the meteoritic abundances by number densities are therefore traditionally normalised relative to Si: $N_{\rm Si} \equiv 10^6$ \citep{1956RvMP...28...53S}. To anchor the two sets of abundances onto the same scale, the meteoritic and photospheric Si abundances are therefore set equal with our recommended solar Si abundance (Sect. \ref{s:intermediate}): $\lgeps{X} = 1.51 + \log N_{\rm X}$.  In principle, more elements  can be included in the conversion \citep[e.g.][who used 12, 39, and 38 elements, respectively]{1989GeCoA..53..197A, 2009LanB...4B...44L, 2014pacs.book...15P}, but for simplicity we only employ Si, which was also done by \citet{2005ASPC..336...25A, 2009ARA&A..47..481A} and \citet{2019arXiv191200844L}; we would have obtained the same result had we adopted the same normalisation elements as \citet{2014pacs.book...15P}.

Several literature compilations of the chemical compositions of CI chondrites exist, with \citet{1988RSPTA.325..535W}, \citet{1989GeCoA..53..197A}, \citet{2003ApJ...591.1220L}, \citet{2009LanB...4B...44L}, \citet{2014pacs.book...15P}, and \citet{2019arXiv191200844L} being particularly highly regarded and authoritative. These studies provide a comprehensive curation of a large number of original references of laboratory data on meteorites. Here (see Table \ref{t:abund}), we adopt the CI chondritic abundances recommended by \citet{2014pacs.book...15P} for the 83 elements for which data from the Orgueil meteorite exist. 
To convert the meteoritic abundance measurements by mass to the standard meteoritic abundance scale by number, we used the atomic weights recommended by \citet{meija2016atomic} with a few adjustments to accommodate more representative Solar System isotopic ratios as described in Sect. \ref{s:isotopes}.
The quoted CI abundance uncertainties are only statistical sampling errors.

\begin{figure}[t!]
    \begin{center}
        \includegraphics[width=9.5cm]{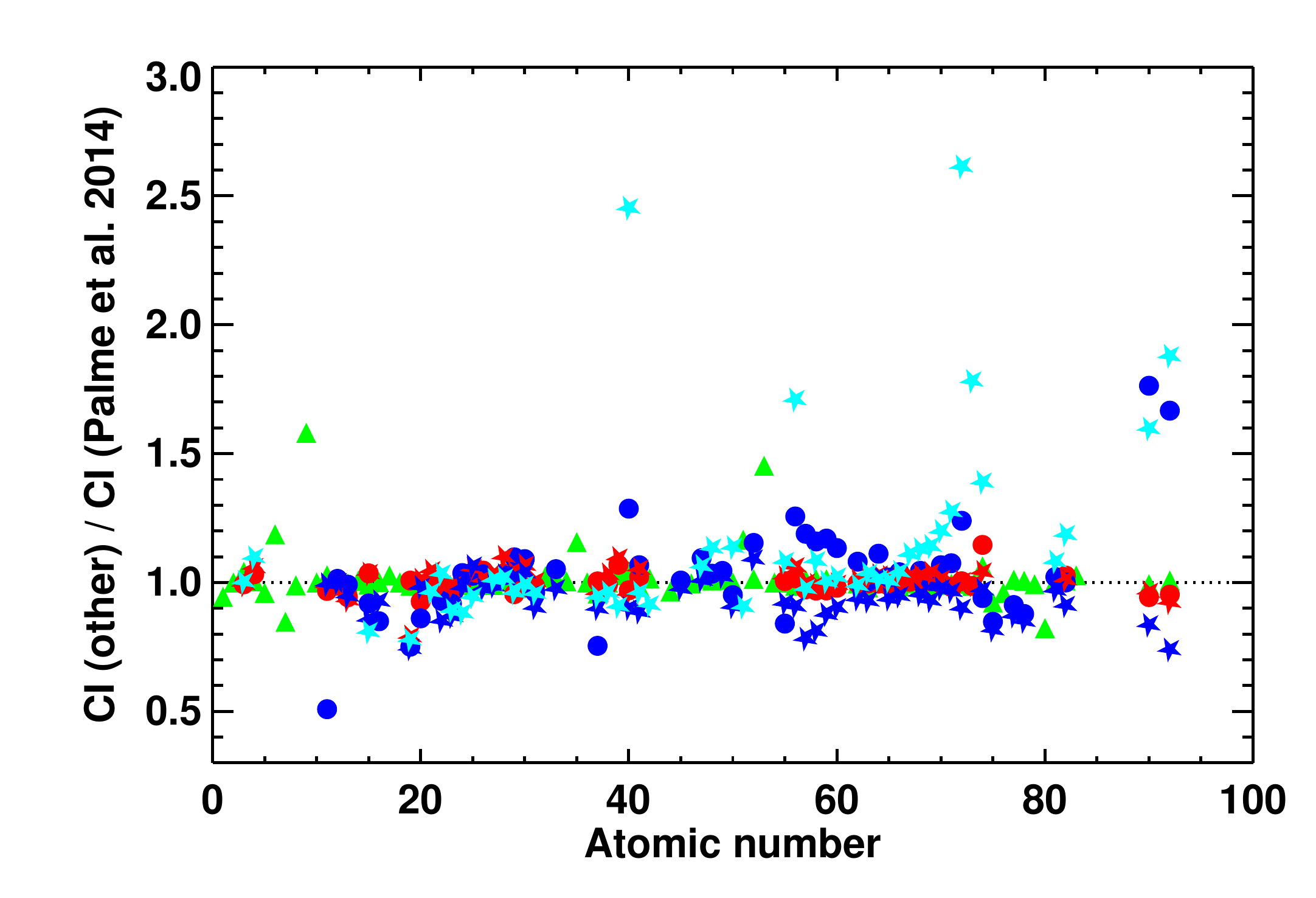}
        \caption{Comparison between the recommended CI chondritic abundances by \citet{2014pacs.book...15P}, which we have adopted here, with other recent CI analyses (red symbols: \citealt{2012GeCoA..83...79B}; blue: \citealt{2018GeCoA.239...17B}; turquoise: \citealt{2020GeCoA.268...73K}; green: \citet{2019arXiv191200844L}. Circle symbols correspond to data for the Orgueil meteorite, star symbols are for the Ivuna CI chondrite, and triangles for mean of all CI chondrites. All data are published abundances (ppm by mass) relative to \citet{2014pacs.book...15P}. 
        }
        \label{f:CI_chondrites}
    \end{center}
\end{figure}

We did not attempt to factor in the recent measurements by \citet{2018GeCoA.239...17B} on account of the surprisingly large variation between their analyses of Orgueil and Ivuna, as well as with prior CI measurements for several elements, including Na, K, Zr, the rare Earth elements, and Th and U (Fig. \ref{f:CI_chondrites}). The Ivuna analysis by \citet{2020GeCoA.268...73K} agrees well with the \citet{2014pacs.book...15P}  recommended abundances overall, although some striking discrepancies exist, including for S, P, K, Zr, Ba, Hf, Ta, W, and Th and U (Fig. \ref{f:CI_chondrites}). These differences can largely be traced to increased mobility in fluids for some elements during the aqueous alterations of the CI chondrites \citep{2014pacs.book...15P,2018GeCoA.239...17B}. 
The recommended values of \citet{2014pacs.book...15P} agree very well with the newer compilation by \citet{2019arXiv191200844L}, with a few exceptions. Notable changes have occurred for C, N, F, Br, Sb, I, and Hg, as seen in Fig. \ref{f:CI_chondrites}, but those elements are also among the most uncertain. \citet{2019arXiv191200844L} performed a weighted average of the available analyses of all CI chondrites, but since they did not provide references to the original data, we instead opted for the compilation of \citet{2014pacs.book...15P}.

\subsection{CI chondritic and photospheric abundances}
\label{s:sunvsci}

\begin{figure}[t!]
    \begin{center}
        \includegraphics[width=9.5cm]{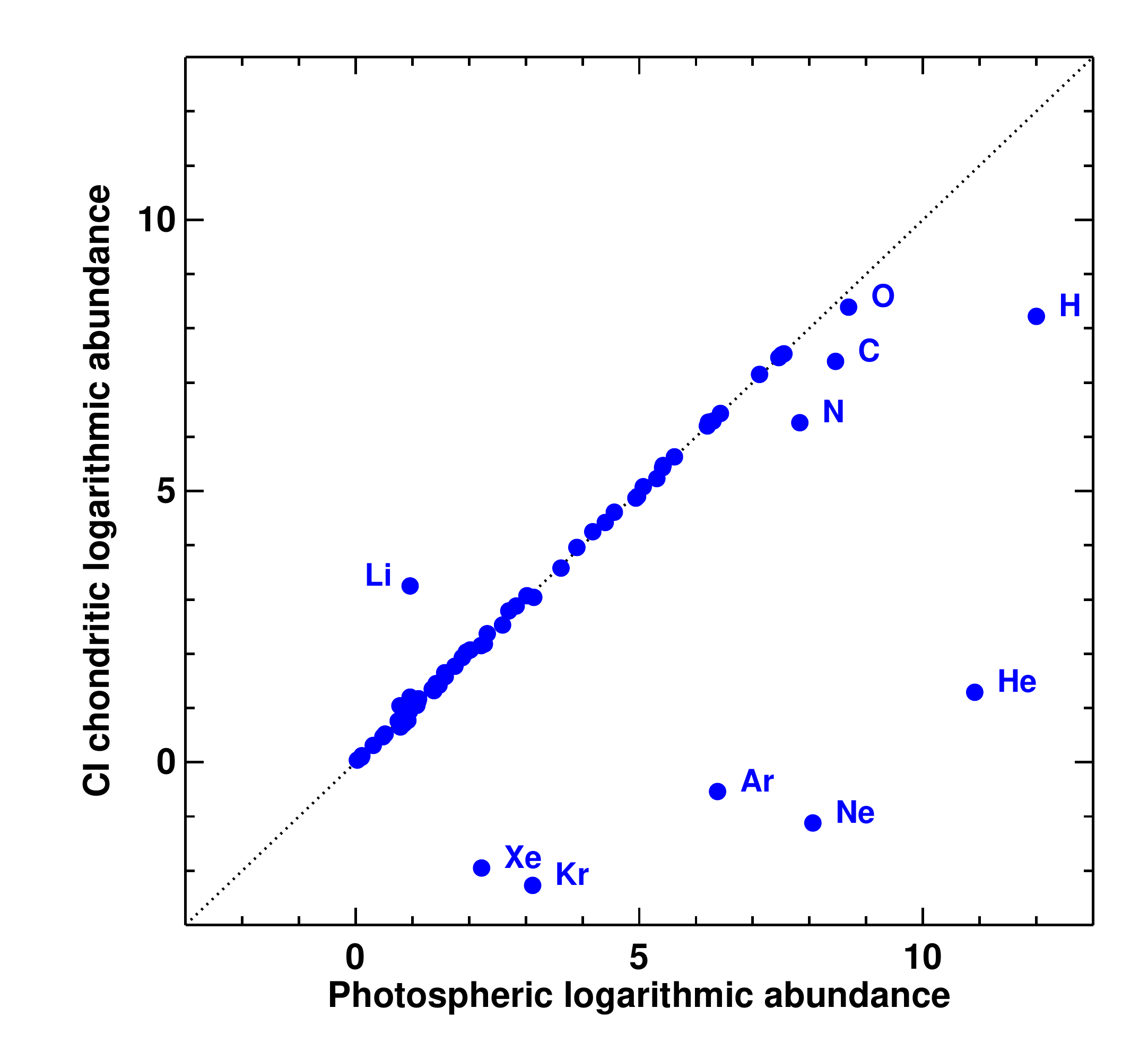}
        \caption{Comparison between our recommended abundances in present-day solar photosphere and CI chondrites. Most elements agree well, with the exception of the highly volatile elements, which are depleted in meteorites, and Li, which has been largely destroyed in the Sun due to nuclear burning.
        }
        \label{f:SunvsCI}
    \end{center}
\end{figure}

Figure \ref{f:SunvsCI} shows a comparison of our recommended present-day solar photospheric abundances with those from CI chondrites on the astronomical logarithmic abundance scale. Clearly, the agreement is very good overall, with a few striking and well-known differences. The noble gases (He, Ne, Ar, Kr, Xe) are highly depleted in CI chondrites by factors ranging from $>10^4$ (Xe) to $>10^9$ (He, Ne), while the other highly volatile gases H, C, N, and O are  depleted  by factors of $10^4$, 10, 40, and 2, respectively. 

The solar photospheric Li abundance is a factor of almost 200 lower than in CI chondrites due to nuclear processing in the Sun, much more than predicted by standard solar models.
It implies additional mixing below the solar convection zone, bringing gas down to high enough temperatures ($\approx 2.5 \cdot 10^6$\,K) for nuclear destruction to take place and subsequently returning it to the convection zone \citep[e.g.][]{2005Sci...309.2189C}. Several possible physical processes for this mixing have been proposed (e.g. internal gravity waves, rotation, convective overshooting, and atomic diffusion or a combination thereof), but all available modelling still requires parametrisation and hence remains uncertain. 
Solar twins show a slow and monotonically increasing Li depletion with age \citep{2019MNRAS.485.4052C}, with the Sun being unusually deficient in Li by $\approx 0.3$\,dex given its age for reasons not yet understood. 
There is a long and chequered history surrounding the question of whether  Li depletion in the Sun and similar stars is related to planet formation 
\citep[e.g.][]{2008A&A...489L..53B,2009Natur.462..189I,2010A&A...519A..87B,2010A&A...521A..44B}.

Given the existence of additional mixing below the convection zone as evident from Li,  the photospheric abundances of Be and B take on particular importance, since they burn at slightly higher temperatures than Li ($\approx 3.5 \cdot 10^6$\,K and $\approx 5 \cdot 10^6$\,K, respectively). 
The fact that solar twins spanning a wide age range all have very similar photospheric Be abundances suggests that the Sun has not experienced significant Be depletion in the convection zone, even though the solar Be abundance is perhaps $\approx 0.05$\,dex lower than expected for its age \citep{2015A&A...576L..10T}.  A modest B deficiency could, in principle, be hidden within the large error bars, but the absence of a significant Be depletion argues against it. The photospheric Li, Be, and B abundances imply shallow extra mixing below the convection zone present over Gyr timescales in the Sun. 

\begin{figure}[t!]
    \begin{center}
        \includegraphics[width=9.5cm]{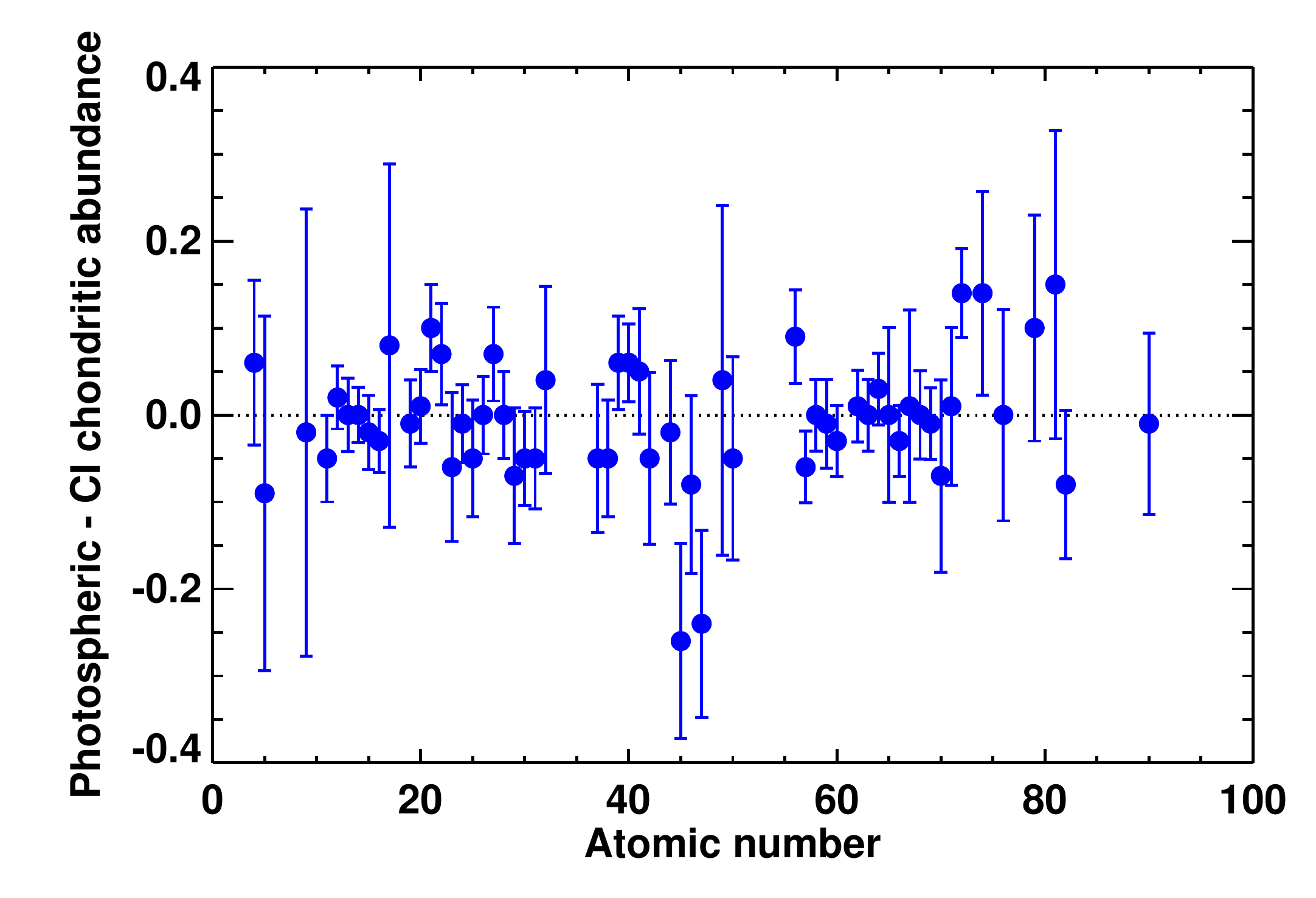}
        \caption{Differences of the photospheric and CI chondritic logarithmic abundances as functions of atomic number. In most cases, the elemental abundances agree within the combined uncertainties; the highly volatile elements and Li are not shown in the figure due to meteoritic and solar depletion, respectively. 
        The two most discrepant elements are Rh and Ag, both of which are suspected to have underestimated photospheric abundances in the absence of non-LTE calculations. 
        }
        \label{f:Sun-CIvsZ}
    \end{center}
\end{figure}

A more detailed comparison of the photospheric and CI chondritic abundances as a function of atomic number is shown in Fig. \ref{f:Sun-CIvsZ}. With the exception of the depleted elements mentioned above, essentially all elements agree within their combined uncertainties. 
Using the inverse of the combined photospheric and meteoritic abundance uncertainties in quadrature as weights, the mean average difference of the 58 non-depleted elements (i.e. excluding Li and the highly volatiles) amounts to $0.000\pm0.007$ (s.d.$=0.052$)\,dex. We remind the reader that the meteoritic abundance scale has been normalised to the photospheric Si abundance, which is here only quoted to within 0.01\,dex, and hence this perfect agreement is somewhat fortuitous. This comparison does, however, support our choice of transition probabilities for Si (Sect. \ref{s:intermediate}).  Likewise, it is noted that atomic diffusion and gravitational settling taking place in the Sun over the past 4.57\,Gyr will retain this overall  agreement since all elements bar H and He diffuse in nearly identical amounts (Sect. \ref{s:protosolar}). Of these 58 elements, four are nominal $> 2\sigma$ outliers: Sc, Rh, Ag, and Hf; we suspect that the photospheric abundances from the minority ionisation stages \ion{Rh}{i} and \ion{Ag}{i} to have been significantly underestimated in the absence of non-LTE calculations (Sect. \ref{s:heavy}). Overall, CI chondrites closely resemble the solar photospheric abundances, justifying their historical use as a compositional proxy. Caveats are in place, however, as discussed below.

\subsection{Chemical signatures of planet formation}
\label{s:Tcond}

\begin{figure}[t!]
    \begin{center}
        \includegraphics[width=9.5cm]{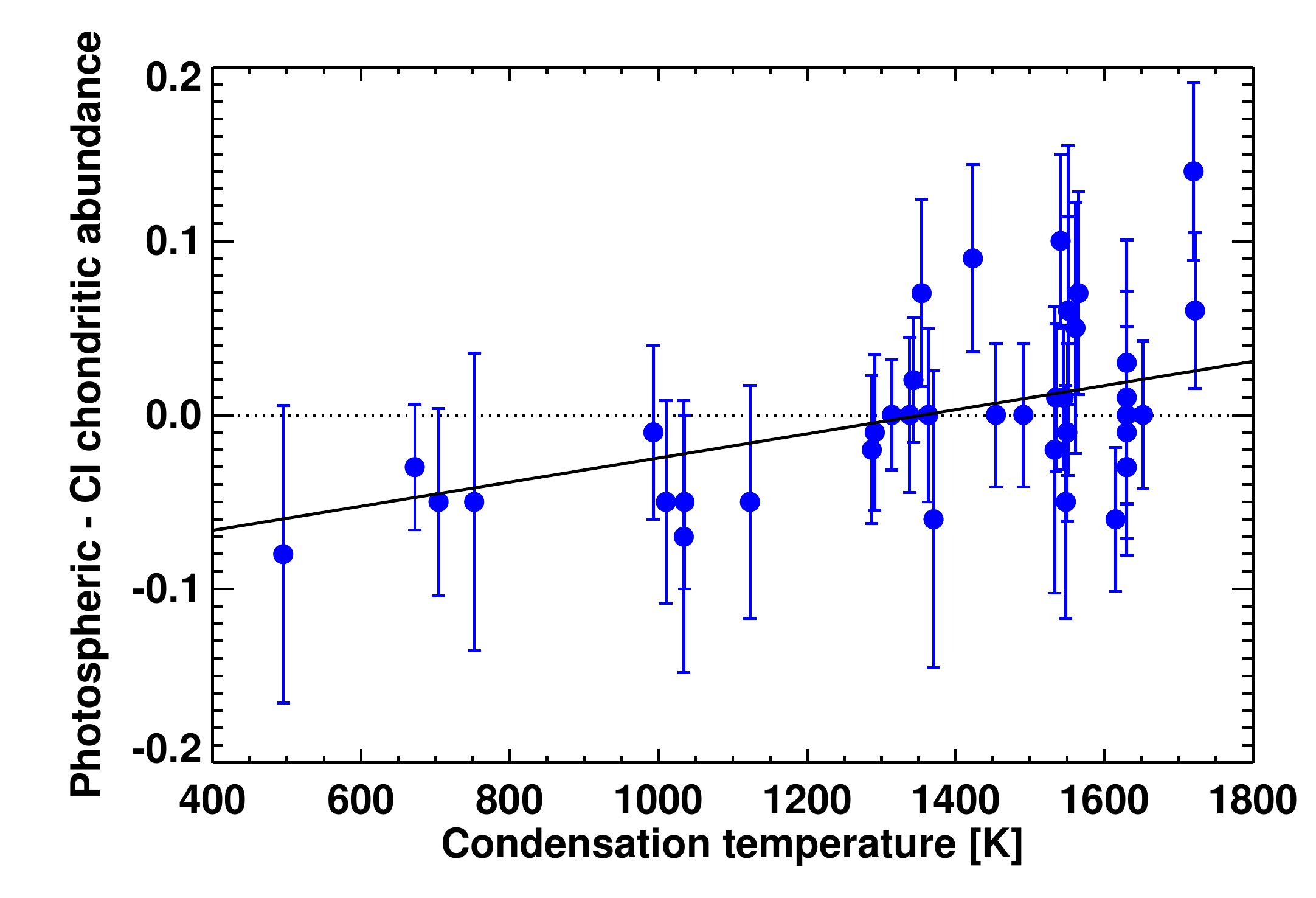}
        \caption{Differences of the photospheric and CI chondritic logarithmic abundances as a function of 50\% condensation temperature at a pressure of $10^{-4}$\,bar for a solar composition gas \citep{2019ApJ...881...55W}, showing a tendency for the refractory elements to be enhanced in the Sun compared with the meteorites. Only elements with a combined uncertainty of $<0.1$\,dex in the abundance difference are shown in the figure. It should be noted that C, N, and O (condensation temperature $<200$\,K) fall outside the plot with the meteoritic abundances depleted compared to the Sun, as is also the case for the noble gases. The solid line is a weighted least-squares fit to the data with a slope of $(6.9\pm2.7) \cdot 10^{-5}$\,dex K$^{-1}$.
        }
        \label{f:Sun-CIvsTcond}
    \end{center}
\end{figure}

The overall impressive consistency between the CI chondrites and the solar photosphere notwithstanding, it is worthwhile to ask whether the two abundance sets are truly the same. Or in other words, are CI chondrites a true reflection of the pristine chemical composition of the proto-solar nebula as has long been assumed \citep[e.g.][]{1956RvMP...28...53S}? 
Figure \ref{f:Sun-CIvsTcond} shows the differences between the photospheric and the CI meteoritic logarithmic abundances as a function of 50\% equilibrium condensation temperature ($T_{\rm cond}$) computed for solar composition gas at a pressure of $10^{-4}$\,bar (\citealt{2019ApJ...881...55W}, see also \citealt{2003ApJ...591.1220L}); for clarity, only elements with a combined abundance uncertainty of $<0.10$\,dex are included. With the anchoring of the two abundance scales through Si ($T_{\rm cond}=1314$\,K), 
it appears as though the moderately volatile  elements ($300 \la T_{\rm cond} \la 1250$\,K; here: Pb, S, Zn, Rb, K, Ga, Cu, Na, Mn, in order of increasing $T_{\rm cond}$)  are {\em \emph{enriched}} in the CI chondrites relative to the Sun: the weighted mean difference is $-0.042\pm0.006$ (s.d.$=0.018$)\,dex. Not surprisingly, the main component elements ($1250 \la T_{\rm cond} \la 1370$\,K; here: Mg, P, Si, Cr, V, Fe, Co, Ni) agree excellently, due to the fact that we enforced the Si abundances to agree: $+0.004\pm0.009$ (s.d.$=0.025$)\,dex.  The refractory elements ($T_{\rm cond} \ga 1370$\,K, here 24 elements) may be slightly {\em \emph{depleted}} in CI chondrites, but the relatively large scatter prevents a firm conclusion: $+0.015\pm0.010$ (s.d.$=0.048$)\,dex. We caution that the absolute level of enhancement/depletion depends on the chosen reference element. It is well known that the CI chondrites are volatile rich compared to the other carbonaceous chondrites, but here we thus find evidence that  this is  also the case in comparison to the Sun. The opposite is of course true for the highly volatile ($T_{\rm cond} \la 250$\,K) elements (H, He, C, N, O, Ne, Ar, Xe, Kr), which are greatly depleted in all meteorites, including the CI chondrites (Sect. \ref{s:ci}). This disposition for volatiles and refractories in the Sun and CI chondrites was first noticed by \citet{1997MNRAS.285..403G} when discussing the metal-richness of the first discovered exoplanet host stars, and it is here substantiated by our more accurate photospheric abundances.

There is significant scatter in Fig. \ref{f:Sun-CIvsTcond} around any possible trend, mostly driven by the photospheric abundance uncertainties; higher precision is naturally obtained when instead intercomparing relative abundances of similar objects analysed homogeneously, such as solar/stellar twins or chondrites. A weighted linear least-squares fit implies $\Delta \lgeps\ = (6.9\pm2.7) \cdot 10^{-5} T_{\rm cond} + (-0.09\pm0.04),$ and hence the correlation does not quite reach $3 \sigma$ significance; with the condensation temperatures of \citet{2003ApJ...591.1220L}, the slope would be $(7.3\pm2.6) \cdot 10^{-5}$\,dex\,K$^{-1}$. Including all elements except Li and the highly volatile elements hardly changes the best fit. The quality of the current abundance data does not motivate a more sophisticated function, such as piece-wise linear and step functions, as attempted in related cases involving dust condensation \citep[e.g.][]{2009ApJ...704L..66M,2014MNRAS.443L..99G}.

Compared to CI chondrites, other types of carbonaceous chondrites show increasing fractionation with condensation temperature \citep[CI$\rightarrow$CM$\rightarrow$CR$\rightarrow$CO$\rightarrow$CV$\rightarrow$CB$\rightarrow$CH,][]{1988RSPTA.325..535W,2014mcp..book....1K}. Since \citet{2014pacs.book...15P} did not provide data for other classes of chondrites, we made use of the published abundances by \citet{2005PNAS..10213755B} for 42 elements, re-normalised and converted to the astronomical logarithmic abundance scale. 
An advantage with the \citet{2005PNAS..10213755B} dataset is that they tabulate the compositions both for the bulk of the chondritic material as well as the fine-grained matrix of the meteorites, which are important for our discussion below. The abundances for other chondrite classes are given normalised to CI (and the element Yb). While the relative Si abundances are provided for the matrix components, they are missing for the bulk material for non-CI chondrites for reasons unclear to us. As a result, when comparing the bulk compositions we have had to normalise to the astronomical abundance scale using Mg as the reference element, which introduces a 0.02\,dex shift; as will become clear, this minor offset is unimportant for our conclusions. 

\begin{figure}[t!]
    \begin{center}
        \includegraphics[width=9.5cm]{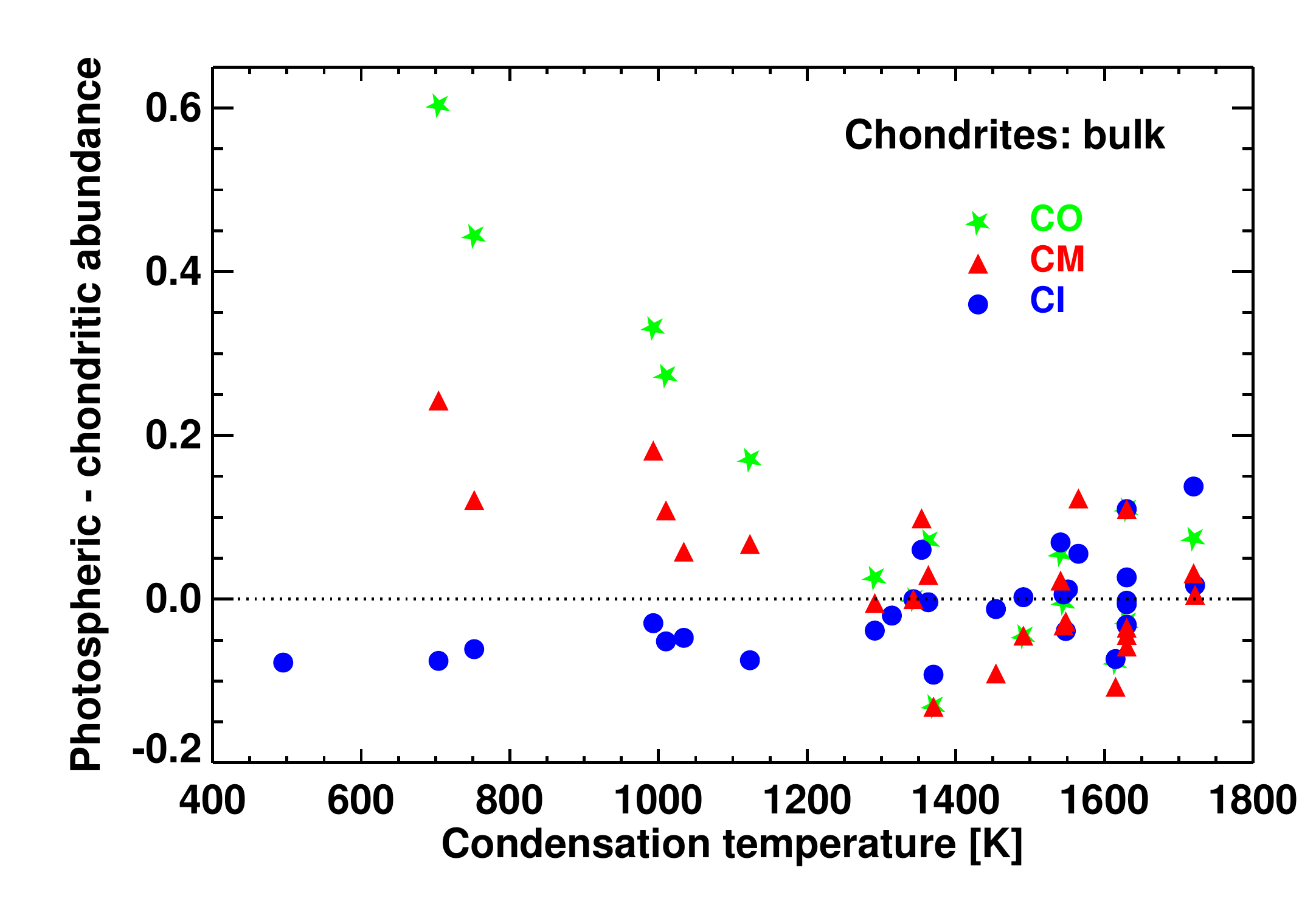}
        \includegraphics[width=9.5cm]{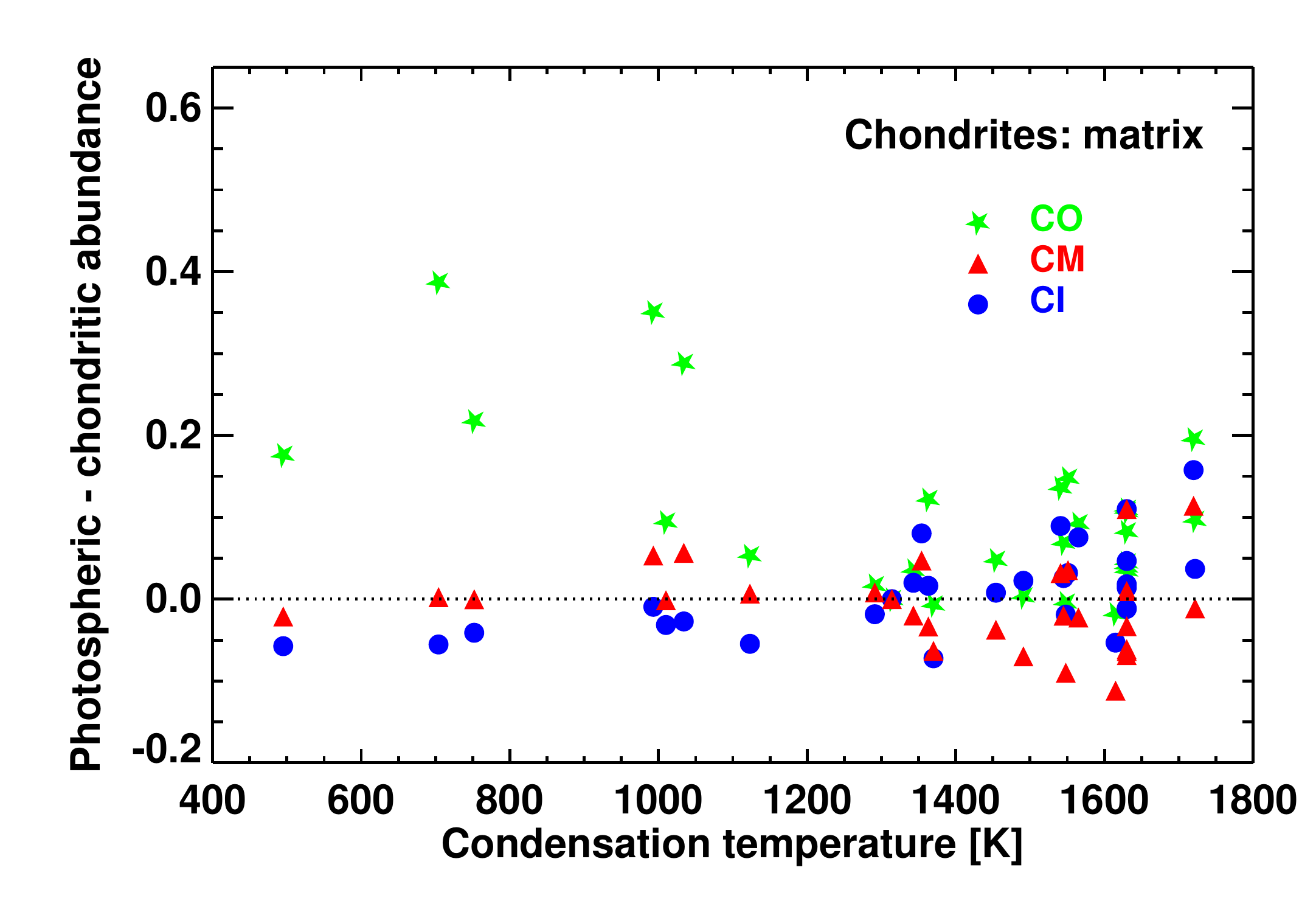}
        \caption{{\em Upper panel:} Difference between the solar photospheric and bulk abundances for CI (blue circles), CM (red triangles), and CO (green stars) chondrites as measured by \citet{2005PNAS..10213755B}. CM and CO chondrites show increasing fractionation compared to the Sun, especially for the volatile elements. {\em Lower panel:} Same, but for the fine-grained matrix of chondrites. Surprisingly, the solar abundances more closely resemble the CM matrix than the CI equivalent, which shows a trend with condensation temperature. As explained in the text, for the bulk comparison Mg has been used as the reference element to connect the meteoritic and photospheric abundance scales since the Si abundances are not provided by \citet{2005PNAS..10213755B} for non-CI chondrite classes, which introduces a 0.02\,dex offset. For the matrix compositions, the usual reference element Si has been used.
        }
        \label{f:Sun-chondritesvsTcond}
    \end{center}
\end{figure}

The resulting CI abundances from \citet{2005PNAS..10213755B} are not identical to those adopted in Sect. \ref{s:ci}, but the differences are in most cases minor, as seen in Fig. \ref{f:Sun-chondritesvsTcond}. Importantly, the correlation with condensation temperature is still apparent even if data for some elements, including the moderately volatile Na and S, are lacking; the weighted difference between the Sun and the \citet{2005PNAS..10213755B} CI data for the moderately volatile elements is $-0.040\pm0.007$ (s.d.$=0.018$)\,dex, which is almost identical to the case with the \citet{2014pacs.book...15P} CI abundances adopted elsewhere in our study. In terms of their bulk abundances, the CI chondrites are clearly the most solar-like, the trend with condensation temperature notwithstanding, followed by the CM and CO meteorites, which show increasing depletion of the volatile elements, while the refractory elements display less of a variation relative to the Sun and CI (Fig. \ref{f:Sun-chondritesvsTcond}); the highly volatile elements (H, C, N, O, and noble gases) not included in the figure are even more depleted in the non-CI chondrites.

An important distinction between the chondrite classes is that CI meteorites are only comprised of the fine-grained matrix, while other chondrites also consist of chondrules and refractory inclusions with a volatile-depleted/refractory-enriched composition relative to their matrices; the fraction of matrix material varies between the classes as well as between meteorites of the same type. Chondrules have experienced rapid heating to temperatures $>1400$\,K, followed by similarly quick cooling, likely in the inner part of the proto-planetary disc, but the details of their formation and subsequent incorporation into the matrix is still hotly debated; the matrix material has not experienced any significant heating events. The chemical complementarity between the chondrules and matrix that has been previously argued for \citep[e.g.][]{2005PNAS..10213755B,2011M&PSA..74.5302P} has more recently been questioned, implying that the chondrules and matrix did not necessarily form from the same reservoir \citep{2019PNAS..11618860V,2019GeCoA.254..277A}. To facilitate a fair comparison with the solar abundances, one should therefore consider the chondritic matrices, which are clearly more pristine than the chondrules, rather than their bulk compositions.

The lower panel of Fig. \ref{f:Sun-chondritesvsTcond} compares the matrix compositions of CI, CM, and CO chondrites relative to the Sun using data from \citet{2005PNAS..10213755B}. Since the chondritic matrix contains a higher abundance of volatile elements than chondrules, the matrices are more similar to the Sun. Remarkably, the CM matrix is overall a very good match to the present-day solar photosphere, closer in fact than the CI chondrites. In particular, the moderately volatile elements agree better with a weighted mean difference of $+0.014\pm0.010$ (s.d.$=0.027$)\,dex, while the  refractory elements show similar differences for CI and CM matrix although of opposite sign; in both cases, the main component elements agree excellently with the Sun. The weighted linear least-squares fit for the CM matrix data has a slope of $(-4.6\pm3.6) \cdot 10^{-5}$\,dex\,K$^{-1}$ , while the corresponding slope for CI chondrites is $(8.6\pm3.6) \cdot 10^{-5}$\,dex\,K$^{-1}$ with the \citet{2005PNAS..10213755B} elemental abundances.
We conclude that CI chondrites are not a perfect compositional match with the Sun, as has been assumed for more than half a century. Instead, the solar abundances more closely resemble those of the matrices of CM chondrites, although not the bulk CM composition. This could reflect spatially separate origins of the CM matrix and chondrules with differing chemical make-up, with some chemical complementarity possibly emerging later following transport to where amalgamation occur \citep{2019PNAS..11618860V}. 
A more detailed comparison will be required to elucidate how, where and when the different chondrite constituents formed and their relation to the solar composition in this new scenario 
\citep[e.g.][]{2007AREPS..35..577S,2019GeCoA.254..277A,2019PNAS..11618860V}. In the meantime, we advise against using meteoritic abundances to estimate the Solar System composition, which is often done due to their generally higher precision \citep[e.g.][]{2017ApJ...835..202V, 2019arXiv191200844L,2019Icar..328..287W}.

The correlations with chondrites shown in Fig. \ref{f:Sun-chondritesvsTcond} are reminiscent of the peculiar solar abundances relative to solar twins; 
the Sun is unusually {\em \emph{volatile rich}} and \emph{{\em refractory poor} }with an abundance amplitude of about $0.08$\,dex \citep{2009ApJ...704L..66M,2010A&A...521A..33R}. In other words, the abundance differences relative to the average of solar twins is almost exactly a mirror image of the differences with the CI chondrites. Only $\approx 10$\% of the studied stars share the detailed chemical fingerprint with the Sun, which furthermore cannot be explained by differences in age or Galactic birthplace of the stars \citep{2018ApJ...865...68B, 2018A&ARv..26....6N}. The correlation with condensation temperature suggests that the signature is related to planet formation. The initial suggestion that the abundance pattern is imprinted by the formation of terrestrial planets in the Solar System \citep{2009ApJ...704L..66M,2010ApJ...724...92C} is likely not correct since the solar convection zone was much deeper during the proto-planetary disc phase compared to today, thus requiring much more than a few M$_\oplus$ of refractory-rich material locked up in rocky planets. 
Furthermore, transit data from Kepler imply that most solar-like stars harbour super-Earths or mini-Neptunes more massive than the terrestrial planets in the Solar System 
\citep{2019MNRAS.483.4479Z},
which would make the Sun stand out in the opposite manner compositionally. The alternative explanation that other stars have on average experienced more accretion of rocky planets 
is also unlikely to hold up due to the rapid erasure of such metal-rich accretion from the stellar surface convection zone due to thermohaline mixing \citep{2012ApJ...744..123T}.

Recently, a more promising scenario was proposed by \citet{2020MNRAS.493.5079B}, which involves the early formation of giant planets trapping $>100$\,M$_\oplus$ of refractory-rich dust external to their orbits while volatile-rich gas continues to accrete onto the proto-star. They found that such gas-dust separation could imprint an abundance pattern similar to that observed when the Sun is deficient in refractory elements by $\approx 0.08$\,dex (20\%). The prediction is thus that most solar twins lack a gas giant planet, allowing the gas and dust accretion to continue largely unimpeded. The incidence of the peculiar solar abundance pattern is consistent with the frequency of giant planets being $\approx 7$\% 
\citep{2020MNRAS.492..377W}. Long-term precision radial velocity monitoring of both stars most similar and most chemically disparate to the Sun should be undertaken to search for Jupiter-like planets; HIP11915 is also a chemically solar twin that hosts a Jupiter twin \citep{2021MNRAS.502L.104Y}, although the planet hosts HIP5301 and HIP15227 seemingly do not fit the pattern \citep{2018ApJ...865...68B}  -- a much larger sample is obviously needed. The connection with the chondritic compositions discussed above also needs to be deciphered within the framework of an early formation of Jupiter.

\section{Isotopic abundances}
\label{s:isotopes}

Isotopic abundances can only be measured from solar spectroscopy for C and O, and even then only with significant uncertainties \citep{2006A&A...456..675S,2013ApJ...765...46A,2018NatCo...9..908L}\footnote{The  \ion{Li}{i} 670.8\,nm line in the Sun is unfortunately too weak to facilitate a determination of the photospheric $^6$Li/$^7$Li isotopic ratio, as has been possible in some stars 
\citep[e.g.][]{2013A&A...554A..96L,2017A&A...604A..44M}.}. Instead, alternative methods are used to infer the solar values, including sample return missions of the solar wind such as NASA's Genesis probe 
\citep{2003SSRv..105..509B,2003SSRv..105..535J} and laboratory measurements of various terrestrial, lunar, meteoritic, cometary, and asteroid samples. Our advocated isotopic abundances can
be found in \tab{t:isotopes}.

The present-day solar D abundance can not be directly determined due to near-complete burning of D to $^3$He during the pre-main-sequence evolution. The increase from the proto-solar $^3$He/$^4$He ratio as indicated by the value measured in Jupiter ($1.66\pm0.05 \cdot 10^{-4}$, \citealt{1998SSRv...84..251M}) and the local interstellar cloud ($1.62\pm0.29 \cdot 10^{-4}$, \citealt{2006ApJ...639..246B}), to the present-day ratio in the solar wind is thus a reflection of the proto-solar D/H ratio. The isotope fluences determined from the Genesis probe suggest $^3$He/$^4$He$=(4.5-4.8) \cdot 10^{-4}$ depending on the solar wind regime \citep{2012ApJ...759..121H}, which indicates the presence of isotopic fractionation during the formation and acceleration of the solar wind. Two methods to account for fractionation are presented in \citet{2012ApJ...759..121H}:  theoretical corrections based on the inefficient Coulomb drag model or those based on empirical correlations between $^3$He/$^4$He and H/$^4$He in the solar wind, which is similar to our determination of the solar Ne and Ar absolute abundances. Substantially different present-day $^3$He/$^4$He ratios and thus proto-solar D/H abundances are obtained: $1.24 \cdot 10^{-5}$ (theoretical) and $1.67 \cdot 10^{-5}$ (empirical). We take the average of the two values together with the Jovian D/H ($2.1 \cdot 10^{-5}$, \citealt{2001A&A...370..610L}) to obtain a proto-solar D/H ratio of $(1.67 \pm 0.25) \cdot 10^{-5}$. 
We caution that the here-derived proto-solar D/H value is in contradiction with the present-day D abundance in the local interstellar medium of D/H$\ge (2.0\pm0.1) \cdot 10^{-5}$ 
\citep{2006ApJ...647.1106L,2010MNRAS.406.1108P} since astration of D in stars should result in a monotonic decrease  with time. The here-deduced proto-solar D would also lead to uncomfortably large astration factors from the primordial value of $(2.53 \pm 0.03) \cdot 10^{-5}$ as inferred from high-redshift, metal-poor damped Ly-alpha systems
\citep{2017ApJ...851...25W,2018ApJ...855..102C}. The proto-solar D/H isotopic ratio clearly deserves further scrutiny.

The $^{12}$C/$^{13}$C ratio does not appear to vary significantly between different Solar System objects. We therefore adopt the reference terrestrial value of $89.3\pm0.2$ \citep{meija2016isotopic}, which is consistent with the solar photospheric determination \citep{2006A&A...456..675S,2018NatCo...9..908L} and the preliminary Genesis results for the solar wind \citep{2013SSRv..175...93W}.

\begin{figure}[t!]
    \begin{center}
        \includegraphics[width=9.5cm]{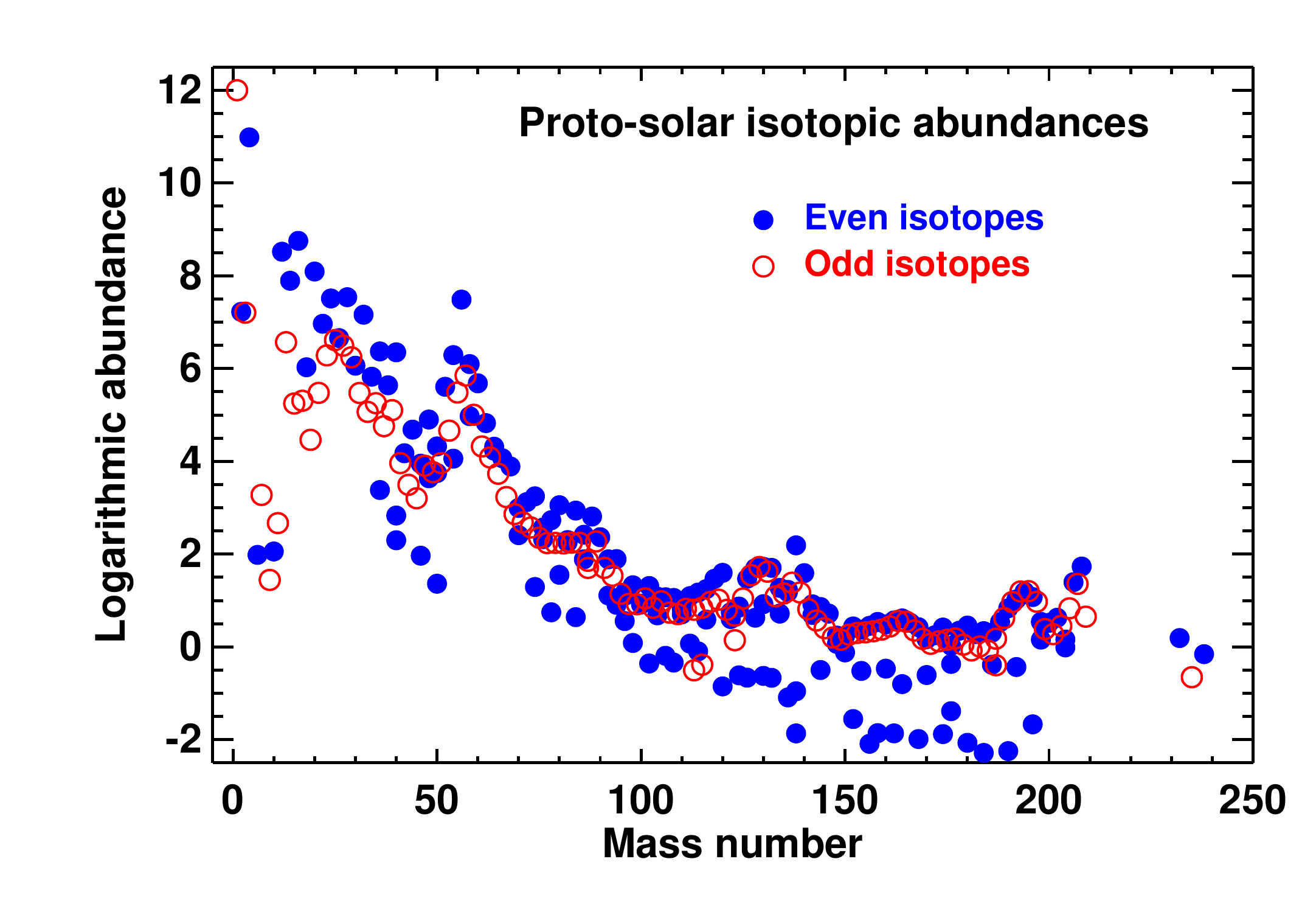}
        \caption{Proto-solar isotopic abundances as a function of mass number after taking into account atomic diffusion for He ($+0.07$\,dex) and heavier elements ($+0.06$\,dex) and radioactive decay. Even isotopes are shown as filled blue circles and odd isotopes as open red circles. 
        }
        \label{f:isotopes}
    \end{center}
\end{figure} 

The $^{14}$N/$^{15}$N ratio varies tremendously between Solar System sources for reasons that are still poorly understood. The Genesis team has measured the isotopic ratio in the solar wind \citep{2010GeCoA..74..340M}, which with an isotopic fractionation correction \citep{2017ApJ...851L..12L} implies $^{14}$N/$^{15}$N$=453\pm66$. A similar value, $435\pm57$, is obtained for Jupiter \citep{2001ApJ...553L..77O}. We take the weighted mean of these two measurements: $^{14}$N/$^{15}$N$=443\pm13$.

The isotopic composition of O in the solar wind was measured by \citet{2011Sci...332.1528M} using the Genesis sample return mission. After correction for isotopic fractionation in the solar wind, they estimated solar isotopic ratios of $^{16}$O/$^{18}$O$=530$ and $^{16}$O/$^{17}$O$=2798$, which we adopt here. The most recent spectroscopic analyses of CO lines in the Sun with 3D atmosphere models are consistent with these ratios, albeit with relatively large uncertainties \citep{2018NatCo...9..908L}. Like almost all rocky material in the inner Solar System, the Earth is enriched in $^{18}$O and $^{17}$O in comparison to  $^{16}$O (isotopic ratios of 499 and 2632, respectively, \citealt{meija2016isotopic}), likely as a result of CO self-shielding during the proto-planetary disc phase of the Solar System's formation \citep{2002Natur.415..860C,2011Sci...332.1528M}.

\begin{figure*}
    \begin{center}
        \includegraphics[scale=0.39]{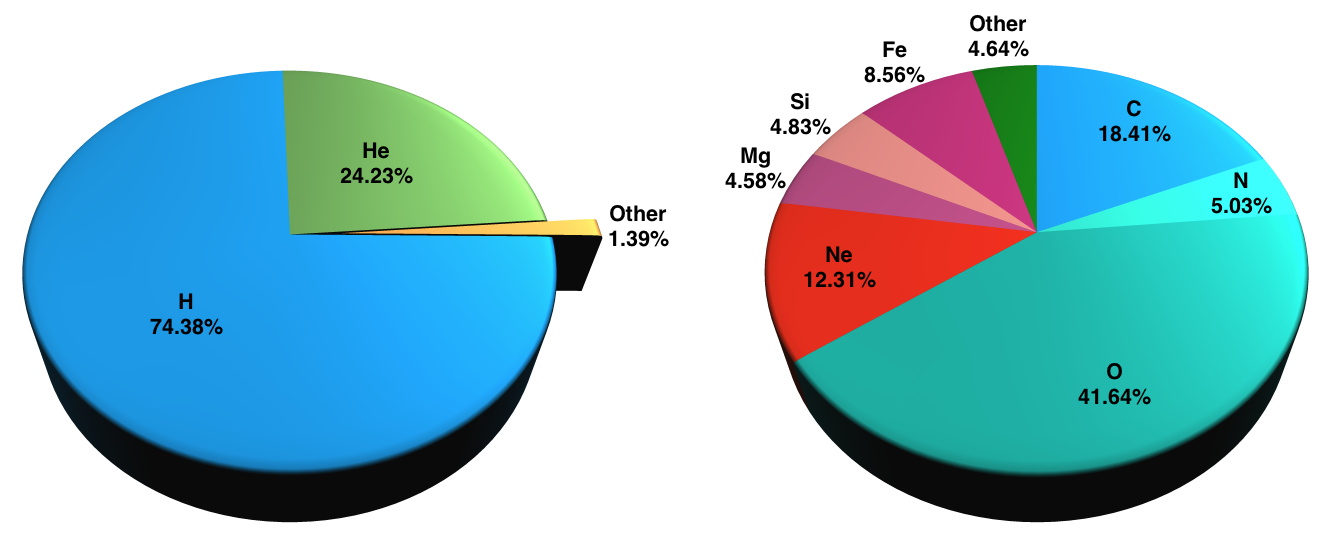}
        \caption{Present-day photospheric mass fractions the most abundant elements. {\em Left panel:} Mass fractions of H, He, and heavy elements. {\em Right panel:} Relative mass fractions of the most abundant heavy elements with the highly volatile C, N, O, and Ne contributing more than $77$\% of all heavy elements by mass. 
        }
        \label{f:massfraction}
    \end{center}
\end{figure*} 

Genesis solar wind data also allow the isotopic abundances of the other noble gases Ne, Ar, Kr, and Xe to be determined. For Ne and Ar, the isotopic fractionations are significantly smaller than for He, while for Kr and Xe they can be neglected altogether \citep{2012ApJ...759..121H}. Using the empirical correlations to correct for the fractionation, 
the present-day solar isotopic ratios are $^{20}$Ne/$^{22}$Ne$=13.36\pm0.09$, $^{21}$Ne/$^{22}$Ne$=0.0324$, and $^{36}$Ar/$^{38}$Ar$=5.37\pm0.03$ 
\citep{2012ApJ...759..121H,2014GeCoA.127..326M}; almost identical values are obtained with the inefficient Coulomb drag theoretical model \citep{2000JGR...105...47B}. To our knowledge, the $^{36}$Ar/$^{40}$Ar ratio has not been determined from the Genesis samples and hence the proto-solar $^{40}$Ar abundance is highly uncertain \citep{1989GeCoA..53..197A}; the terrestrial abundance is completely dominated by the radioactive decay of $^{40}$K \citep{meija2016isotopic}, making it the most abundant isotope. For Kr and Xe, we embrace the Genesis results from \citet{2014GeCoA.127..326M} and \citet{2020GeCoA.276..289M}, respectively. 

For the non-volatile elements, the Genesis solar wind samples have not yet delivered final isotopic abundances, although this should be feasible for elements like Mg, Si, S, and Fe in due course \citep{2019M&PS...54.1092B}. Mg is particularly interesting since it enables a critical test of the solar wind isotopic fractionation models \citep[e.g.][]{2000JGR...105...47B,2017ApJ...851L..12L},
which are in reasonable agreement with the most recent Genesis results \citep{2020M&PS...55..352J}. Fe isotopic measurements of the Genesis data on the other hand would test the notion of a chemically homogeneous proto-solar nebula, that is, whether the Sun and, for example, the most chemically pristine meteorites are expected to have identical composition as normally assumed (Sect. \ref{s:meteorites}). 

For the elements not discussed above, we take the present-day relative isotopic fractions from the recommendations by the International Union of Pure and Applied Chemistry (\citealt{meija2016isotopic} with subsequent updates for Yb, Ta, Ir, and Hf\footnote{\url{https://www.ciaaw.org/isotopic-abundances.htm}}) for representative terrestrial samples (the mean when a range is given). The proto-solar isotopic fractions given in Table \ref{t:isotopes} are assumed to be the same as the present-day values, with the exception of radioactive elements and their daughter products; we adopt the recommended isotopic half-lives from IAEA Nuclear Data Services\footnote{\url{https://www-nds.iaea.org/relnsd/NdsEnsdf/QueryForm.html}} and a Solar System age of 4.568\,Gyr \citep{2010NatGe...3..637B,2012Sci...338..651C}.

\section{Present-day and proto-solar Solar System abundances}
\label{s:protosolar}

With the Sun providing $>99.85$\% of the total mass in the Solar System, we can safely adopt the solar abundances as the basis of the chemical composition of the solar system overall. The present-day photospheric elemental abundances, however, are not the same the Sun was born with due to three main effects: nuclear processing, radio-active decay, and diffusion. The first effect only impacts Li (and possibly Be), which has been largely destroyed in the convection zone, as is evident from the comparison with the most pristine meteorites (Sect. \ref{s:sunvsci}). No nuclear products from H-burning in the solar core have reached the solar surface layers, with the exception of D-burning in the deep convection zone during the pre-main sequence (Sect. \ref{s:isotopes}). Secondly, elements such as Th and U are radioactively unstable, and hence their proto-solar abundances 4.57\,Gyr ago were larger than measured today; as noted in Sect. \ref{s:isotopes}, we account for this using the recommended half-lives. Finally, all elements heavier than H have partly settled from the surface convection zone to the radiative interior from the combined effects of microscopic thermal diffusion, gravitational settling, and radiative acceleration, normally together referred to as atomic diffusion \citep{1994ApJ...421..828T}. 

Quantitative predictions for elemental diffusion are hampered by the possibility of additional mixing below the convection zone due to still poorly understood processes like convective overshooting, rotation, internal gravity waves, and turbulence 
\citep[e.g.][]{1967ApJ...150..571G,1995ApJ...441..865C,2017ApJ...845L...6B,2019ARA&A..57...35A,2021A&A...646A..48D}, which is required to explain the solar Li abundance and interior rotation profile. Different groups thus require slightly varying proto-solar abundances depending on the ingredients of their stellar evolution codes when calibrating to the solar luminosity, effective temperature and surface composition at the solar age. Here we adopt the predictions from the standard solar model of \citet{2017ApJ...835..202V} computed with the \citet{2009ARA&A..47..481A} chemical composition. In terms of number densities relative to H, the proto-solar He abundance is higher by 0.070\,dex compared with the present-day photospheric value, while for all heavier elements the difference is 0.064\,dex. Because the \citet{2017ApJ...835..202V}  solar model does not include any additional mixing beyond atomic diffusion as required to explain the solar Li depletion, these diffusion corrections may be slightly over-estimated; more recent solar models which simultaneously agree better with the Li depletion and helioseismology (see below) have diffusion abundance corrections about half of this: $\approx 0.03$\,dex for all elements other than H
(\citealt{eggenberger_submitted}).

Regarding elements for which no photospheric abundance determination is feasible from spectroscopy or solar wind measurements (Table \ref{t:abund}), we base their present-day photospheric abundances on the CI chondritic abundances. Those are first corrected for the trend in abundance difference with condensation temperature (\fig{f:Sun-CIvsTcond}, Sect. \ref{s:Tcond}) before applying the effects of atomic diffusion to compute the proto-solar abundances. The proto-solar isotopic abundances are given in Table \ref{t:isotopes}.

In terms of mass fractions of H, He, and heavy elements, the elemental abundances presented in Table \ref{t:abund} correspond to present-day photospheric values of $X_{\rm surface}=0.7438\pm0.0054$, $Y_{\rm surface}=0.2423\pm 0.0054$, $Z_{\rm surface}=0.0139\pm 0.0006$, and $Z_{\rm surface}/X_{\rm surface}=0.0187\pm 0.0009$; Fig. \ref{f:massfraction} illustrates the present-day photospheric mass fractions of the most abundant elements. The quoted uncertainty on $Y_{\rm surface}$ stems from the helioseismic determination (Sect. \ref{s:noble}), while for the heavier elements we assumed that their abundance errors to be independent.
The here inferred solar surface metallicity is slightly higher than (but within the uncertainties of) the value ($Z_{\rm surface}=0.0134$) derived in \citet{2009ARA&A..47..481A}, mainly as a result of the revised Ne abundance following the adoption of the Genesis solar wind measurements (Sect. \ref{s:noble}). 
With the \citet{2017ApJ...835..202V} solar model for atomic diffusion, the proto-solar mass fractions would be $X_{\rm initial}=0.7121$, $Y_{\rm initial}=0.2725$, and $Z_{\rm initial}=0.0154,$ while with a smaller abundance correction of 0.03\,dex for all elements heavier than H \citep{eggenberger_submitted} the values are $X_{\rm initial}=0.7304$, $Y_{\rm initial}=0.2550$, and $Z_{\rm initial}=0.0146$. 
Table \ref{t:massfractions} lists the inferred present-day photospheric and proto-solar mass fractions based on some widely used compilations of the solar chemical composition over the past three decades.

\begin{table*}[t!]
\caption{Present-day photospheric and proto-solar mass fractions of hydrogen (X), helium (Y), and metals (Z) for a number of widely used compilations of the solar chemical composition.
\label{t:massfractions}}
\centering
\smallskip
\begin{tabular}{lcccc|cccc}
\hline
\noalign{\smallskip}
\multirow{2}{*}{Source} & \multicolumn{4}{c|}{Present-day photospheric} & \multicolumn{4}{c}{Proto-solar}  \\
 & $X_{\rm surface}$ & $Y_{\rm surface}$ & $Z_{\rm surface}$ & $Z_{\rm surface}/X_{\rm surface}$ & $X_{\rm initial}$ & $Y_{\rm initial}$ & $Z_{\rm initial}$ & $Z_{\rm initial}/X_{\rm initial}$ \\
\hline
\noalign{\smallskip}
\noalign{\smallskip}
{\bf Present work}                              & 0.7438        & 0.2423        & 0.0139        & 0.0187 & 0.7121        & 0.2725        & 0.0154        & 0.0216 \\  
\citealt{2019arXiv191200844L}                   & 0.7389        & 0.2463        & 0.0148        & 0.0200 & 0.7061        & 0.2766        & 0.0173        & 0.0245 \\  
\citealt{2011SoPh..268..255C}\tablefootmark{a}                  & 0.7321        & 0.2526        & 0.0153        & 0.0209 & & & & \\  
{\citealt{2009ARA&A..47..481A}}   & 0.7381        & 0.2485        & 0.0134        & 0.0181 & 0.7154        & 0.2703        & 0.0142        & 0.0199\\
\citealt{2009LanB...4B...44L}     & 0.7390        & 0.2469        & 0.0141        & 0.0191  & 0.7112        & 0.2735        & 0.0153        & 0.0215 \\  
\citealt{2005ASPC..336...25A}           & 0.7392        & 0.2485        & 0.0122        & 0.0165 & 0.7166        & 0.2704        & 0.0130        & 0.0181 \\
\citealt{2003ApJ...591.1220L}           & 0.7491        & 0.2377        & 0.0133        & 0.0177  & 0.7111        & 0.2741        & 0.0149        & 0.0210 \\  
\citealt{1998SSRv...85..161G}   & 0.7347        & 0.2483        & 0.0169        & 0.0231 & 0.7062 & 0.2750 &  0.0188 & 0.0266 \\
\citealt{1993oee..conf...15G}   & 0.7028        & 0.2800        & 0.0172        & 0.0245 & & & & \\ 
\citealt{1989GeCoA..53..197A}   & 0.7065        & 0.2741        & 0.0194        & 0.0274 & & & & \\ 
\noalign{\smallskip}
\hline 
\end{tabular}
\tablefoot{\tablefoottext{a}{ \citet{2011SoPh..268..255C} determined the abundances of Li, C, N, O, P, S, K, Fe, Eu, Hf, Os, and Th and adopted the recommended values of \citet{2009LanB...4B...44L} for all other elements. They did not quote any proto-solar mass fractions. }}
\end{table*}

\section{Solar abundances, standard solar models, and helioseismology}
\label{s:helioseismology}

Care must be exercised when interpreting proto-solar mass fractions in light of the long-standing and exasperating discrepancy between helioseismology and predictions from standard solar models constructed with present-day surface abundances similar to those presented here: the so-called solar modelling problem\footnote{While sometimes referred to as the solar abundance problem, we argue that this is a misnomer since it is not at all clear that the solar composition is at fault.} \citep[e.g.][]{2004ApJ...606L..85B,2005ApJ...621L..85B,2017ApJ...835..202V,2021LRSP...18....2C}. In particular, the comparatively low photospheric C, N, and O abundances (and associated Ne abundance) from 3D-based spectroscopic analyses \citep[e.g.][]{2001ApJ...556L..63A, 2002ApJ...573L.137A,2004A&A...417..751A,2005A&A...431..693A,2009ARA&A..47..481A,2018A&A...616A..89A,2019A&A...624A.111A,2020A&A...636A.120A} have ruined the good agreements with helioseismology that were obtained with global solar models computed with the older higher solar metallicity 
(e.g. \citealt{1989GeCoA..53..197A}: $Z_{\rm surface}/X_{\rm surface}=0.274$;
\citealt{1993oee..conf...15G}: $Z_{\rm surface}/X_{\rm surface}=0.246$; \citealt{1998SSRv...85..161G}: $Z_{\rm surface}/X_{\rm surface}=0.231$). 
The predicted surface He abundance, depth of the convection zone, and the sound speed as a function of solar radius from a solar calibration are now significantly more discrepant compared to the inversion of the observed solar p-mode oscillation frequencies; the solar surface composition presented here goes in the right direction but will not resolve the mismatch. 
We note, however, that the available seismic data in fact favours a low metallicity ($Z_{\rm surface}<0.014$) in the convection zone 
\citep{2013MNRAS.430.1636V,2017MNRAS.472..751B}, suggesting that current standard solar models are missing important physics or are constructed with faulty input data in the radiative zone below the convection zone.

A huge body of work has been devoted to reconciling the spectroscopic and interior solar models triggered by our 3D-based solar abundances over the past two decades.  Arguably the most straightforward explanation would be underestimated opacities, especially near the bottom of the convection zone where the discrepancies are most acute, to compensate for the lowering of the abundances of the dominant elements. It would require the computed data sets of atomic opacities \citep[e.g.][]{2005MNRAS.360..458B,2015ApJS..220....2M,2016ApJ...817..116C} all to be underestimated by about 20\% in the relevant temperature-pressure regime \citep{2009ApJ...705L.123S,2009A&A...494..205C}. In fact recent experimental opacity measurements are systematically higher than calculations with the missing opacity for Fe alone accounting for about half of what is required to resolve the solar modelling problem \citep{2015Natur.517...56B,2019PhRvL.122w5001N}. With new experiments and updated calculations with previously overlooked atomic processes underway \citep[e.g.][]{2016PhRvL.116w5003N,2018ApJ...856..135K} it remains to be seen whether the final atomic opacities will be sufficient. Important independent constraints on the solar interior composition will also come from measurements of the solar neutrino fluxes, especially those produced in the CNO-cycle that were recently reported by the Borexino experiment \citep{2020Natur.587..577B}. Finally, encouraging signs are appearing from new solar models with improved mixing below the convection zone aimed at explaining the depletion of Li and the interior rotation profile, which standard models cannot do 
\citep[e.g.][]{2019ApJ...873...18Y,2019ApJ...881..103Z,2019A&A...621A..33B,2019FrASS...6...42B,eggenberger_submitted}.  We remain cautiously optimistic that the persistent solar modelling problem will soon have been resolved without necessitating any major revision of the photospheric solar abundances presented here.

\section{Concluding remarks}
\label{s:conclusions}

The elemental abundances derived and presented here are a further stepping stone towards the advancement of our understanding of the solar chemical composition, which is of such crucial importance for astronomy as a whole. 
As demonstrated above, major improvements in terms of the solar spectroscopic analysis have been accomplished, which should make the inferred elemental abundances both more precise and more accurate. It is now perfectly feasible to perform detailed 3D non-LTE line formation calculations using highly realistic 3D radiative-hydrodynamical simulations of the solar surface convection and atmosphere. This has been achieved for 13 elements -- Li, C, N, O, Na, Mg, Al, Si, K, Ca, Mn, Fe, and Ba -- all using comprehensive model atoms with reliable atomic data, including the previously notoriously uncertain cross-sections for H collisions. Compared to \citet{2009ARA&A..47..481A}, in which only the abundances of Li and O were based on a consistent 3D non-LTE analysis, this certainly represents substantial progress. The abundances of a further 45 elements have been determined on the basis of a 3D LTE spectroscopic analysis. In total, we present the solar photospheric abundances of 68 elements when also including the noble gases, for which helioseismic and solar wind measurements are available, and four trace elements evaluated from sunspot observations. Importantly, the abundance uncertainties for these 68 elements have been carefully evaluated, including both statistical and systematic uncertainties. These have been augmented with meteoritic data (corrected for an identified trend with condensation temperature) for a further 15 elements, thus providing Solar System abundances for all 83 long-lived elements. 

Based on our state-of-the-art 3D non-LTE analysis, we can safely confirm that the old high solar metallicity of, for example, \citet{1993oee..conf...15G} and \citet{1998SSRv...85..161G}, which are still used to construct standard solar models, should finally be discarded -- there are simply no arguments in favour of those outdated 1D LTE results spectroscopically. 
Our study is, however, definitely not the final word on the issue of the solar chemical composition, with much work remaining to reach comparable abundance precision to that of the most pristine meteorites.  Arguably, the most pressing need is to extend non-LTE calculations to additional elements: only 26 elements have their solar abundances determined from a non-LTE analysis, half of which are only in 1D. Of particularly high priority is to carry out 3D non-LTE computations for the remaining Fe-peak elements (including revisiting the important case of Mn, Sect. \ref{s:fepeak}) as well as other elements of key astronomical significance, such as Be (to constrain solar depletion, Sect. \ref{s:sunvsci}), S, Zn, Rb (volatile elements to confirm differences with CI chondrites, Sect. \ref{s:Tcond}), P, Cu, and Sr (stellar nucleosynthesis yields, \citealt{2020ApJ...900..179K}). Several of these elements already show substantial non-LTE effects in existing 1D non-LTE calculations, which can be expected to be amplified in 3D. 

As always, solar spectroscopy is reliant on having access to accurate atomic data. Over the past decade, the situation regarding experimental transition probabilities has improved significantly for the Fe-peak elements in particular, thanks to the dedicated and valiant efforts by a relatively small group of atomic physicists around the world, as discussed in Sect. \ref{s:fepeak}. Several other elements have also seen significant updates, however, the need for further work is essentially endless. Here it suffices to mention the case of Si, the element anchoring the meteoritic abundance scale to the solar photospheric values (Sect. \ref{s:intermediate}). Besides the need for refined transition probabilities of spectral lines used as diagnostics, there is also a demand for more and better atomic data for lines blending the primary abundance indicators. Since those lines are typically weaker and arise from higher-excitation levels, they are often not even properly identified or, when included in existing atomic calculations, have faulty $gf$-values. 

For non-LTE modelling, significantly more atomic data are necessary, including all radiative (bound-bound and bound-free) and collisional (with electrons and hydrogen) transitions coupling the different atomic levels. Great progress has recently been achieved in this regard, especially related to inelastic H collisions \citep[e.g.][]{2016A&ARv..24....9B,2018ApJ...867...87B,2021ApJ...908..245B}, which has made the non-LTE results much more trustworthy. Crucially, the relevant atomic data now exist for a 3D non-LTE study for almost all of the elements highlighted above; mainly what is lacking is (wo)man-power to take on the necessary but time-consuming 3D non-LTE analyses. What has not yet been fully explored for the Sun is how departures from LTE in one element affect those of other elements or feed back onto the solar atmospheric structure; preliminary calculations suggest that this may well be of importance \citep[][]{2020A&A...637A..80O}. Furthermore, we caution that magnetic fields have not been included in the 3D solar atmosphere model employed throughout this study. While early work based on imposed vertical magnetic fields implied that the effects could be substantial \citep{2010ApJ...724.1536F}, more recent 3D LTE calculations based on a 3D MHD solar model with a self-consistent small-scale magnetic dynamo \citep{2014ApJ...789..132R} reveal that the impact on solar abundance determination is negligible ($< 0.01$\,dex) at least for C, N, O, and Fe \citep{2015A&A...579A.112S,2016A&A...586A.145S}; this may not hold however for elements in which non-LTE effects are important and thus deserves further scrutiny. 
Similarly, there is no indication that the inclusion of a chromosphere or other magnetic activity in the solar model would significantly impact the derived abundances for the weak photospheric lines employed here using disc-centre observations of the quiet Sun \citep{Bergemann_2021_O}.

Besides the obvious benefits from having a better definition of a widely used astronomical yardstick, the continuously improved 3D-based solar abundances reported here and in our earlier works \citep[e.g.][]{2005ASPC..336...25A,2009ARA&A..47..481A} have uncovered two surprising peculiarities regarding the solar chemical composition, neither of which have been explained. The inconsistency between the composition at the surface, as measured from spectroscopy, and the interior, as inferred from helioseismology, has now been around for almost two decades in spite of tremendous efforts in reconciling the two with further improvements to the solar modelling. {\em \emph{It is crucial to remember that this is not a solar problem but a stellar problem}.} If indeed the explanation is missing opacity 
as suggested by recent experiments (Sect. \ref{s:helioseismology}), it would imply that current stellar interior and evolution models are not fully trustworthy. Since such stellar models underpin much of the field of astronomy, it would have profound implications. The second solar abundance feature uncovered here, namely that the supposedly most primitive meteorites -- the CI chondrites -- and the Sun in fact do not share the same composition, implies that a revision of the earliest history of the Solar System is required, given that their equivalence has been taken for granted for more than half a century (Sect. \ref{s:Tcond}). The existence of a similar abundance trend of the condensation temperature between the Sun and solar twins suggests that this may well give us a deeper understanding of planet formation around stars in general.

\begin{acknowledgements}
We thank all of our wonderful colleagues around the world for highly rewarding collaborations over many years, numerous stimulating discussions about the topics described herein and their implications for astronomy and physics, and for providing necessary input atomic/molecular data, all of which have greatly benefited this work. 
MA gratefully acknowledges generous funding from the Australian Research Council through a Laureate Fellowship (FL110100012) and a Discovery Project (DP150100250). 
AMA acknowledges support from the Swedish Research
Council (VR 2016-03765 and 2020-03940).
This research was supported by computational resources provided by the Australian Government through the National Computational Infrastructure (NCI) under the National Computational Merit Allocation Scheme and the ANU Merit Allocation Scheme (project y89). 
Some of the computations were also enabled by resources provided by 
the Swedish National Infrastructure for Computing (SNIC) at 
the Multidisciplinary Center for Advanced Computational Science (UPPMAX) 
and at the High Performance Computing Center North (HPC2N) partially 
funded by the Swedish Research Council through grant agreement no. 
2018-05973.

\end{acknowledgements}

\bibliographystyle{aa} 
\bibliography{refs}

\begin{appendix}

\onecolumn

\section{Line lists for Na, Mg, K, Ca, and Fe}

\begin{table*}[h]
\begin{center}
\caption{Line list for the Na, Mg, K, and Ca lines used for 3D non-LTE spectral line formation calculations to derive the solar photospheric abundances}
\label{t:linelist_NaCa}
\begin{tabular}{rrrrr|cccc|cccc}
\hline
\noalign{\smallskip}
\multirow{2}{*}{$\lambda_{\text{air}}/\nm$} & \multirow{2}{*}{$E_{\text{lower}}/\mathrm{eV}$} & \multirow{2}{*}{$E_{\text{upper}}/\mathrm{eV}$} & \multirow{2}{*}{$\lggf$} & \multirow{2}{*}{$W/\mathrm{pm}$} &  \multicolumn{4}{c|}{Non-LTE} & \multicolumn{4}{c}{LTE} \\
 & & & & & 3D & \mtd{} & \marcs{} & HM & 3D & \mtd{} & \marcs{} & HM  \\
\noalign{\smallskip}
\hline
\noalign{\smallskip}
\multicolumn{13}{c}{\ion{Na}{i}} \\
\noalign{\smallskip}
    475.18 &  2.10443 &  4.71289 &  -2.077 &   1.14 &  6.223 &  6.225 &  6.199 &  6.243 &  6.250 &  6.259 &  6.231 &  6.280 \\
    514.88 &  2.10230 &  4.50963 &  -2.044 &   1.32 &  6.231 &  6.235 &  6.206 &  6.251 &  6.259 &  6.271 &  6.239 &  6.288 \\
    615.42 &  2.10230 &  4.11636 &  -1.547 &   3.98 &  6.250 &  6.255 &  6.215 &  6.266 &  6.294 &  6.307 &  6.262 &  6.320 \\
    616.07 &  2.10443 &  4.11636 &  -1.246 &   5.83 &  6.178 &  6.179 &  6.135 &  6.190 &  6.229 &  6.240 &  6.193 &  6.256 \\
   1074.64 &  3.19135 &  4.34476 &  -1.294 &   1.33 &  6.221 &  6.221 &  6.188 &  6.223 &  6.226 &  6.227 &  6.194 &  6.231 \\
\noalign{\smallskip}
\hline
\noalign{\smallskip}
\multicolumn{13}{c}{\ion{Mg}{i}} \\
\noalign{\smallskip}
    631.87 &  5.10783 &  7.06946 &  -2.020 &   4.13 &  7.546 &  7.546 &  7.509 &  7.559 &  7.545 &  7.547 &  7.509 &  7.559 \\
    631.92 &  5.10783 &  7.06930 &  -2.242 &   2.60 &  7.549 &  7.547 &  7.516 &  7.558 &  7.548 &  7.547 &  7.516 &  7.558 \\
    871.27 &  5.93195 &  7.35459 &  -1.152 &   6.80 &  7.565 &  7.560 &  7.517 &  7.566 &  7.566 &  7.561 &  7.518 &  7.568 \\
    871.78 &  5.93279 &  7.35459 &  -0.930 &  10.00 &  7.581 &  7.575 &  7.524 &  7.584 &  7.583 &  7.577 &  7.526 &  7.587 \\
    892.36 &  5.39373 &  6.78275 &  -1.679 &   6.33 &  7.608 &  7.605 &  7.562 &  7.614 &  7.607 &  7.605 &  7.562 &  7.614 \\
    998.32 &  5.93154 &  7.17313 &  -2.177 &   1.00 &  7.558 &  7.553 &  7.525 &  7.556 &  7.559 &  7.554 &  7.526 &  7.558 \\
   1031.25 &  6.11821 &  7.32015 &  -1.718 &   1.83 &  7.515 &  7.508 &  7.479 &  7.513 &  7.515 &  7.509 &  7.479 &  7.513 \\
   1587.95 &  5.94592 &  6.72648 &  -1.226 &  16.80 &  7.570 &  7.565 &  7.517 &  7.594 &  7.571 &  7.566 &  7.518 &  7.595 \\
\noalign{\smallskip}
\multicolumn{13}{c}{\ion{Mg}{ii}} \\
\noalign{\smallskip}
    787.70 &  9.99554 & 11.56910 &   0.389 &   1.90 &  7.540 &  7.542 &  7.526 &  7.572 &  7.568 &  7.570 &  7.542 &  7.606 \\
    789.64 &  9.99933 & 11.56904 &   0.645 &   3.00 &  7.635 &  7.655 &  7.640 &  7.681 &  7.682 &  7.702 &  7.669 &  7.731 \\
    921.82 &  8.65471 &  9.99933 &   0.269 &   7.40 &  7.531 &  7.501 &  7.481 &  7.504 &  7.642 &  7.592 &  7.559 &  7.607 \\
    924.43 &  8.65471 &  9.99554 &  -0.034 &   5.15 &  7.489 &  7.465 &  7.447 &  7.478 &  7.567 &  7.535 &  7.505 &  7.554 \\
   1009.22 & 11.62968 & 12.85786 &   1.283 &   1.33 &  7.495 &  7.541 &  7.523 &  7.598 &  7.490 &  7.536 &  7.518 &  7.596 \\
   1091.42 &  8.86365 &  9.99933 &   0.036 &   5.22 &  7.427 &  7.401 &  7.388 &  7.419 &  7.518 &  7.481 &  7.451 &  7.509 \\
\noalign{\smallskip}
\hline
\noalign{\smallskip}
\multicolumn{13}{c}{\ion{K}{i}} \\
\noalign{\smallskip}
    693.88 &  1.61711 &  3.40345 &  -1.145 &   0.43 &  5.045 &  5.044 &  5.010 &  5.050 &  5.071 &  5.081 &  5.046 &  5.087 \\
   1176.97 &  1.61711 &  2.67025 &  -0.481 &   3.50 &  5.082 &  5.070 &  5.029 &  5.079 &  5.135 &  5.144 &  5.102 &  5.153 \\
   1252.22 &  1.61711 &  2.60696 &  -0.128 &   7.60 &  5.093 &  5.075 &  5.025 &  5.090 &  5.157 &  5.163 &  5.110 &  5.178 \\
\noalign{\smallskip}
\hline
\noalign{\smallskip}
\multicolumn{13}{c}{\ion{Ca}{i}} \\
\noalign{\smallskip}
    451.23 &  2.52568 &  5.27263 &  -1.900 &   2.20 &  6.296 &  6.304 &  6.272 &  6.327 &  6.285 &  6.304 &  6.270 &  6.327 \\
    526.04 &  2.52126 &  4.87755 &  -1.719 &   3.00 &  6.265 &  6.271 &  6.230 &  6.290 &  6.253 &  6.272 &  6.230 &  6.292 \\
    586.76 &  2.93251 &  5.04497 &  -1.570 &   2.30 &  6.278 &  6.284 &  6.244 &  6.298 &  6.269 &  6.285 &  6.244 &  6.300 \\
    616.38 &  2.52126 &  4.53221 &  -1.286 &   6.20 &  6.295 &  6.331 &  6.282 &  6.350 &  6.286 &  6.339 &  6.290 &  6.362 \\
    616.64 &  2.52126 &  4.53134 &  -1.142 &   7.13 &  6.281 &  6.271 &  6.210 &  6.294 &  6.275 &  6.284 &  6.221 &  6.309 \\
    616.93\tablefootmark{a} &  2.52433 &  4.53347 &  -0.308 &  21.66 &  6.293 &  6.259 &  6.185 &  6.287 &  6.304 &  6.295 &  6.215 &  6.325 \\
    645.56 &  2.52299 &  4.44302 &  -1.340 &   5.65 &  6.319 &  6.309 &  6.261 &  6.326 &  6.311 &  6.322 &  6.272 &  6.340 \\
    647.17 &  2.52568 &  4.44095 &  -0.686 &   9.30 &  6.297 &  6.241 &  6.177 &  6.263 &  6.317 &  6.294 &  6.224 &  6.320 \\
    649.96 &  2.52299 &  4.43001 &  -0.818 &   8.75 &  6.334 &  6.286 &  6.225 &  6.307 &  6.348 &  6.330 &  6.264 &  6.355 \\
\noalign{\smallskip}
\multicolumn{13}{c}{\ion{Ca}{ii}} \\
\noalign{\smallskip}
    500.15 &  7.50514 &  9.98340 &  -0.507 &   1.35 &  6.231 &  6.235 &  6.238 &  6.290 &  6.237 &  6.241 &  6.243 &  6.299 \\
    645.69 &  8.43798 & 10.35764 &   0.412 &   1.85 &  6.296 &  6.291 &  6.269 &  6.329 &  6.296 &  6.293 &  6.269 &  6.331 \\
    732.39 &  0.00000 &  1.69241 &  -7.536 &   1.00 &  6.372 &  6.373 &  6.344 &  6.391 &  6.372 &  6.373 &  6.344 &  6.391 \\
    824.88 &  7.51484 &  9.01749 &   0.556 &   6.70 &  6.302 &  6.261 &  6.228 &  6.267 &  6.364 &  6.324 &  6.279 &  6.339 \\
    825.47 &  7.51484 &  9.01641 &  -0.398 &   1.80 &  6.296 &  6.284 &  6.270 &  6.303 &  6.322 &  6.310 &  6.291 &  6.333 \\
\noalign{\smallskip}
\hline
\noalign{\smallskip}
\end{tabular}
\end{center}
\tablefoot{\tablefoottext{a}{Doublet consisting of the $616.9042\,\nm$ and $616.9563\,\nm$ components. The quoted $gf$-value and equivalent width are the combined values.}}
\end{table*}

\begin{table*}[h]
\begin{center}
\caption{Line list for the Fe lines used for 3D non-LTE spectral line formation calculations to derive the solar photospheric abundances}
\label{t:linelist_Fe}
\begin{tabular}{rrrrr|cccc|cccc}
\hline
\noalign{\smallskip}
\multirow{2}{*}{$\lambda_{\text{air}}/\nm$} & \multirow{2}{*}{$E_{\text{lower}}/\mathrm{eV}$} & \multirow{2}{*}{$E_{\text{upper}}/\mathrm{eV}$} & \multirow{2}{*}{$\lggf$} & \multirow{2}{*}{$W/\mathrm{pm}$} &  \multicolumn{4}{c|}{Non-LTE} & \multicolumn{4}{c}{LTE} \\
 & & & & & 3D & \mtd{} & \marcs{} & HM & 3D & \mtd{} & \marcs{} & HM  \\
\noalign{\smallskip}
\hline
\noalign{\smallskip}
\multicolumn{13}{c}{\ion{Fe}{i}} \\
\noalign{\smallskip}
    444.55 &  0.08729 &  2.87550 &  -5.412 &   3.80 &  7.455 &  7.496 &  7.453 &  7.542 &  7.424 &  7.511 &  7.462 &  7.558 \\
    457.42 &  3.21119 &  5.92093 &  -2.350 &   3.87 &  7.451 &  7.448 &  7.408 &  7.475 &  7.440 &  7.450 &  7.408 &  7.479 \\
    524.70 &  0.08729 &  2.44956 &  -4.961 &   6.40 &  7.483 &  7.477 &  7.421 &  7.523 &  7.434 &  7.498 &  7.436 &  7.550 \\
    537.96 &  3.69460 &  5.99868 &  -1.420 &   6.17 &  7.433 &  7.383 &  7.329 &  7.413 &  7.429 &  7.399 &  7.340 &  7.433 \\
    548.31 &  4.15435 &  6.41493 &  -1.390 &   4.37 &  7.417 &  7.399 &  7.353 &  7.421 &  7.407 &  7.401 &  7.352 &  7.426 \\
    553.85 &  4.21758 &  6.45554 &  -1.540 &   3.73 &  7.488 &  7.477 &  7.436 &  7.495 &  7.480 &  7.478 &  7.435 &  7.498 \\
    560.02 &  4.26045 &  6.47375 &  -1.420 &   3.65 &  7.402 &  7.386 &  7.343 &  7.405 &  7.388 &  7.385 &  7.339 &  7.407 \\
    561.86 &  4.20888 &  6.41493 &  -1.250 &   5.00 &  7.444 &  7.418 &  7.369 &  7.440 &  7.432 &  7.421 &  7.368 &  7.446 \\
    566.13 &  4.28435 &  6.47375 &  -1.756 &   2.22 &  7.437 &  7.429 &  7.393 &  7.444 &  7.424 &  7.427 &  7.389 &  7.445 \\
    570.55 &  4.30128 &  6.47375 &  -1.355 &   3.96 &  7.428 &  7.411 &  7.367 &  7.430 &  7.418 &  7.413 &  7.366 &  7.434 \\
    577.51 &  4.22036 &  6.36665 &  -1.080 &   6.19 &  7.486 &  7.449 &  7.396 &  7.473 &  7.481 &  7.459 &  7.401 &  7.486 \\
    577.85 &  2.58811 &  4.73314 &  -3.440 &   2.04 &  7.430 &  7.446 &  7.406 &  7.467 &  7.416 &  7.451 &  7.409 &  7.474 \\
    578.47 &  3.39651 &  5.53924 &  -2.532 &   2.58 &  7.443 &  7.444 &  7.405 &  7.462 &  7.429 &  7.445 &  7.403 &  7.465 \\
    585.51 &  4.60759 &  6.72456 &  -1.478 &   2.20 &  7.441 &  7.432 &  7.395 &  7.445 &  7.430 &  7.430 &  7.392 &  7.446 \\
    595.67 &  0.85900 &  2.93985 &  -4.552 &   5.02 &  7.451 &  7.467 &  7.419 &  7.500 &  7.424 &  7.483 &  7.431 &  7.522 \\
    615.16 &  2.17595 &  4.19086 &  -3.282 &   4.88 &  7.459 &  7.455 &  7.409 &  7.481 &  7.443 &  7.466 &  7.417 &  7.494 \\
    624.06 &  2.22271 &  4.20888 &  -3.287 &   4.76 &  7.478 &  7.477 &  7.432 &  7.503 &  7.465 &  7.488 &  7.441 &  7.519 \\
    631.15 &  2.83159 &  4.79547 &  -3.141 &   2.66 &  7.492 &  7.502 &  7.463 &  7.524 &  7.481 &  7.509 &  7.467 &  7.532 \\
    649.89 &  0.95816 &  2.86539 &  -4.695 &   4.16 &  7.469 &  7.499 &  7.455 &  7.536 &  7.445 &  7.517 &  7.466 &  7.555 \\
    651.84 &  2.83159 &  4.73314 &  -2.448 &   5.72 &  7.405 &  7.378 &  7.325 &  7.405 &  7.392 &  7.392 &  7.334 &  7.421 \\
    657.42 &  0.99011 &  2.87550 &  -5.010 &   2.54 &  7.465 &  7.517 &  7.472 &  7.549 &  7.442 &  7.532 &  7.482 &  7.565 \\
    669.91 &  4.59311 &  6.44335 &  -2.101 &   0.78 &  7.487 &  7.481 &  7.448 &  7.488 &  7.476 &  7.480 &  7.445 &  7.490 \\
    679.33 &  4.07581 &  5.90041 &  -2.326 &   1.25 &  7.445 &  7.443 &  7.408 &  7.453 &  7.434 &  7.444 &  7.407 &  7.456 \\
    683.70 &  4.59311 &  6.40604 &  -1.687 &   1.77 &  7.486 &  7.474 &  7.440 &  7.482 &  7.475 &  7.475 &  7.438 &  7.485 \\
    684.37 &  4.54851 &  6.35967 &  -0.730 &   6.50 &  7.409 &  7.362 &  7.304 &  7.380 &  7.401 &  7.373 &  7.309 &  7.396 \\
    685.48 &  4.59311 &  6.40133 &  -1.926 &   1.29 &  7.434 &  7.439 &  7.395 &  7.451 &  7.423 &  7.438 &  7.391 &  7.451 \\
    685.81 &  4.60759 &  6.41493 &  -0.900 &   5.50 &  7.464 &  7.431 &  7.380 &  7.445 &  7.456 &  7.437 &  7.382 &  7.455 \\
    699.99 &  4.10337 &  5.87412 &  -1.380 &   5.79 &  7.482 &  7.456 &  7.405 &  7.472 &  7.477 &  7.466 &  7.412 &  7.486 \\
    700.80 &  4.17770 &  5.94640 &  -1.770 &   2.95 &  7.435 &  7.426 &  7.386 &  7.438 &  7.425 &  7.429 &  7.386 &  7.443 \\
    721.97 &  4.07581 &  5.79265 &  -1.430 &   4.90 &  7.399 &  7.366 &  7.318 &  7.381 &  7.393 &  7.379 &  7.326 &  7.397 \\
    740.17 &  4.18636 &  5.86098 &  -1.500 &   4.16 &  7.423 &  7.399 &  7.355 &  7.411 &  7.415 &  7.408 &  7.361 &  7.423 \\
    744.30 &  4.18636 &  5.85168 &  -1.640 &   3.59 &  7.450 &  7.431 &  7.390 &  7.441 &  7.439 &  7.436 &  7.392 &  7.449 \\
    791.29 &  0.85900 &  2.42543 &  -4.848 &   4.57 &  7.480 &  7.517 &  7.470 &  7.552 &  7.452 &  7.540 &  7.484 &  7.575 \\
    829.35 &  3.30092 &  4.79547 &  -2.203 &   5.85 &  7.486 &  7.461 &  7.416 &  7.480 &  7.475 &  7.475 &  7.425 &  7.495 \\
    857.18 &  5.00952 &  6.45554 &  -1.110 &   3.17 &  7.499 &  7.489 &  7.452 &  7.495 &  7.501 &  7.496 &  7.457 &  7.505 \\
    859.88 &  4.38646 &  5.82794 &  -1.200 &   5.67 &  7.419 &  7.394 &  7.343 &  7.407 &  7.413 &  7.404 &  7.350 &  7.421 \\
    887.60 &  5.02028 &  6.41674 &  -1.050 &   3.55 &  7.502 &  7.489 &  7.451 &  7.495 &  7.504 &  7.497 &  7.457 &  7.506 \\
    890.60 &  5.06378 &  6.45554 &  -1.220 &   2.47 &  7.492 &  7.484 &  7.448 &  7.488 &  7.492 &  7.490 &  7.453 &  7.496 \\
    910.36 &  4.17770 &  5.53924 &  -2.190 &   1.92 &  7.481 &  7.479 &  7.444 &  7.487 &  7.468 &  7.479 &  7.441 &  7.489 \\
    978.66 &  4.60759 &  5.87412 &  -1.840 &   1.82 &  7.466 &  7.460 &  7.425 &  7.466 &  7.456 &  7.461 &  7.424 &  7.469 \\
\noalign{\smallskip}
\multicolumn{13}{c}{\ion{Fe}{ii}} \\
\noalign{\smallskip}
    441.68 &  2.77846 &  5.58477 &  -2.570 &   8.07 &  7.419 &  7.316 &  7.283 &  7.370 &  7.427 &  7.319 &  7.286 &  7.374 \\
    450.83 &  2.85552 &  5.60489 &  -2.420 &   8.85 &  7.488 &  7.382 &  7.346 &  7.434 &  7.498 &  7.386 &  7.350 &  7.440 \\
    462.05 &  2.82812 &  5.51071 &  -3.210 &   5.40 &  7.429 &  7.362 &  7.335 &  7.406 &  7.433 &  7.363 &  7.336 &  7.407 \\
    465.70 &  2.89102 &  5.55261 &  -3.600 &   3.60 &  7.435 &  7.395 &  7.381 &  7.434 &  7.437 &  7.396 &  7.381 &  7.434 \\
    523.46 &  3.22131 &  5.58920 &  -2.180 &   8.80 &  7.484 &  7.361 &  7.318 &  7.403 &  7.498 &  7.368 &  7.323 &  7.410 \\
    526.48 &  3.23046 &  5.58477 &  -3.130 &   4.62 &  7.496 &  7.438 &  7.413 &  7.469 &  7.499 &  7.439 &  7.414 &  7.470 \\
    541.41 &  3.22131 &  5.51071 &  -3.580 &   2.73 &  7.471 &  7.436 &  7.421 &  7.464 &  7.473 &  7.436 &  7.421 &  7.464 \\
    643.27 &  2.89102 &  4.81790 &  -3.570 &   4.30 &  7.492 &  7.434 &  7.409 &  7.454 &  7.494 &  7.434 &  7.408 &  7.453 \\
    651.61 &  2.89102 &  4.79324 &  -3.310 &   5.69 &  7.545 &  7.454 &  7.422 &  7.474 &  7.548 &  7.453 &  7.422 &  7.473 \\
    722.24 &  3.88870 &  5.60489 &  -3.260 &   1.87 &  7.455 &  7.419 &  7.406 &  7.436 &  7.457 &  7.420 &  7.407 &  7.436 \\
    722.45 &  3.88919 &  5.60489 &  -3.200 &   2.10 &  7.469 &  7.431 &  7.417 &  7.447 &  7.470 &  7.431 &  7.417 &  7.447 \\
    751.58 &  3.90342 &  5.55261 &  -3.390 &   1.47 &  7.447 &  7.414 &  7.403 &  7.429 &  7.448 &  7.414 &  7.403 &  7.430 \\
    771.17 &  3.90342 &  5.51071 &  -2.500 &   5.04 &  7.457 &  7.376 &  7.345 &  7.388 &  7.462 &  7.378 &  7.347 &  7.391 \\
\noalign{\smallskip}
\hline
\noalign{\smallskip}
\end{tabular}
\end{center}
\end{table*}

\clearpage

\section{Isotopic abundances}

\longtab[1]{
\begin{longtable}{rlrrr|rlrrr|rlrrr}
\caption{Representative isotopic abundance fractions and logarithmic abundances by number ($\lgeps{H} \equiv 12.00$) at the time of birth of the Sun. The proto-solar abundances account for atomic diffusion of $0.07$\,dex for He and $0.064$\,dex for all heavier elements as predicted by \citet{2017ApJ...835..202V}, as well as radio-active decay. The elements for which no solar abundance is available are based on CI chondrites adjusted for the trend in condensation temperature between the Sun and meteorities (Sect. \ref{s:Tcond}).
\label{t:isotopes}}
\centering
\tabularnewline
\hline
\noalign{\smallskip}
Z & Elem. & A & fraction \% & $\log \epsilon$ & Z & Elem. & A & fraction \% & $\log \epsilon$ & Z & Elem. & A & fraction \% & $\log \epsilon$ \\
\noalign{\smallskip}
\hline
\noalign{\smallskip}
\endfirsthead
\caption{continued.}\\
\noalign{\smallskip}
\hline
\endhead
\hline
\endfoot
1       &        H      &       1       &       99.998  &       12.00   &         24      &       Cr      &       50      &       4.345   &       4.32    &         41      &       Nb      &       93      &       100.000 &       1.53    \\
1       &        H      &       2       &       0.002   &       7.22    &         24      &       Cr      &       52      &       83.789  &       5.61    &         42      &       Mo      &       92      &       14.649  &       1.11    \\
2       &       He      &       3       &       0.017   &       7.20    &         24      &       Cr      &       53      &       9.501   &       4.66    &         42      &       Mo      &       94      &       9.187   &       0.91    \\
2       &       He      &       4       &       99.983  &       10.98   &         24      &       Cr      &       54      &       2.365   &       4.06    &         42      &       Mo      &       95      &       15.873  &       1.14    \\
3       &       Li      &       6       &       4.850   &       1.98    &         25      &       Mn      &       55      &       100.000 &       5.48    &         42      &       Mo      &       96      &       16.673  &       1.17    \\
3       &       Li      &       7       &       95.150  &       3.28    &         26      &       Fe      &       54      &       5.845   &       6.29    &         42      &       Mo      &       97      &       9.582   &       0.93    \\
4       &       Be      &       9       &       100.000 &       1.44    &         26      &       Fe      &       56      &       91.754  &       7.49    &         42      &       Mo      &       98      &       24.292  &       1.33    \\
5       &        B      &       10      &       19.650  &       2.06    &         26      &       Fe      &       57      &       2.119   &       5.85    &         42      &       Mo      &       100     &       9.744   &       0.93    \\
5       &        B      &       11      &       80.350  &       2.67    &         26      &       Fe      &       58      &       0.282   &       4.97    &         44      &       Ru      &       96      &       5.540   &       0.56    \\
6       &        C      &       12      &       98.893  &       8.52    &         27      &       Co      &       59      &       100.000 &       5.00    &         44      &       Ru      &       98      &       1.870   &       0.09    \\
6       &        C      &       13      &       1.107   &       6.57    &         28      &       Ni      &       58      &       68.077  &       6.10    &         44      &       Ru      &       99      &       12.760  &       0.92    \\
7       &        N      &       14      &       99.775  &       7.89    &         28      &       Ni      &       60      &       26.223  &       5.68    &         44      &       Ru      &       100     &       12.600  &       0.91    \\
7       &        N      &       15      &       0.225   &       5.25    &         28      &       Ni      &       61      &       1.140   &       4.32    &         44      &       Ru      &       101     &       17.060  &       1.05    \\
8       &        O      &       16      &       99.776  &       8.75    &         28      &       Ni      &       62      &       3.635   &       4.82    &         44      &       Ru      &       102     &       31.550  &       1.31    \\
8       &        O      &       17      &       0.036   &       5.31    &         28      &       Ni      &       64      &       0.926   &       4.23    &         44      &       Ru      &       104     &       18.620  &       1.08    \\
8       &        O      &       18      &       0.188   &       6.03    &         29      &       Cu      &       63      &       69.150  &       4.08    &         44      &       Rh      &       103     &       100.000 &       0.84    \\
9       &        F      &       19      &       100.000 &       4.46    &         29      &       Cu      &       65      &       30.850  &       3.73    &         45      &       Pd      &       102     &       1.020   &       -0.36   \\
10      &       Ne      &       20      &       92.810  &       8.09    &         30      &       Zn      &       64      &       49.170  &       4.32    &         46      &       Pd      &       104     &       11.140  &       0.68    \\
10      &       Ne      &       21      &       0.242   &       5.51    &         30      &       Zn      &       66      &       27.730  &       4.07    &         46      &       Pd      &       105     &       22.330  &       0.98    \\
10      &       Ne      &       22      &       6.948   &       6.97    &         30      &       Zn      &       67      &       4.040   &       3.23    &         46      &       Pd      &       106     &       27.330  &       1.07    \\
11      &       Na      &       23      &       100.000 &       6.28    &         30      &       Zn      &       68      &       18.450  &       3.89    &         46      &       Pd      &       108     &       26.460  &       1.06    \\
12      &       Mg      &       24      &       78.965  &       7.51    &         30      &       Zn      &       70      &       0.610   &       2.41    &         46      &       Pd      &       110     &       11.720  &       0.70    \\
12      &       Mg      &       25      &       10.011  &       6.61    &         31      &       Ga      &       69      &       60.108  &       2.86    &         46      &       Ag      &       107     &       51.839  &       0.74    \\
12      &       Mg      &       26      &       11.025  &       6.66    &         31      &       Ga      &       71      &       39.892  &       2.68    &         47      &       Ag      &       109     &       48.161  &       0.71    \\
13      &       Al      &       27      &       100.000 &       6.49    &         32      &       Ge      &       70      &       20.520  &       3.00    &         47      &       Cd      &       106     &       1.245   &       -0.19   \\
14      &       Si      &       28      &       92.254  &       7.54    &         32      &       Ge      &       72      &       27.450  &       3.12    &         48      &       Cd      &       108     &       0.888   &       -0.34   \\
14      &       Si      &       29      &       4.672   &       6.24    &         32      &       Ge      &       73      &       7.760   &       2.57    &         48      &       Cd      &       110     &       12.470  &       0.81    \\
14      &       Si      &       30      &       3.073   &       6.06    &         32      &       Ge      &       74      &       36.520  &       3.25    &         48      &       Cd      &       111     &       12.795  &       0.82    \\
15      &        P      &       31      &       100.000 &       5.47    &         32      &       Ge      &       76      &       7.750   &       2.57    &         48      &       Cd      &       112     &       24.109  &       1.10    \\
16      &        S      &       32      &       94.850  &       7.16    &         33      &       As      &       75      &       100.000 &       2.35    &         48      &       Cd      &       113     &       12.227  &       0.80    \\
16      &        S      &       33      &       0.763   &       5.07    &         34      &       Se      &       74      &       0.860   &       1.29    &         48      &       Cd      &       114     &       28.754  &       1.17    \\
16      &        S      &       34      &       4.365   &       5.82    &         34      &       Se      &       76      &       9.230   &       2.32    &         48      &       Cd      &       116     &       7.512   &       0.59    \\
16      &        S      &       36      &       0.016   &       3.38    &         34      &       Se      &       77      &       7.600   &       2.24    &         48      &       In      &       113     &       4.281   &       -0.50   \\
17      &       Cl      &       35      &       75.800  &       5.25    &         34      &       Se      &       78      &       23.690  &       2.73    &         49      &       In      &       115     &       95.719  &       0.84    \\
17      &       Cl      &       37      &       24.200  &       4.76    &         34      &       Se      &       80      &       49.800  &       3.05    &         49      &       Sn      &       112     &       0.970   &       0.07    \\
18      &       Ar      &       36      &       84.281  &       6.37    &         34      &       Se      &       82      &       8.820   &       2.30    &         50      &       Sn      &       114     &       0.660   &       0.10    \\
18      &       Ar      &       38      &       15.695  &       5.64    &         35      &       Br      &       79      &       50.650  &       2.24    &         50      &       Sn      &       115     &       0.340   &       0.38    \\
18      &       Ar      &       40      &       0.024   &       2.83    &         35      &       Br      &       81      &       49.350  &       2.23    &         50      &       Sn      &       116     &       14.540  &       1.25    \\
19      &        K      &       39      &       93.133  &       5.10    &         36      &       Kr      &       78      &       0.365   &       0.75    &         50      &       Sn      &       117     &       7.680   &       0.97    \\
19      &        K\tablefootmark{*}     &       40      &       0.146   &         2.30    &       36      &       Kr      &       80      &       2.344   &         1.55    &       50      &       Sn      &       118     &       24.220  &         1.47    \\
19      &        K      &       41      &       6.721   &       3.96    &         36      &       Kr      &       82      &       11.686  &       2.25    &         50      &       Sn      &       119     &       8.590   &       1.02    \\
20      &       Ca      &       40      &       96.941  &       6.35    &         36      &       Kr      &       83      &       11.573  &       2.25    &         50      &       Sn      &       120     &       32.580  &       1.60    \\
20      &       Ca      &       42      &       0.647   &       4.17    &         36      &       Kr      &       84      &       56.895  &       2.94    &         50      &       Sn      &       122     &       4.630   &       0.75    \\
20      &       Ca      &       43      &       0.135   &       3.49    &         36      &       Kr      &       86      &       17.137  &       2.42    &         50      &       Sn      &       124     &       5.790   &       0.85    \\
20      &       Ca      &       44      &       2.086   &       4.68    &         37      &       Rb      &       85      &       70.875  &       2.24    &         51      &       Sb      &       121     &       57.210  &       0.80    \\
20      &       Ca      &       46      &       0.004   &       1.97    &         37      &       Rb\tablefootmark{*}     &       87      &       29.125  &         1.86    &       51      &       Sb      &       123     &       42.790  &         0.67    \\
20      &       Ca      &       48      &       0.187   &       3.64    &         38      &       Sr      &       84      &       0.563   &       0.64    &         52      &       Te      &       120     &       0.090   &       -0.85   \\
21      &       Sc      &       45      &       100.000 &       3.20    &         38      &       Sr      &       86      &       9.916   &       1.89    &         52      &       Te      &       122     &       2.550   &       0.60    \\
22      &       Ti      &       46      &       8.250   &       3.95    &         38      &       Sr      &       87      &       6.472   &       1.70    &         52      &       Te      &       123     &       0.890   &       0.14    \\
22      &       Ti      &       47      &       7.440   &       3.91    &         38      &       Sr      &       88      &       83.049  &       2.81    &         52      &       Te      &       124     &       4.740   &       0.87    \\
22      &       Ti      &       48      &       73.720  &       4.90    &         39      &        Y      &       89      &       100.000 &       2.27    &         52      &       Te      &       125     &       7.070   &       1.04    \\
22      &       Ti      &       49      &       5.410   &       3.77    &         40      &       Zr      &       90      &       51.450  &       2.37    &         52      &       Te      &       126     &       18.840  &       1.47    \\
22      &       Ti      &       50      &       5.180   &       3.75    &         40      &       Zr      &       91      &       11.220  &       1.70    &         52      &       Te      &       128     &       31.740  &       1.70    \\
23      &        V      &       50      &       0.250   &       1.36    &         40      &       Zr      &       92      &       17.150  &       1.89    &         52      &       Te      &       130     &       34.080  &       1.73    \\
23      &        V      &       51      &       99.750  &       3.96    &         40      &       Zr      &       94      &       17.380  &       1.89    &         53      &        I      &       127     &       100.000 &       1.55    \\
        &               &               &               &               &         40      &       Zr      &       96      &       2.800   &       1.10    &                 &               &               &               &               \\
        &               &               &               &               &                 &               &               &               &               &                 &               &               &               &               \\
        &               &               &               &               &                 &               &               &               &               &                 &               &               &               &               \\
54      &       Xe      &       124     &       0.127   &       -0.61   &         64      &       Gd      &       152     &       0.200   &       -1.55   &         74      &        W      &       180     &       0.120   &       -2.07   \\
54      &       Xe      &       126     &       0.113   &       -0.66   &         64      &       Gd      &       154     &       2.180   &       -0.52   &         74      &        W      &       182     &       26.500  &       0.28    \\
54      &       Xe      &       128     &       2.216   &       0.63    &         64      &       Gd      &       155     &       14.800  &       0.31    &         74      &        W      &       183     &       14.310  &       0.01    \\
54      &       Xe      &       129     &       27.438  &       1.72    &         64      &       Gd      &       156     &       20.470  &       0.46    &         74      &        W      &       184     &       30.640  &       0.34    \\
54      &       Xe      &       130     &       4.346   &       0.92    &         64      &       Gd      &       157     &       15.650  &       0.34    &         74      &        W      &       186     &       28.430  &       0.31    \\
54      &       Xe      &       131     &       21.771  &       1.62    &         64      &       Gd      &       158     &       24.840  &       0.54    &         75      &       Re      &       185     &       35.707  &       -0.08   \\
54      &       Xe      &       132     &       26.347  &       1.70    &         64      &       Gd      &       160     &       21.860  &       0.48    &         75      &       Re\tablefootmark{*}     &       187     &       64.293  &         0.18    \\
54      &       Xe      &       134     &       9.743   &       1.27    &         65      &       Tb      &       159     &       100.000 &       0.37    &         76      &       Os      &       184     &       0.020   &       -2.28   \\
54      &       Xe      &       136     &       7.900   &       1.18    &         66      &       Dy      &       156     &       0.056   &       -2.09   &         76      &       Os      &       186     &       1.597   &       -0.38   \\
55      &       Cs      &       133     &       100.000 &       1.09    &         66      &       Dy      &       158     &       0.095   &       -1.86   &         76      &       Os      &       187     &       1.559   &       -0.40   \\
56      &       Ba      &       130     &       0.110   &       -0.62   &         66      &       Dy      &       160     &       2.329   &       -0.47   &         76      &       Os      &       188     &       13.294  &       0.54    \\
56      &       Ba      &       132     &       0.100   &       -0.67   &         66      &       Dy      &       161     &       18.889  &       0.44    &         76      &       Os      &       189     &       16.216  &       0.62    \\
56      &       Ba      &       134     &       2.420   &       0.72    &         66      &       Dy      &       162     &       25.475  &       0.57    &         76      &       Os      &       190     &       26.368  &       0.83    \\
56      &       Ba      &       135     &       6.590   &       1.15    &         66      &       Dy      &       163     &       24.896  &       0.56    &         76      &       Os      &       192     &       40.947  &       1.02    \\
56      &       Ba      &       136     &       7.850   &       1.23    &         66      &       Dy      &       164     &       28.260  &       0.62    &         77      &       Ir      &       191     &       37.230  &       0.97    \\
56      &       Ba      &       137     &       11.230  &       1.38    &         67      &       Ho      &       165     &       100.000 &       0.54    &         77      &       Ir      &       193     &       62.770  &       1.19    \\
56      &       Ba      &       138     &       71.700  &       2.19    &         68      &       Er      &       162     &       0.139   &       -1.86   &         78      &       Pt\tablefootmark{*}     &       190     &       0.012   &         -2.25   \\
57      &       La\tablefootmark{*}     &       138     &       0.092   &         -1.86   &       68      &       Er      &       164     &       1.601   &         -0.80   &       78      &       Pt      &       192     &       0.782   &         -0.43   \\
57      &       La      &       139     &       99.908  &       1.17    &         68      &       Er      &       166     &       33.503  &       0.52    &         78      &       Pt      &       194     &       32.864  &       1.19    \\
58      &       Ce      &       136     &       0.185   &       -1.09   &         68      &       Er      &       167     &       22.869  &       0.35    &         78      &       Pt      &       195     &       33.775  &       1.20    \\
58      &       Ce      &       138     &       0.251   &       -0.96   &         68      &       Er      &       168     &       26.978  &       0.43    &         78      &       Pt      &       196     &       25.211  &       1.07    \\
58      &       Ce      &       140     &       88.450  &       1.59    &         68      &       Er      &       170     &       14.910  &       0.17    &         78      &       Pt      &       198     &       7.356   &       0.54    \\
58      &       Ce      &       142     &       11.114  &       0.69    &         69      &       Tm      &       169     &       100.000 &       0.16    &         79      &       Au      &       197     &       100.000 &       0.97    \\
59      &       Pr      &       141     &       100.000 &       0.81    &         70      &       Yb      &       168     &       0.126   &       -1.99   &         80      &       Hg      &       196     &       0.150   &       -1.67   \\
60      &       Nd      &       142     &       27.152  &       0.92    &         70      &       Yb      &       170     &       3.023   &       -0.61   &         80      &       Hg      &       198     &       10.040  &       0.16    \\
60      &       Nd      &       143     &       12.174  &       0.57    &         70      &       Yb      &       171     &       14.216  &       0.07    &         80      &       Hg      &       199     &       16.940  &       0.38    \\
60      &       Nd      &       144     &       23.798  &       0.86    &         70      &       Yb      &       172     &       21.754  &       0.25    &         80      &       Hg      &       200     &       23.140  &       0.52    \\
60      &       Nd      &       145     &       8.293   &       0.40    &         70      &       Yb      &       173     &       16.098  &       0.12    &         80      &       Hg      &       201     &       13.170  &       0.28    \\
60      &       Nd      &       146     &       17.189  &       0.72    &         70      &       Yb      &       174     &       31.896  &       0.42    &         80      &       Hg      &       202     &       29.740  &       0.63    \\
60      &       Nd      &       148     &       5.756   &       0.24    &         70      &       Yb      &       176     &       12.887  &       0.02    &         80      &       Hg      &       204     &       6.820   &       -0.01   \\
60      &       Nd      &       150     &       5.638   &       0.24    &         71      &       Lu      &       175     &       97.180  &       0.15    &         81      &       Tl      &       203     &       29.515  &       0.45    \\
62      &       Sm      &       144     &       3.066   &       -0.50   &         71      &       Lu\tablefootmark{*}     &       176     &       2.820   &         -1.38   &       81      &       Tl      &       205     &       70.485  &         0.83    \\
62      &       Sm\tablefootmark{*}     &       147     &       15.384  &         0.20    &       72      &       Hf      &       174     &       0.161   &         -1.88   &       82      &       Pb      &       204     &       1.400   &         0.16    \\
62      &       Sm      &       148     &       11.199  &       0.07    &         72      &       Hf      &       176     &       5.240   &       -0.37   &         82      &       Pb      &       206     &       24.100  &       1.40    \\
62      &       Sm      &       149     &       13.758  &       0.15    &         72      &       Hf      &       177     &       18.580  &       0.18    &         82      &       Pb      &       207     &       22.100  &       1.36    \\
62      &       Sm      &       150     &       7.337   &       -0.12   &         72      &       Hf      &       178     &       27.280  &       0.35    &         82      &       Pb      &       208     &       52.400  &       1.73    \\
62      &       Sm      &       152     &       26.619  &       0.44    &         72      &       Hf      &       179     &       13.630  &       0.05    &         83      &       Bi      &       209     &       100.000 &       0.65    \\
62      &       Sm      &       154     &       22.637  &       0.37    &         72      &       Hf      &       180     &       35.120  &       0.46    &         90      &       Th\tablefootmark{*}     &       232     &       100.000 &         0.19    \\
63      &       Eu      &       151     &       47.810  &       0.26    &         73      &       Ta      &       180     &       0.012   &       -4.00   &         92      &        U\tablefootmark{*}     &       234     &       0.002   &         -4.73   \\
63      &       Eu      &       153     &       52.190  &       0.30    &         73      &       Ta      &       181     &       99.988  &       -0.07   &         92      &        U\tablefootmark{*}     &       235     &       24.167  &         -0.65   \\
                                        &               &               &                 &               &               &               &               &                 &               &       92      &        U\tablefootmark{*}         &       238     &       75.831  &       -0.16   \\
\noalign{\smallskip}
\hline
\end{longtable}
\tablefoot{\tablefoottext{*}{Long-lived radio-active element}}
}

\end{appendix}

\label{lastpage}
\end{document}